\documentclass[fleqn,usenatbib]{mnras}

\usepackage{newtxtext,newtxmath}

\usepackage[T1]{fontenc}
\usepackage{ae,aecompl}
\usepackage{xcolor}
\usepackage{ulem}
\usepackage{soul}
\usepackage{amsmath}

\usepackage{subfig}

\usepackage{graphicx}	
\usepackage{amsmath}	
\usepackage{verbatim}
\usepackage{bm}
\usepackage{multirow}

\usepackage{booktabs}

\usepackage{tikz}

\graphicspath{{./plots/}}

\makeatletter
\DeclareRobustCommand{\I}{%
	\mbox{\check@mathfonts\fontsize\sf@size\z@\selectfont I}%
}
\DeclareRobustCommand{\V}{%
	\mbox{\check@mathfonts\fontsize\sf@size\z@\selectfont V}%
}
\makeatother

\title[The dearth of high mass DAZs]{The dearth of high-mass hydrogen-atmosphere metal-polluted white dwarfs within 40\,pc}

\author[T. Cunningham et. al]
{Tim Cunningham,$^{1}$\thanks{E-mail: tim.cunningham@cfa.harvard.edu }\thanks{NASA Hubble Fellow.}
Pier-Emmanuel Tremblay,$^{2}$ 
Mairi O'Brien,$^{2}$
Evan B. Bauer,$^{1}$
Mark A. Hollands,$^{2}$ 
\newauthor{Detlev Koester,$^{3}$
Scott J. Kenyon$^{1}$, 
David Charbonneau$^{1}$, 
Dimitri Veras$^{2}$,
Muhammad Furqaan Yusaf$^{4}$}
\\
$^{1}$Center for Astrophysics | Harvard \& Smithsonian, 60 Garden St., Cambridge, MA 02138, USA\\
$^{2}$Department of Physics, University of Warwick, Coventry, CV4 7AL, UK\\
$^{3}$Institut f\"ur Theoretische Physik und Astrophysik, University of Kiel, 24098 Kiel, Germany\\
$^{3}$King’s College London, University of London, Strand, London, WC2R 2LS, UK\\
}

\date{Accepted XXX. Received YYY; in original form ZZZ}

\pubyear{2024}

\begin{document}
\label{firstpage}
\pagerange{\pageref{firstpage}--\pageref{lastpage}}
\maketitle

\begin{abstract}
We present a population synthesis model which addresses the different mass distributions of the metal-polluted and non-metal-polluted hydrogen-atmosphere white dwarfs identified in volume-limited samples. Specifically,  metal-pollution has been observed to be rare in white dwarfs more massive than $\approx$0.7\,$M_{\odot}$. Our population synthesis model invokes episodic accretion of planetary debris onto a synthetic population of white dwarfs. We find that the observed difference can be explained in the regime where most debris disks last for $10^4$--$10^6$ years. This is broadly consistent with observational estimates that disk lifetimes are on the order 10$^5$--10$^7$ years. We also invoke an alternate model which explores an upper limit on planetary system formation and survival around the intermediate-mass progenitors of the more massive white dwarfs. In this scenario, we find an upper limit on the polluted white dwarf mass of $M_{\rm wd}<0.72^{+0.07}_{-0.03}$\,M$_{\odot}$. This implies an empirical maximum progenitor mass of $M_{\rm ZAMS}^{\rm max}=2.9^{+0.7}_{-0.3}$\,M$_{\odot}$. This value is consistent with the maximum reliable host star mass of currently known exoplanet systems. We conclude by imposing these two results on the sample of He-atmosphere white dwarfs within 40\,pc. We find that both scenarios are capable of providing a consistent solution to the full sample of H- and He-atmosphere white dwarfs. 

\end{abstract}




\section{Introduction}
\label{sec:intro}

White dwarfs are the stellar remnants of main sequence stars with masses between 1--8\,$M_{\odot}$. The high surface gravities of white dwarf atmospheres leads to rapid elemental sedimentation, as heavy elements sink below the photosphere on short timescales. Thus, white dwarfs should have pristine atmospheres of hydrogen or helium. However, a large fraction of white dwarfs have a substantial source of heavy elements which they accrete as they cool \citep{koester14,zuckerman03,zuckerman10,hollands18b,koester09,coutu19}, with 25--50\% of white dwarfs showing the presence of planetary debris in their atmospheres \citep{koester14,OuldRouis2024}. The origin of this material is the remnants of the planetary system they hosted as main sequence stars \citep{veras16,jura2008} and numerous studies have shown that measurements of photospheric metal abundances can yield the parent body composition of accreted rocky material \citep{farihi2016review,harrison2018,hollands18b}.  To date, there are over 1700 white dwarfs which show evidence of photospheric metal lines; evidence of the accretion of planetary debris \citep{williams2024}.

Models describing the transport of metals from the white dwarf surface can be coupled with measured photospheric abundances to infer a mass flux of a given metal through the atmosphere. This mass flux serves as a proxy for the accretion rate of material onto the surface. These models are the solution to diffusion equations which account for gravitational settling, chemical diffusion, thermal diffusion and radiative diffusion \citep{paquette86a,dupuis1992,koester09}. When white dwarfs are young (warm), energy transport in their atmospheres is dominated by radiation; sinking timescales of metals are days \citep{koester09}. As white dwarfs cool, energy transport in their outer layers becomes dominated by convection, rather than radiation. The convective motions dominate over the microscopic diffusion of metals, suspending metals at the surface with an $e$-folding time set by the microscopic diffusion physics at the base of the convection zone. For H-atmosphere white dwarfs, convection sets in around effective temperatures of 18\,000\,K, whilst for He-atmosphere white dwarfs convection starts earlier, at effective temperatures of 30\,000--50\,000\,K \citep{tassoul90,cukanovaite19}. As the white dwarf cools, its convection zone grows in mass, and diffusion timescales at its base increase. By the time a H-atmosphere white dwarf has cooled to 6000\,K, its diffusion timescale is predicted to reach $\sim$10$^4$\,years, whilst He-atmosphere white dwarfs of comparable temperature are predicted to have diffusion timescales of $\sim$$10^6$\,years. In addition to atmospheric composition and surface temperature, surface gravity -- and thus white dwarf mass -- also affects diffusion timescales. Higher-mass white dwarfs tend to have shorter diffusion timescales \citep{koester2020}\footnote{\url{https://www1.astrophysik.uni-kiel.de/~koester/astrophysics/astrophysics.html}}.
Finally, depending on the atmospheric composition and stellar parameters, additional transport processes may significantly lengthen these diffusion times, such as convective overshoot and radiative levitation \citep{cunningham19,OuldRouis2024}. Thus, for a given mass flux through the atmosphere, the measured photospheric abundances can vary significantly as a function of white dwarf parameters.

White dwarfs typically have hydrogen or helium atmospheres, with $\approx$80\% having hydrogen atmospheres and $\approx$20\% having helium atmospheres \citep{cunningham20,bedard2020}. Metal-pollution has been identified in both hydrogen and helium atmosphere white dwarfs. Using optical spectroscopy and data from \textit{Gaia}, \citet{mccleery2020} and \citet{obrien2024} recently examined a sample of all white dwarfs within 40\,pc of the Sun. They found that the mean mass of hydrogen-atmosphere white dwarfs 
(mean mass 0.66\,M$_{\odot}$) 
is significantly higher than that of their metal-polluted counterparts 
(mean mass 0.59\,M$_{\odot}$). 
This disparity primarily arises from the lack of almost any metal-polluted, hydrogen-atmosphere white dwarfs with masses exceeding 0.8\,M$_{\odot}$, whereas pristine hydrogen atmosphere white dwarfs 
with masses as large as 1.3\,M$_{\odot}$ are found in the sample. 
With only one exception\footnote{Within 40\,pc, the only metal-polluted hydrogen atmosphere white dwarf with a mass exceeding 0.8\,M$_{\odot}$ is WD\,J085021.30$-$584806.21, with a mass of $M_{\rm wd}=1.11\pm0.06$\,M$_{\odot}$ and spectral type DZA, where the uncertainty is the systematic between different fitting techniques. Spectral types DZA can have atmospheres dominated by either hydrogen or helium. This white dwarf is determined to have a hydrogen atmosphere from spectroscopic fits \citep{OBrien2023}}, there are no hydrogen atmosphere metal-polluted white dwarfs within 40\,pc with a mass exceeding 0.8\,M$_{\odot}$.
Using ultraviolet (UV) spectroscopy, \citet{OuldRouis2024} more recently identified a similar high-mass pollution deficit from a sample of warm H-atmosphere white dwarfs ($T_{\rm eff}>$12\,000\,K) observed with the Cosmic Origins Spectrograph \citep[COS;][]{green2012COS} onboard the Hubble Space Telescope (HST). The authors identify a significant scarcity of metal-polluted white dwarfs above 0.7\,M$_{\odot}$, whilst at lower masses the actively accreting percentage was constrained to be $44\pm6$\%.

Theoretical studies show that solid material with sizes larger than $\sim$centimeters and orbital separations of tens of au from the host star probably survive the evolution of a main sequence star into a white dwarf, whilst smaller, closer-in grains are likely to be blown away by stellar winds \citep{zotos2020,bonsor2010}, or broken apart by the YORP (Yarkovsky-O'Keefe-Radviesvki-Paddock) effect \citep{veras2014,veras2020-yorp}. The likelihood of specifically intermediate-mass and high-mass main sequence stars forming and retaining planets has been investigated by means of theoretical studies \citep{veras2020-mass,kunitomo2021,johnston2024}. For instance, \citet{veras2020-mass} predicted survival limits for major and minor planets around the most massive white dwarf progenitors; O- and B-type stars with zero-age main sequence masses of 6--8\,M$_{\odot}$. They found that major and minor planets beyond 3--6\,au are likely to survive until the white dwarf phase, either as intact planets, or fragmented debris.

Here, we 
explore whether the scarcity of higher-mass metal-polluted white dwarfs can be attributed to biases in white dwarf atmospheric physics. Specifically, metals are predicted to sink more rapidly in the atmospheres of higher-mass white dwarfs, which 
could naturally make metal-polluted white dwarfs more rare. Alternatively, we consider whether a mass dependence on the formation and evolutionary scenarios for planetary systems may provide a more complete explanation. In Section\,\ref{sec:sample} we describe the 40\,pc white dwarf sample used in this work. In Section\,\ref{sec:PopSyn} we describe the population synthesis techniques we use to model the effects of atmospheric transport and episodic accretion on a representative synthetic population. In Section\,\ref{sec:results} we present the results of our population synthesis and episodic accretion model, and discuss the interpretation with respect to the observed sample and wider population of white dwarfs. In Section\,\ref{sec:helium} we compare our results to those of the He-atmosphere population. In Section\,\ref{sec:conclusions} we present our conclusions.

\section{Sample}
\label{sec:sample}
We make use of the volume-complete spectroscopic sample of white dwarfs within 40\,pc of the Sun, as described in \citep{limoges15,tremblay2020,mccleery2020,Gentile2021,obrien2024,cunningham2024}. The sample is built upon a catalogue of photometric and astrometric white dwarf candidates identified in \textit{Gaia} \citep{ngf19,Gentile2021}. The spectroscopically-complete sample comprises a large archival dataset and extensive targeted observations of white dwarfs in both the northern and southern hemispheres. \citet{OBrien2023,obrien2024} recently demonstrated that the sample includes medium-resolution optical spectroscopy with a resolving power of at least $R\approx1200$ and a signal-to-noise of S/N\,$>$\,30 for at least 99\% of the 1083 \textit{Gaia} white dwarf candidates within 40\,pc. The volume-limited 40pc sample provides an unbiased sample of white dwarfs in the Solar neighbourhood. Compared to UV samples and magnitude-limited samples, the sample used in this work allows the study of evolved planetary systems across a wide range of cooling ages, from 20\,Myr--10\,Gyr \cite{obrien2024}.

We focus on white dwarfs with hydrogen-dominated atmospheres. We make use of the corrected photometric \textit{Gaia} atmospheric parameters derived from \citet{obrien2024}, where the correction accounts for the ``low-mass problem'' in cool white dwarfs ($T_{\rm eff}$\,$\lessapprox$\,$6000$\,K), possibly due to an incorrect Lyman\,$\alpha$ red wing opacity. In all cases, we apply a mass cut of $M_{\rm wd}>0.53$\,M$_{\odot}$ to remove likely unresolved double degenerates. We also impose a minimum effective temperature of $T_{\rm eff}$\,$>$\,5000\,K. Below this threshold, determining atmospheric composition becomes considerably more challenging due to the minimal strength of hydrogen absorption lines. Finally, all DAZ white dwarfs have continuum gas opacities dominated by hydrogen and the major electron donor is hydrogen, hence trace metals have a negligible effect on the atmospheric parameter determinations from \textit{Gaia} data.

\begin{figure}
	\centering

 	\subfloat{\includegraphics[width=1.\columnwidth]{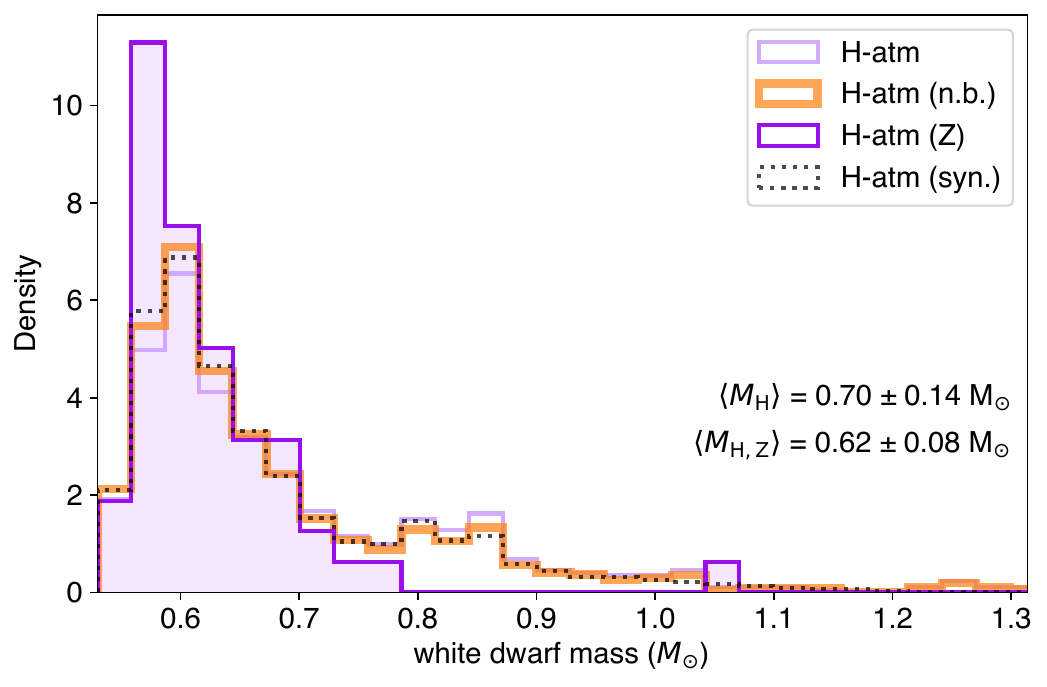}} \\
 	\subfloat{\includegraphics[width=1.\columnwidth]{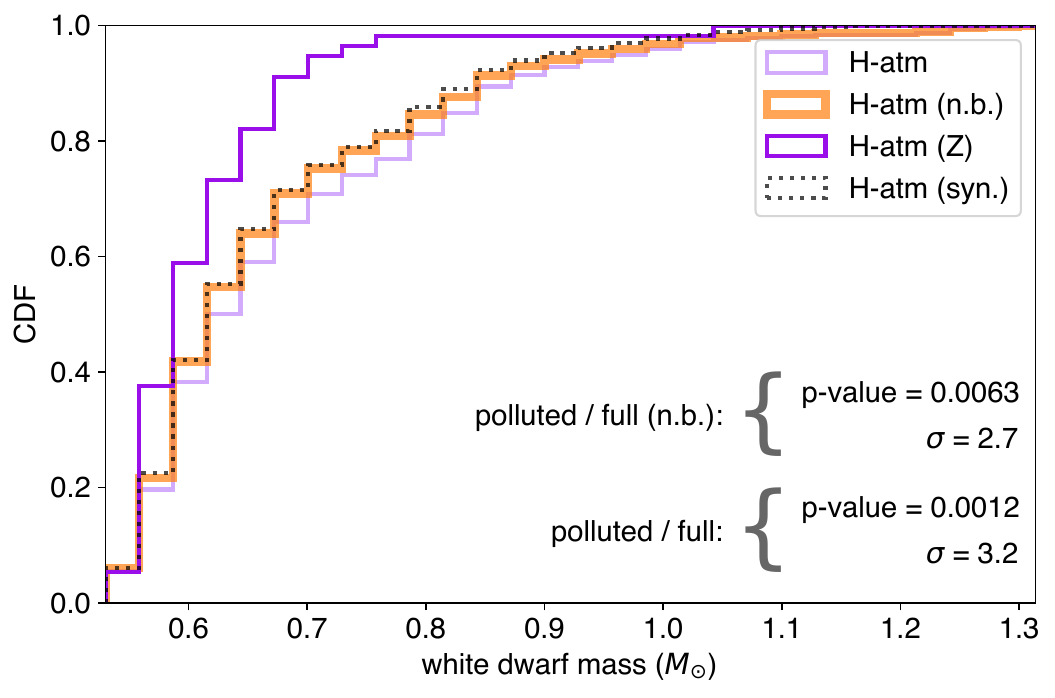}}
	\caption{\textbf{Top}: Histogram of the observed 40\,pc sample of DA (H-atm) and DAZ/DZA (metal-polluted, H-atm) white dwarfs, in light lilac and filled-purple, respectively. Following the methodology of \citet{cunningham2024}, the orange histogram is the result of merger removal from the H-atmosphere distribution; labelled ``H-atm (n.b. - no binaries)''. \textbf{Bottom}: Cumulative distribution functions for the three samples in the top panel (same colours). We show the results of a two-sided KS test in the panel, when comparing the polluted distribution to the H-atm and H-atm (n.b.) distributions. We also show the synthetic DA distribution with the black dotted curve.}
	\label{fg:mass-hist}
\end{figure}

Among the 56 H-atmosphere metal-polluted white dwarfs in the 40\,pc sample, 6 are of spectral type DZA, whilst the remaining 50 are DAZ. We consider this DAZ/DZA sample to be the H-atmosphere metal-polluted sample studied in this work. Fig.\,\ref{fg:mass-hist} shows the mass distribution of DA white dwarfs, the distribution corrected for mergers following the methodology of \citet{cunningham2024} using the binary population synthesis models of \citet{temmink2020}, and the mass distribution of metal-polluted hydrogen-rich DAZ/DZA white dwarfs. It was first identified by \citet{mccleery2020} that the distributions of polluted and non-polluted hydrogen-atmosphere white dwarfs exhibited a significant statistical disagreement. In the lower panel of Fig.\,\ref{fg:mass-hist} we show the cumulative distribution functions (CDF) for the three samples. We perform a two-sided Kolmogorov-Smirnov (KS) test to estimate the probability that the samples are drawn from the same underlying distribution. We find that for the default case, the two samples differ in excess of 3.2$\sigma$, whilst in the binary-removed case, the disagreement is reduced to 2.7$\sigma$. The main origin of this disagreement is the significant deficit of high mass ($M_{\rm wd}>0.8$\,$M_{\odot}$) DAZ/DZA white dwarfs. This absence is evident in the top panel, where the high-mass bump in the DA distribution at 0.8\,$M_{\odot}$ is non-existent in the DAZ/DZA distribution.

\section{Population Synthesis Techniques}
\label{sec:PopSyn}

We create a synthetic population of white dwarfs to describe the observed population of H-atmosphere white dwarfs following the methodology described by \citet{cunningham2024}. This involves creating an initial population of main sequence stars, using the following assumptions:
1) constant stellar formation rate, which has recently been shown to be appropriate for the 40\,pc sample \citep{cukanovaite2023}, 2) \citet{salpeter1955} initial mass function (IMF), 3) main sequence lifetimes determined for Solar metallicity from the models of \citet{hurley2000} with a 4) Galactic disk age of 10\,Gyr \citep{cukanovaite2023}. We then adopt the self-consistent initial-final mass relation (IFMR) of \citet{cunningham2024} to generate a sample of synthetic white dwarfs. For the purposes of this study, we do not take into account the uncertainties on the IFMR or the age of the Galactic disk as none of these ingredients have much impact on the incidence of metal pollution.

\subsection{Stellar mergers}

Between 20--45\% of white dwarfs are expected to be the product of stellar mergers, either two white dwarfs, a white dwarf and a main-sequence star, or two main-sequence stars, with the total percentage of mergers broadly correlated with mass \citep{temmink2020}. Following the methodology of \citet{cunningham2024}, we probabilistically alter the observed mass distribution of DA white dwarfs to remove stellar mergers, as predicted by the population synthesis models of \citet{temmink2020}. The motivation for this correction lies in the fact that a stellar merger implies the formation and evolution of a planetary system within a binary star system. The dynamical evolution of such a planetary system, both before and after the merger, is anticipated to differ significantly from that of a planetary system orbiting a single white dwarf. As such, it is possible that stellar mergers could be the cause for fewer high-mass DAZ/DZA white dwarfs and thus it is a bias for which we must account. Fig.\,\ref{fg:mass-hist} shows the distribution of DA (H-atm) white dwarfs, before and after this stochastic merger removal procedure. It is evident that the high-mass bump has been reduced in amplitude, but still remains in disagreement with the DAZ/DZA mass distribution. Thus, the predicted stellar merger fraction does not provide an adequate explanation for the observations. In the following, we include the merger correction as default in the simulations.

\begin{figure}
    \centering
    \begin{tikzpicture}

    \draw[->] (0,3) -- (0,2+3);
    \draw[->] (0,3) -- (7,3);

    \draw
      (0.3, 0+3) |- (1.3, 1.4+3);
    \draw[]
      (1.3, 1.4+3) -- (1.3,0+3);
      
    \draw
      (0.3+3.8, 0+3) |- (1.3+3.8, 1.4+3);
    \draw[]
      (1.3+3.8, 1.4+3) -- (1.3+3.8,0+3);

    \draw[->] (0,0) -- (0,2.6);
    \draw[->] (0,0) -- (7,0);

    \draw[dashed]
      (1.3, 2) -- (3,0);
    \draw[dotted]
      (1.3, 2) -- (1.3,0);
    \draw[dotted]
      (2.05, 1.1) -- (2.05,0);
   \draw[dotted]
      (0.3, 0) -- (0.3,2.4);

    \draw (0.3,0) .. controls (0.5,1.3) and (0.6,1.8) .. (1.3,2);
    \draw[dotted]
      (0.52, 0) -- (0.52,1.1);

    \draw[dashed]
      (1.3+3.8, 2) -- (3+3.8,0);
    \draw[dotted]
      (1.3+3.8, 2) -- (1.3+3.8,0);
    \draw[dotted]
      (2.05+3.8, 1.1) -- (2.05+3.8,0);
    \draw (0.3+3.8,0) .. controls (0.5+3.8,1.3) and (0.6+3.8,1.8) .. (1.3+3.8,2);
    \draw[dotted]
      (0.52+3.8, 0) -- (0.52+3.8,1.1);
    \draw[dotted]
      (0.3+3.8, 0) -- (0.3+3.8,2.4);
      
    \node[] at (0.33,-0.3) {$t_0$};
    \node[] at (0.56,-0.3) {$t_{\rm z}^*$};
    \node[] at (1.33,-0.3) {$t_{\rm d}$};
    \node[] at (2.05,-0.3) {$t_{\rm z}$};
    \node[] at (0.33+3.8,-0.3) {$t_{\rm p}$};
    \node[] at (1.33+3.8,-0.3) {$t_{\rm p} + t_{\rm d}$};

    \node[] at (0.33,3-0.3) {$t_0$};
    \node[] at (1.33,3-0.3) {$t_{\rm d}$};
    \node[] at (0.33+3.8,3-0.3) {$t_{\rm p}$};
    \node[] at (1.33+3.8,3-0.3) {$t_{\rm p} + t_{\rm d}$};

    \draw[dotted]
      (0,2.3) -- (3+3.8,2.3);
    \draw[dotted]
      (0,2) -- (3+3.8,2);
    \draw[dotted]
      (0,1.1) -- (7,1.1);
    \node[] at (-0.35,1.1) {$\chi_{{\rm det},i}$};
    \node[] at (-0.35,2.3) {$\chi_{{\rm ss},i}$};
    \node[] at (-0.4,2.0) {$\chi_{{\rm max},i}$};
    \node[] at (2.2,1.6) {$e^{-t/\tau_i}$};
    \node[] at (-0.3,1.4+3) {$\dot{M_i}$};
    
    \node[] at (3.5,-0.6) {\large time};
    \node[rotate=90] at (-0.99, 1.4) {\large log(surface fraction)};

    \node[rotate=90] at (-0.99, 1+3) {\large accretion rate};
    
    \end{tikzpicture}
    \caption{Schematic of our simple accretion model. \textbf{Top}: The accretion turns on at a constant accretion rate, $\dot{M}$, for a time, $t_{\rm d}$, and turns off for the remainder of the full accretion period, $t_{\rm p}$. In our model, the accretion is periodic. \textbf{Bottom}: Illustration of how the photospheric abundance of a specific element responds to the periodic accretion. Initially, abundance grows as $1-e^{-t/\tau_i}$, where $\tau_i$ is the sinking timescale for a specific metal, $i$. The abundance becomes detectable at time $t_{\rm z}^*$ when the abundance exceeds the detection limit and stops being detectable at $t_{\rm z}$. From the onset of accretion, the abundance continues to grow exponentially for a few $e$-folding times ($\tau_i$). If the duration of the accretion episode is long enough, the abundance will reach a steady state between accretion and diffusion. Depicted here is the case in which accretion stops before a steady state is reached. At the cessation of accretion, the abundance decays as $e^{-(t-t_{\rm d})/\tau_i}$ until the time $t_{\rm p}$. By imposing a periodic boundary condition ($\chi_{0,i}\equiv\chi_{{\rm p},i}$), the equation can be solved in a closed form. $\chi_{{\rm det},i}$, $\chi_{{\rm max},i}$ and $\chi_{{\rm ss},i}$ are the detectable, maximum and steady-state photospheric abundances for a specific metal, $i$.}
    \label{fig:tikz-model-increasing}
\end{figure}
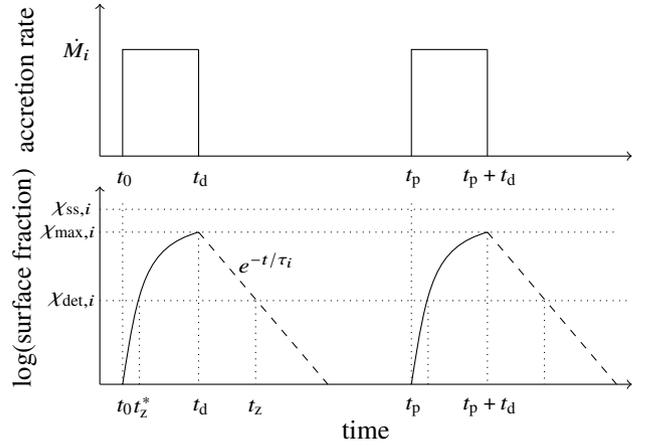

\subsection{Mathematical model}
We adopt a simple toy model of planetary debris accretion across the population of white dwarfs. Our starting assumption is that white dwarfs accrete for a finite time, multiple times over their lifetime. For mathematical ease, we assume these accretion events are periodic. The physical motivation for this model is that debris disks around white dwarfs are thought to live for $10^5$--$10^7$ years \citep{girven12,cunningham2021}.

The model is indicated in Fig.\,\ref{fig:tikz-model-increasing}. Accretion is turned on at $t=t_0$ and persists with constant accretion rate until $t_{\rm d}$. At the cessation of accretion, the surface abundances decay exponentially, with a $e$-folding time set by the metal sinking timescale, $\tau_i$, where the $i$ subscript denotes a specific metal (e.g. Ca, Mg, Fe, etc.). In our model, the metals in a white dwarf atmosphere are considered detectable as long as the surface abundance fraction, $\chi_i$, exceeds the detection limit, $\chi_{{\rm det},i}$. The next accretion event occurs at time $t_{\rm p}$, which formally defines the total accretion period in our model. The advantage of this simple model is that a closed form exists, allowing for an analytic determination of the fraction of an accretion period over which the white dwarf will present detectable metals. This provides a simple metric for the probability that any given white dwarf should have detectable photospheric metals borne of planetary debris accretion. 
In our model, we assume that all accretion episodes for a given white dwarf turn on at the same, constant, accretion rate $\dot{M}$. 

Following the framework developed by \citet{paquette86a, dupuis1992, koester09}, the solution to the differential equation governing the evolution of the surface abundance of a specific metal is given as follows:
\begin{equation}
\chi_i(t) = \chi_i(t_0) e^{-t/\tau_i} + \frac{\tau_i \dot{M_i}}{M_{{\rm cvz}}} \left(1-e^{-t/\tau_i} \right)\, ,
\label{eq:diff-eq}
\end{equation}
where $M_{{\rm cvz}}$ is the mass of the 
convection zone, and $\chi_i$, $\tau_i$, $\dot{M_i}$ refer to the values for a given metal, $i$.

In our periodic accretion model, we define the time-averaged probability of a given white dwarf having detectable metals as the time for which the surface abundance exceeds a detection threshold, divided by the total accretion period. This can be written as
\begin{align}
    P_{\rm det} &\equiv \frac{t_{\rm z}-t_{\rm z}^*}{t_{\rm p}} \\
    &= \frac{\tau_i}{t_{\rm p}}\ln\left( \frac{\left[\chi_{0,i} + \chi_{{\rm ss}, i} \left( e^{t_{\rm d}/\tau_i} - 1\right)\right]\left(\chi_{{\rm det},i}-\chi_{{\rm ss},i}\right)}{\chi_{{\rm det},i}\left(\chi_{0,i} - \chi_{{\rm ss},i}\right)} \right)~.
    \label{eq:prob}
\end{align}
where $t_{\rm d}$ is the disk lifetime (or time at which accretion turns off), $t_{\rm p}$ is the time for the full accretion period, $\chi_{\rm det}$ is the detection limit on surface abundance (see Fig.\,\ref{fg:obs-EW-detect}), $t_{\rm z}^*$ is the time at which the metal first becomes detectable, $t_{\rm z}$ is the time at which the metal stops being detectable, $\chi_{0,i}$ is the abundance when accretion turns on, 
and $\chi_{{\rm ss},i}$ is the abundance once the system has reached a steady state between accretion and diffusion (see Appendix~\ref{appendixA} for full derivation).
Eq\,\eqref{eq:prob} is valid providing that providing that $\chi_{{\rm ss},i}(1-e^{-t_{\rm d}/\tau_i})> \chi_{{\rm det},i}$. The term on the left hand side of the inequality is the maximum abundance ever reached (see Fig.\,\ref{fig:tikz-model-increasing}). If the inequality is not satisfied, the detection probability goes to zero ($P_{\rm det}=0$).
The steady state surface abundance of a given metal, $i$, is given by
\begin{equation}
    \chi_{{\rm ss},i} = \frac{\tau_i \dot{M_i}}{M_{\rm cvz}}~.
\end{equation}
Throughout this study, we treat Ca as the proxy for accretion, motivated by the the fact that the vast majority of the metal-polluted white dwarfs in our sample are identified through Ca absorption. We make the assumption that all material being accreted has bulk Earth composition, such that the steady state Ca abundance can be expressed in terms of the total accretion rate, $\dot{M}$, as
\begin{equation}
    \chi_{{\rm ss, Ca}} = f_{\rm Ca} \times \frac{\tau_{\rm Ca} \dot{M}}{M_{\rm cvz}}~.
    \label{eq:Xss}
\end{equation}
where $f_{\rm Ca}$ is the Ca fraction of the accreted material. For bulk Earth, we adopt $f_{\rm Ca}=0.016$, or 1.6\% \citep{mcdonough1995}.

The detection probability, $P_{\rm det}$ (Eq.\,\ref{eq:prob}), is thus a function of 
six parameters, two of which are determined by stellar models, three free parameters, and an observational detection limit ($\chi_{{\rm det},i}$). The stellar model parameters are the diffusion timescales ($\tau_i$) and convection zone mass ($M_{\rm cvz}$), for which we interpolate publicly-available grids \citep{koester2020}. For each synthetic white dwarf, these two parameters are thus fixed by the 
synthetic stellar parameters (effective temperature and surface gravity).
The three free parameters are the duration of the accretion episode ($t_{\rm d}$; or lifetime of the disk), the period of the total accretion cycle ($t_{\rm p}$), and the total accretion rate ($\dot{M}$). 

\subsection{Accretion rates}
There is strong observational evidence that instantaneous accretion rates span a range of 5 orders of magnitude \citep{farihi2016review}. 
In our model, instead of keeping accretion as a truly free parameter, we adopt a distribution of accretion rates from observational constraints.
Using an unbiased sample of DA and non-DA white dwarfs, \citet{wyatt14} developed a stochastic accretion model to constrain the distribution of accretion rates onto white dwarfs. The authors adopted a log-normal distribution, and found that a reasonable fit to the observed distribution of accretion rates could be achieved with a log-normal mean and standard deviation of $\langle\log \dot{M}\rangle=8.1$ and $\sigma(\log \dot{M})=1.6$, respectively. Our synthetic white dwarfs are given an accretion rate, drawn from this distribution. Combined with the diffusion timescales and convection zone mass (fixed by measured stellar parameters and models), this determines the expected abundance in a steady state scenario (Eq.\,\ref{eq:Xss}). This is broadly a fixed, input parameter, but in Section\,\ref{sec:results} we do consider the sensitivity of our results on the choice of mean accretion rate. Thus this provides an additional loose observational constraint.

\subsection{Observational constraints}
Now that we have set up our simple model, we can test whether the mass distribution of metal-polluted white dwarfs can be explained by the physics of accretion and metal diffusion in white dwarf atmospheres. First we consider four key observational constraints that can guide our modelling. We note that only the first two constraints in the following are used as strict constraints on our model. The two other observational constraints  serve for a qualitative interpretation of the results of the model.

\paragraph*{1) Shape of mass distribution.}
The primary diagnostic underpinning our model is how well the synthetic polluted mass distribution  fits the observed DAZ/DZA mass distribution, for a given set of accretion parameters ($t_{\rm d}$ and $t_{\rm p}$). This is quantified in our simulation using a KS test, which calculates the probability of ruling out the null hypothesis that the two samples are drawn from the same underlying distribution.

\paragraph*{2) Observed polluted fraction.}
From the 40\,pc sample, we find that 8$\pm$1\,\% of H-atmosphere white dwarfs have been detected to have photospheric metals. This polluted percentage is lower than the heuristic 25--50\% metal pollution percentage identified through space-based ultraviolet observations with HST \citep{koester14,OuldRouis2024}. This is partly 
because there are more and stronger metal lines at ultraviolet wavelengths. Given that the observational input to this work is medium- and low-resolution ground-based spectroscopy, we adopt the value of 8$\pm$1\,\% as our percentage of white dwarfs with detected metals. The fraction of synthetic white dwarfs with detectable metals within our model can be computed as 
\begin{equation}
    f_{\rm z} = \frac{1}{N_{\rm wd}}\sum_{k=0}^{N_{\rm wd}} P_{\rm det}^{k}\, ,
\end{equation}
where $N_{\rm wd}$ is the total number of white dwarfs in our synthetic population, and $P_{\rm det}^k$ is the detection probability for each synthetic white dwarf.

\paragraph*{3) Active accretion.}
The most informative proxy for ongoing accretion onto white dwarfs lies in the form of infrared excess. A sample of 117 white dwarfs observed with Spitzer found the frequency of white dwarfs to exhibit a detectable IR excess to be $4.3^{+2.7}_{-1.2}$\% \citep{barber2012}. Later, a reanalysis of a larger sample of hydrogen-atmosphere white dwarfs in Spitzer refined the infrared excess percentage to $1.5^{+1.5}_{-0.5}$\%, based on the sample 196 seemingly isolated white dwarfs in the effective temperature range 14\,000–-31\,000\,K observed with Spitzer \citep{wilson2019}. We choose to interpret this as a heuristic observational constraint on the number of white dwarfs that are actively accreting at any given moment. In our model, if we were to assume that all actively accreting white dwarfs produce detectable IR emission, a synthetic IR fraction could be estimated from the ratio of the disk time, to the total accretion period, such that the ``predicted'' IR fraction is given by $f_{\rm IR}^{\rm pred} \sim t_{\rm d}/t_{\rm p}$. However, disks are seldom detected at white dwarfs cooler than 10\,000\,K \citep{farihi2016review}. We account for this temperature dependence by assuming that in our synthetic population, only white dwarfs above 10\,000\,K which are actively accreting may exhibit a detectable IR excess. 
In our synthetic population, approximately 10\% have effective temperatures above 10\,000\,K. 
Thus, the observable IR excess fraction should be given by
\begin{align}
    f_{\rm IR}^{\rm obs}&\approx 0.1f_{\rm IR}^{\rm pred}\,,\\
    &\approx 0.1\frac{t_{\rm d}}{t_{\rm p}}\,.
\end{align}
Rearranging this expression for the accretion period gives 
\begin{equation}
\label{eq:tp-fIR}
t_{\rm p}\approx\frac{0.1}{f_{\rm IR}^{\rm obs}}t_{\rm d}\,,
\end{equation}
and thus allows a comparison of the observed IR excess fraction with the theoretical detectable IR excess from our population synthesis as a function of accretion period and disk lifetime.

\paragraph*{4) Disk lifetimes.}
Crude observational constraints estimate the lifetime of dusty disks around white dwarfs to be in the between $10^5$--$10^7$\,yr \citep{cunningham2021,girven12}. This estimate was made by comparing the mean inferred accretion rates of H-atmosphere white dwarfs and the mean minimum accreted mass of He-atmosphere polluted white dwarfs. Accretion rates inferred from optical and UV spectroscopic abundance measurements are implicitly time-averaged over the predicted diffusion timescales. Since H-atmosphere white dwarfs have relatively short sinking timescales, the inferred accretion rates in these systems offer the best proxy for ``typical'' current accretion rates of debris disk accretion onto white dwarfs. The inferred accretion rate onto the canonical dusty disk hosting metal-polluted white dwarf -- G29$-$38 -- was recently confirmed with a measurement of the model independent instantaneous accretion rate by means of X-ray observations \citep{cunningham2022}. On the other hand, He-atmosphere white dwarfs have diffusion timescales typically 2 dex longer, which gives them a better ``memory'' of past accretion events, including a minimum accreted mass, but the inferred accretion rates are much less likely to reflect the instantaneous accretion rate at these systems. Dividing the mean He-atmosphere minimum mass by the mean H-atmosphere accretion rate gives an estimate of the disk lifetime.

Although the best available observational constraints favour disk lifetimes of 10$^5$--10$^7$ years, it has been shown from theoretical investigations that there is no lower limit on the expected lifetime of debris disks around white dwarfs \citep{veras2020-disk-lifetime}. 
We aim to test whether a population of polluted white dwarfs can be created which both matches the observed mass distribution, and the aforementioned constraints.

\begin{figure}
	\centering
	  \subfloat{\includegraphics[width=0.4\textwidth]{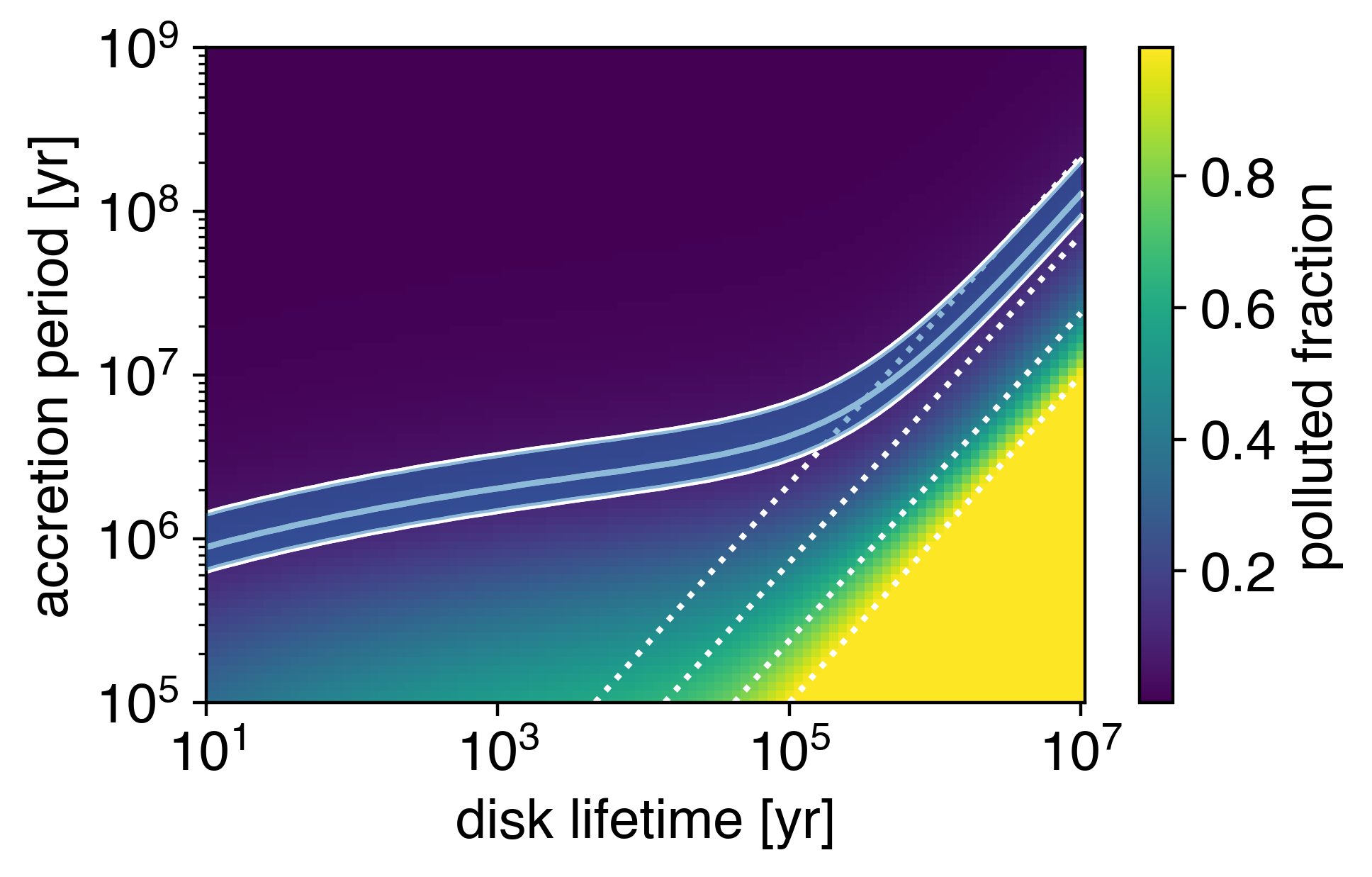}}\hspace{20pt}\\
	  \subfloat{\includegraphics[width=0.4\textwidth]{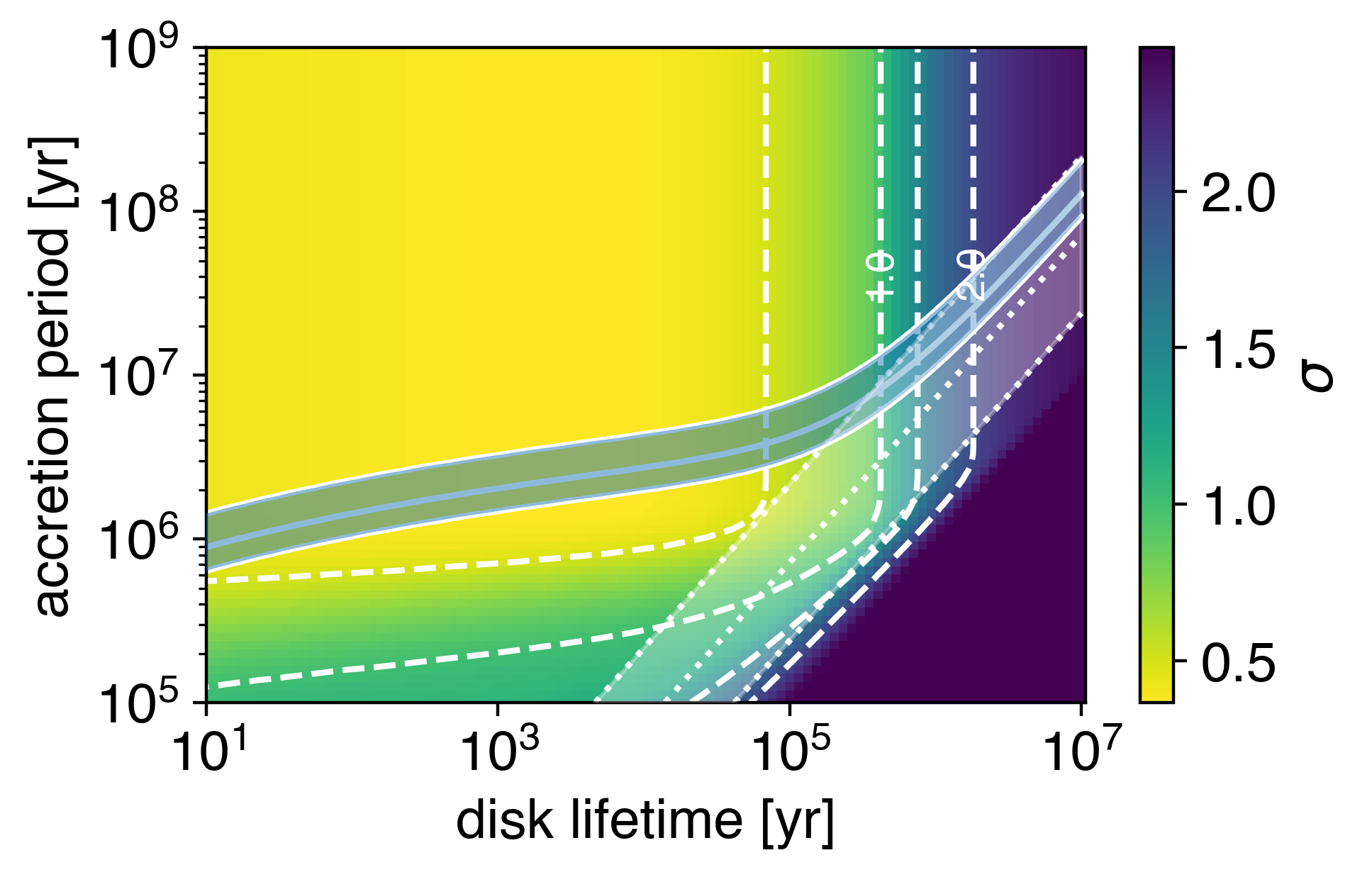}}\\
	  \subfloat{\includegraphics[width=0.4\textwidth]{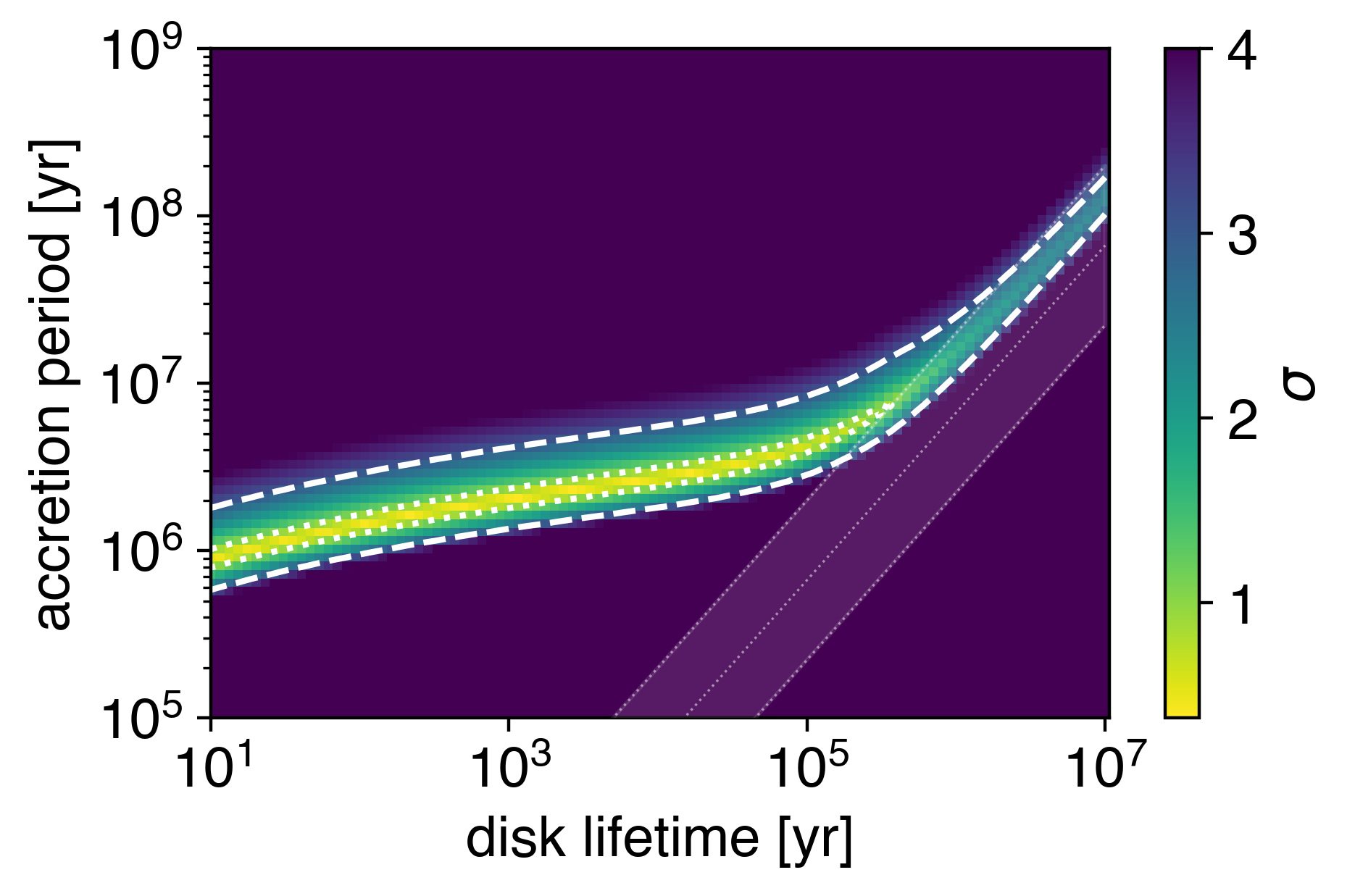}}   
	\caption{\textbf{Upper}: The polluted fraction across the accretion-period 
 and disk-lifetime ($t_{\rm p}$-$t_{\rm d}$) parameter space, following our population synthesis computation. The blue shaded region indicates the 3$\sigma$ observed polluted fraction ($8\pm3$\,\%). The dotted lines indicate the 2$\sigma$ confidence on the fraction of white dwarfs observed to exhibit an infrared excess, as defined by Eq.\,\eqref{eq:tp-fIR} with the empirical observed fraction set to $f_{\rm IR}^{\rm obs}$\,$=$\,$0.015^{+0.03}_{-0.01}$, as per \citet{wilson2019}. 
We take this as an observational constraint on the mean duty cycle, $t_{\rm p}/t_{\rm d}$. 
\textbf{Middle}: From the KS test, the statistical confidence ($\sigma$) for which the synthetic DAZ/DZA populations are ruled out by the observed DAZ/DZA population. 
The dashed white lines show contours of 0.5, 1.0, 1.5, and 2.0$\sigma$.
As in the top panels, the filled blue band represents the 3$\sigma$ limit on the observed polluted fraction. We also show the infrared excess fraction in the white shaded region. 
\textbf{Bottom}: Combined significance of the top two panels. Contours of 1 and 3$\sigma$ are shown in dotted and dashed lines, respectively. Combined significances are computed by multiplying the p-value from the KS tests (akin to the middle panel of Fig.\,\ref{fg:pop}) and the p-value implied from the synthetic polluted fraction (see Eq.\,\ref{eq:PZ}).}
	\label{fg:pop}
\end{figure}

\section{Results and discussion}
\label{sec:results} 

We initially keep two free parameters that we explore in this section: the disk lifetime, $t_{\rm d}$, and the total accretion period, $t_{\rm p}$. For a given $t_{\rm d}$ and $t_{\rm p}$, we compute the detection probability for every white dwarf in the synthetic population. The detection probability is defined by Eq.\,\ref{eq:prob}, which in turn is derived from the closed-form of Eq.\,\ref{eq:diff-eq}, assuming periodic accretion. 
We then compare the synthetic mass distribution, where each member of the population is weighted by its detection probability, $P_{\rm det}$, to the observed mass distributions of DA and DAZ/DZA white dwarfs. We remind the reader that the synthetic and observed DA mass distributions, without taking into account metal pollution, have already been successfully compared in \citet{cunningham2024}. In particular the IFMR is designed to provide a best fit to the observed DA mass distribution, hence the synthetic and observed DAZ/DZA mass distributions can be compared directly. To quantify this comparison, we adopt a weighted 2-sample KS test, in which each member of the synthetic population is weighted by its detection probability as defined by Eq.\,\eqref{eq:prob}. This allows to quantify the probability that the two samples were drawn from the same underlying distribution.

We explore the 2D parameter space with ranges $10$\,$<$\,$t_{\rm d}$\,$<$\,$10^7$\,years and $10^5$\,$<$\,$t_{\rm p}$\,$<$\,$10^{9}$\,years and investigate whether our models finds a solution consistent with the two primary observational constraints -- 1) the shape of the mass distribution and 2) the observed polluted fraction. Later, we consider if our results are consistent with the less stringent observational constraints of 3) infrared fraction and 4) observed disk lifetime.

For the first observational constraint, the middle panel of Fig.\,\ref{fg:pop} shows the statistical significance with which each synthetic polluted mass distribution (where each distribution has been synthesized for a specific combination of accretion parameters $t_{\rm d}$ and ${t_{\rm p}}$) can be distinguished from the observed DAZ/DZA mass distribution. The white dashed lines show contours of 0.5, 1, 1.5, and 2$\sigma$. We find that there is a large part of the parameter space that produces a synthetic population which is statistically consistent with the observations. We find that the synthetic population achieves 1$\sigma$ agreement with the observed DAZ/DZA mass distribution when the disk lifetime is less than a few $10^5$ years and the total accretion period is in excess of $\sim$$10^6$\ years. 

For the second observational constraint, for every position in this 2D parameter space, there is an implicit synthetic metal-polluted fraction. This is defined as the mean of all the detection probabilities, $P_{\rm det}$ (see Eq.\,\ref{eq:prob}). 
The top panel of Fig.\,\ref{fg:pop} shows the polluted fraction across the $(t_{\rm d}$--$t_{\rm p})$ plane. The colour indicates the synthetic metal-polluted fraction, from 0 to 1. The blue shaded region indicates the 3$\sigma$ limits on the observed metal-polluted fraction ($8\pm3$\%) of the 40\,pc sample. The middle panel shows the same blue shaded band, on top of the confidence with which the synthetic population agrees with the observations. We assume that the fraction of systems in our synthetic population capable of becoming polluted is 100\%. In reality, the system fraction could be anything from the measured UV pollution fraction (25--50\%; \citealt{koester14,OuldRouis2024}) up to 100\%. Opting for the lower limit of 25\% would directly translate to an effective observational constraint on the pollution fraction of white dwarfs with a planetary system a factor of 4 times higher ($32$\%). The top panel of Fig.\,\ref{fg:pop} shows that the metal-pollution fraction increases rapidly from the current position of the blue shaded region towards shorter total accretion periods. Intuitively, if fewer white dwarfs are permitted to undergo accretion at some point in their evolution, then the frequency of planetary debris in those that can  have accretion must be higher to explain the observations. Our model recovers this by favouring shorter total accretion periods. Given how rapidly the pollution fraction is increasing below the blue shaded region, we do not expect the planetary system fraction assumption to impact our results. 

In the bottom panel of Fig.\,\ref{fg:pop}, we combine these first two observational constraints into a single statistical estimation of how well each combination of accretion parameters ($t_{\rm d}$ and ${t_{\rm p}}$) describes the observations. Across the 2D parameter space of synthetic polluted fractions (e.g., top panel of Fig.\,\ref{fg:pop}), we define the continuous standard deviation, $\sigma_{\rm z}$, as
\begin{equation}
    \sigma_{\rm z} = \frac{|f_{\rm z} - \langle f_{\rm z}\rangle|}{\sigma^*_{\rm z}} 
\end{equation}
where $f_{\rm z}$ is the synthetic polluted fraction, $\langle f_{\rm z}\rangle$ is the observed polluted fraction, and $\sigma^*_{\rm z}$ is the uncertainty on the observed polluted fraction, for which we adopt the less stringent 3 standard deviation uncertainty (i.e., $\pm3$\%), to account for uncertainty in the system fraction. We then convert that to a probability
\begin{equation}
    P_{\rm z} = 1-\mathrm{erf}\left(\frac{\sigma_{\rm z}}{\sqrt2}\right)~,
    \label{eq:PZ}
\end{equation}
which we multiply by the $p$-value obtained from the KS tests. This provides an overall probability to indicate the preferred accretion model parameters. 
In the bottom panel of Fig.\,\ref{fg:pop} we express this in terms of statistical significance, $\sigma$. The colour indicates the statistical significance, and we indicate bands of 1 and 3\,$\sigma$ in dotted and dashed white lines.
This panel shows that there is a band of solutions favoured at the 1\,$\sigma$ level, with disk lifetimes in the range 10--$10^5$\,years providing a consistent explanation for the observations. 
We note that this solution is relatively insensitive to the total accretion period, for which solutions around $\sim$\,$10^6$\,years are favoured. In the following, we go on to consider whether this solution is consistent with the third and fourth observational constraints.

The third observed quantity to consider is the fraction of white dwarfs to show an infrared excess ($f_{\rm IR}$), which we interpret as a proxy for the fraction of white dwarfs above 10\,000\,K that are currently accreting. The percentage of white dwarfs with IR excess is $1.5^{+1.5}_{-0.5}$ \citep{wilson2019}. In all three panels of Fig.\,\ref{fg:pop}, the diagonal white dotted lines indicate the 2$\sigma$ limits on the IR fraction, defined by Eq.\,\eqref{eq:tp-fIR}. 
Most of the 1\,$\sigma$ solution region is not consistent with this constraint. In other words, the accretion model favours disk lifetimes shorter or comparable to the diffusion timescale, and much longer total accretion periods. 
This leads to a significant under prediction of the infrared excess fraction, assuming that only the polluted white dwarfs in the sample having effective temperatures above 10\,000\,K should show a detectable IR excess. However, we note that the edge of the allowed solutions at long disk lifetimes becomes consistent with this IR fraction constraint.

The fourth observational constraint to consider is that of the disk lifetime, $t_{\rm d}$. 
The favoured solution from the population synthesis and the first two observational constraints is marginally consistent with existing estimates of disk lifetimes ($10^5$--$10^7$ years). However solutions favouring disk lifetimes in excess of a few $10^5$\,years are ruled out at 1$\sigma$, and solutions favouring disk lifetimes in excess of a few 10$^6$\,years are ruled out at 2$\sigma$. 
Hence there is a moderate tension between the population synthesis, which favours short disk lifetimes, and the observational constraints which favour long disk lifetimes.

\begin{figure}
	\centering
	  \subfloat{\includegraphics[width=0.42\textwidth]{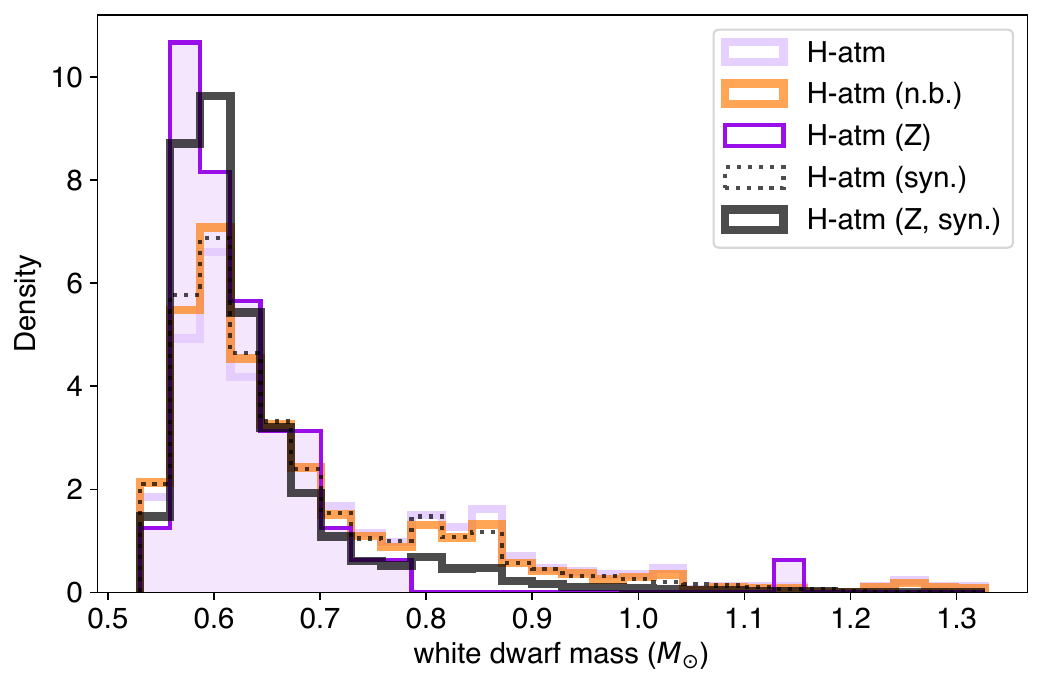}} \\
	  \subfloat{\includegraphics[width=0.42\textwidth]{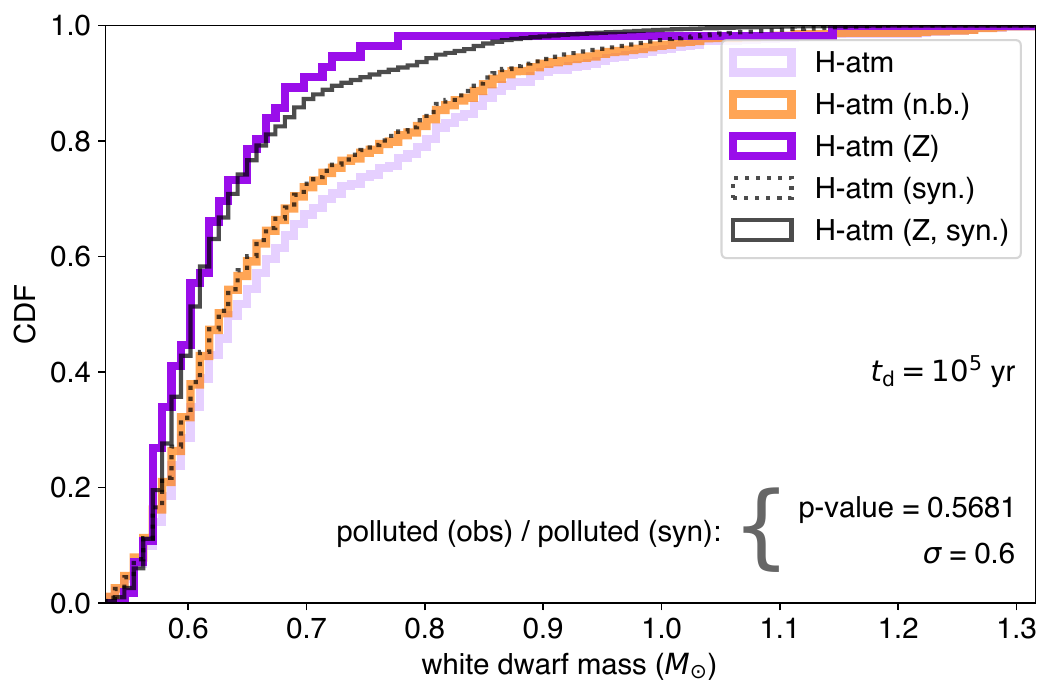}}   
	\caption{Similar to Fig.\,\ref{fg:mass-hist}, now including the results of the population synthesis. \textbf{Top}: We show the same three distributions as in Fig.\,\ref{fg:mass-hist}. We also show the synthetic histogram for our distribution of un-polluted white dwarfs (DA) and the final polluted sample, with disk parameters of $t_{\rm d}=10^5$\,yr and $t_{\rm p}=2\times10^{6}$\,yr.
 \textbf{Bottom:} We show the CDF for the same five distributions. The mass distribution of the polluted synthetic population (polluted according to our accretion model for $t_{\rm d}=10^5$\,yr and $t_{\rm p}=2\times10^6$\,yr) is consistent with the observed polluted mass distribution.} 
	\label{fg:mass-hist-syn}
\end{figure}

\begin{figure}
	\centering
	  \subfloat{\includegraphics[width=0.42\textwidth]{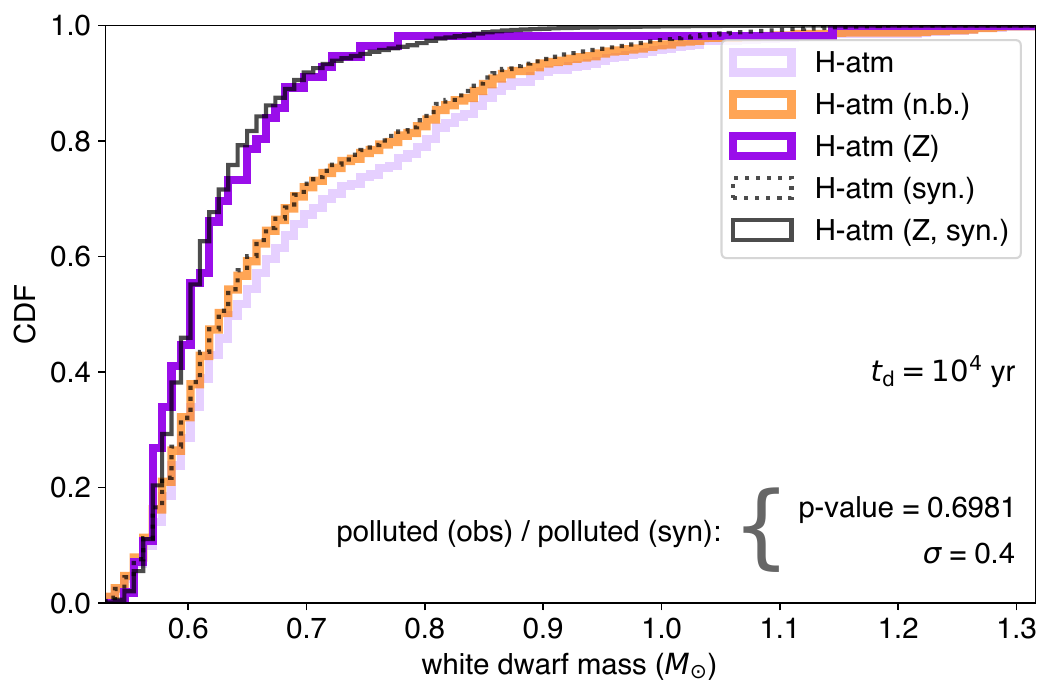}}   \\ 
	  \subfloat{\includegraphics[width=0.42\textwidth]{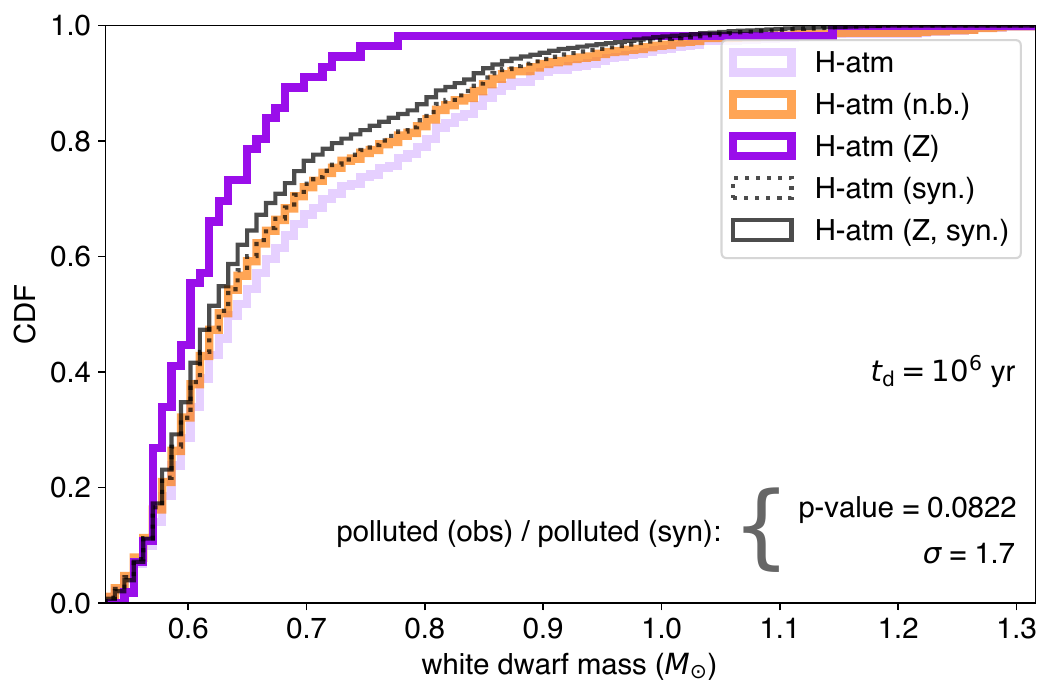}}  
	\caption{Similar to the lower panel of Fig.\,\ref{fg:mass-hist-syn}. In both panels, we fix the total accretion period at $t_{\rm p}=2\times10^6$\,yr, and vary the disk lifetime. The disk lifetimes are $t_{\rm d}=10^4$ and $10^6$\,yr, on the top and bottom panels, respectively.}  
	\label{fg:mass-hist-syn-vary-t1}
\end{figure}

Fig.\,\ref{fg:mass-hist-syn} shows the metal-polluted mass distribution from the synthetic population. For comparison, we also show the same observational mass distributions as shown in Fig.\,\ref{fg:mass-hist}. For the synthetic mass distribution we use a statistically-favoured case with $t_{\rm d}=10^5$\,years and $t_{\rm p}=2\times10^6$\,years. Qualitatively, the majority of the high-mass bump has been removed. Statistically, this distribution is at least 1$\sigma$-consistent with the observed distribution. Fig.\,\ref{fg:mass-hist-syn-vary-t1} shows that a 
statistically-favoured retrieval of the observed DAZ/DZA fraction can 
also
be achieved at shorter disk lifetimes. This was also evident in the middle and lower panels of Fig.\,\ref{fg:pop}. However, as discussed previously, disk lifetimes as short as 10$^{4}$\,years are inconsistent with observational constraints of disk lifetimes being in the range $10^5$--$10^7$\,years.
We now go on to consider the biases and variations in other parameters that could impact our results, and test whether the tensions between the best-fitting solution and the observed disk lifetime and infrared excess fraction can be reconciled.

\begin{figure}
	\centering
	  \subfloat{\includegraphics[width=0.49\textwidth]{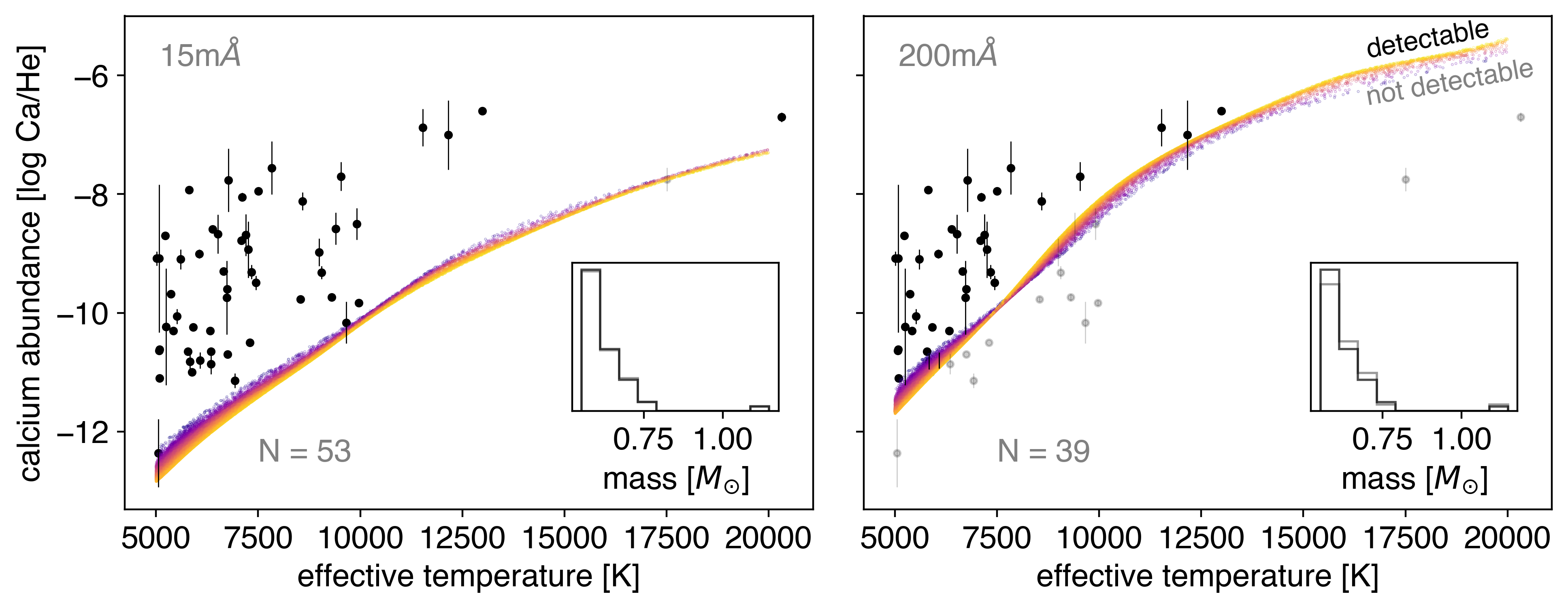}}
	\caption{Measured calcium abundance for the full sample of 56 H-atmosphere metal-polluted white dwarfs within 40pc. Abundances taken from the Planetary-Enriched White Dwarf Database \citep[PEWDD;][]{williams2024} for 54 of the 56. The remaining 2 do not have a published abundance. We also show the calcium abundance required to produce the Ca\,{\sc k} line with an equivalent width (EW) of 15 and 200\,m\AA; left and right, respectively. These lines act as an indication of the expected detection thresholds at high- (15\,m\AA), and medium-resolution (200\,m\AA). The model predictions are sensitive to surface gravity. The color gradient spans the surface gravity range $7.0<\log\,g<9.0$. Measured calcium abundances that would be detectable for a given EW are shown in black, whilst those not detectable are shown in grey. The number, N, of ``detectable'' DAZ/DZAs is given in the panel. The inlays show a normalised mass distribution of the full sample of DAZ/DZAs (black) and just those detectable for a given EW. We find no significant bias is introduced by using the full sample in our statistics. At 200\,m\AA\ the detection limit in the right panel is representative of medium resolution ($R$\,$\approx$\,2000) observations with S/N\,$\approx$\,30. This is comparable to the minimum resolution and S/N achieved across the full 40\,pc sample.}
	\label{fg:obs-EW-detect}
\end{figure}

\subsection{Resolution and signal-to-noise}
The 40\,pc sample of white dwarfs is the largest spectroscopically complete, volume-limited sample to date. However, the origin of the spectroscopy is varied, comprising a mixture of archival spectra, surveys, and targeted observations. The result is that the resolution and signal-to-noise of the spectra are variable across the sample. As discussed in \citet{OBrien2023}, all white dwarfs in the sample have at least spectroscopy with medium-resolution ($R$\,$>$\,1000) and high signal-to-noise (S/N$>$30). However, some white dwarfs were also observed at high-resolution ($R$\,$>$\,25\,000). We acknowledge that the heterogeneous nature of the sample could lead to biases in the observed mass distribution of metal-polluted white dwarfs. This is because the identification of metals in a spectrum generally requires a higher signal-to-noise and higher resolution than is required to identify broad hydrogen lines. Furthermore, because higher mass white dwarfs have broader lines, the detectability may have a mass dependence.

To take this into account we computed a grid of calcium equivalent widths using the model atmosphere code of \citet{koester2010,koester2020}. The grid covers 4 dimensions; namely i) effective temperature, ii) surface gravity, iii) calcium abundance and iv) equivalent width. This grid allows us to make an estimate of the mass dependence of the detection limit. 

In Fig.\,\ref{fg:obs-EW-detect} we show the calcium abundances for the 50 DAZ (and 6 DZA) white dwarfs within 40\,pc as a function of effective temperature. We also show with solid lines the detection thresholds defined as equivalent widths of 15\,m\AA\ (left) and 200\,m\AA\ (right). The range of colors in the detection thresholds correspond to different surface gravities between $\log\,g$=7--9. For each member of the observed sample, we retrieve the expected equivalent width. We define two sub-samples for the observational data; those which produce a Ca\,{\sc k} line of at least i) 15\,m\AA\ and ii) 200\,m\AA. These loosely correspond to samples which would be accessible with i) high-, and ii) medium-resolution spectroscopic observations, respectively.

The inlay in both panels shows the normalised mass distribution for the full sample of DAZ/DZA in black. In grey, the normalised mass distribution is shown for only those stars with detectable calcium at the given equivalent width. We find no appreciable change in the normalised mass distribution as a function of spectral resolution. This is likely due to both the large range of measured calcium abundances, with the majority of the sample sitting well above even the low-resolution (200\,m\AA) detection threshold, and the full sample having been observed with a minimum signal-to-noise (approximately S/N\,=\,30 at $R$\,=\,2000). We thus conclude that our choice to use the full sample, which comprises white dwarfs with metals identified across a range of spectral resolutions, contributes negligible bias to our results.

\begin{figure}
	\centering
	  \subfloat{\includegraphics[width=0.4\textwidth]{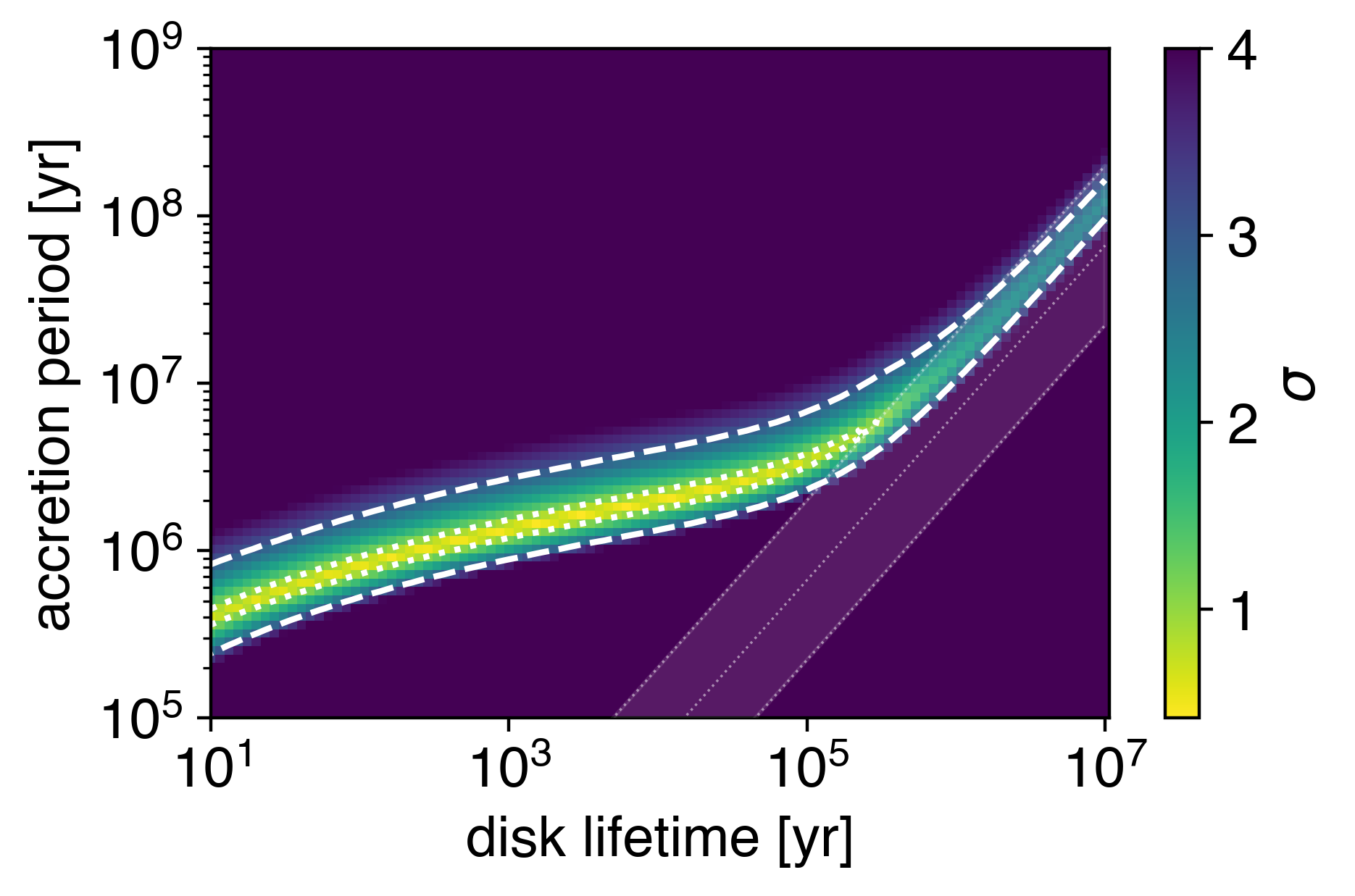}}
	\caption{Similar to the lower panel of Fig.\,\ref{fg:pop}, but with an adopted equivalent width detection limit of 200\,m\AA.}
	\label{fg:pop-vary-EW}
\end{figure}

We also perform the population synthesis for the two resolution cases, and find minimal impact on the results. Fig.\,\ref{fg:pop-vary-EW} shows the solution parameter space for the 200\,m\AA\ limit (compared to 15\,m\AA\ in the default simulation). We can see that the 3$\sigma$ limits on the observed fraction is pushed to slightly lower total accretion periods, but qualitatively the result is unchanged. This is consistent with the expectation from Fig.\,\ref{fg:obs-EW-detect}, in that lower resolution should lead to a smaller polluted fraction and thus shorter times in between accretion periods ought to be favoured.

\begin{figure}
	\centering
 	  \subfloat{\includegraphics[width=0.35\textwidth]{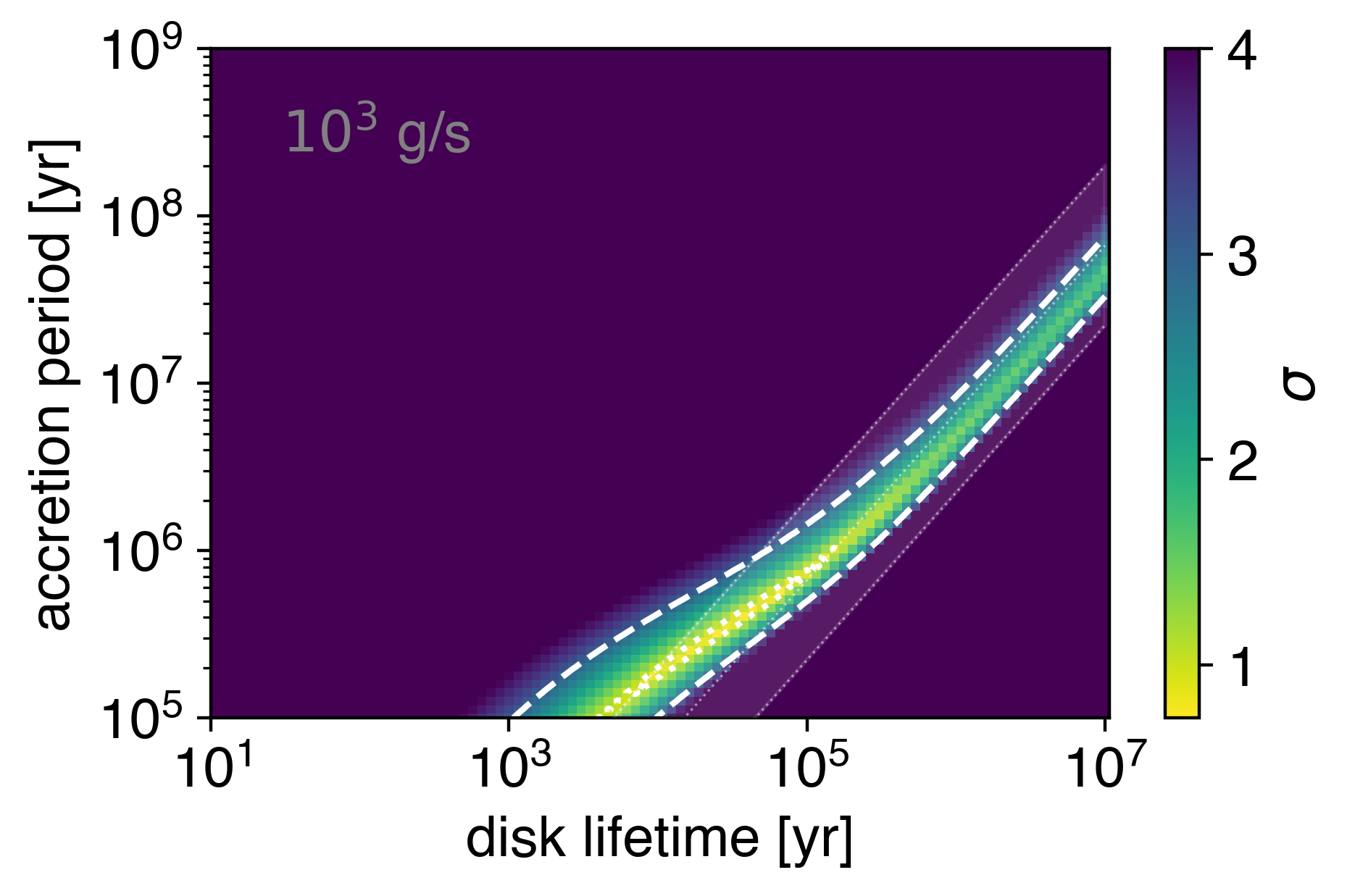}}\\
	  \subfloat{\includegraphics[width=0.35\textwidth]{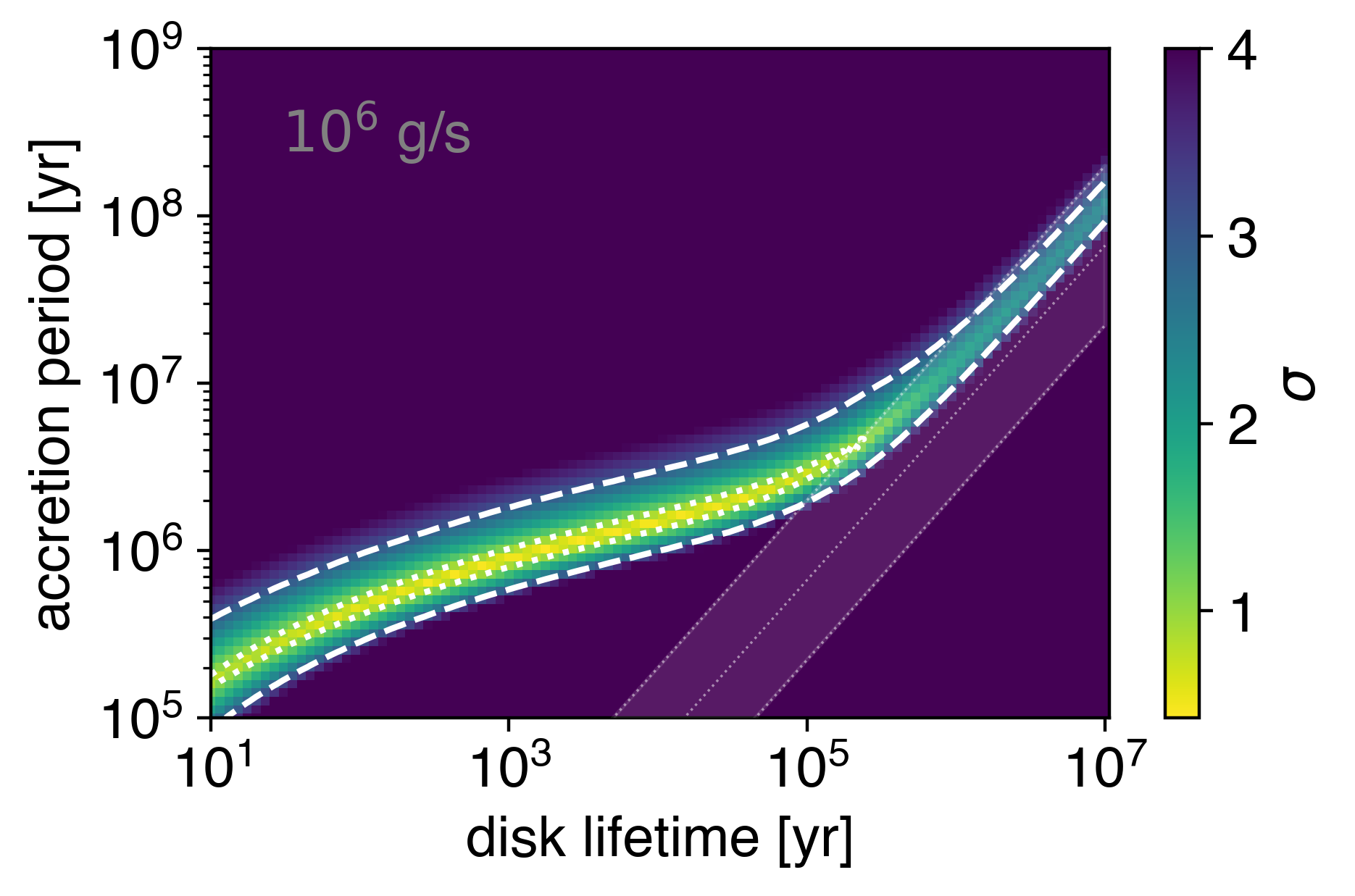}}\\
	  \subfloat{\includegraphics[width=0.35\textwidth]{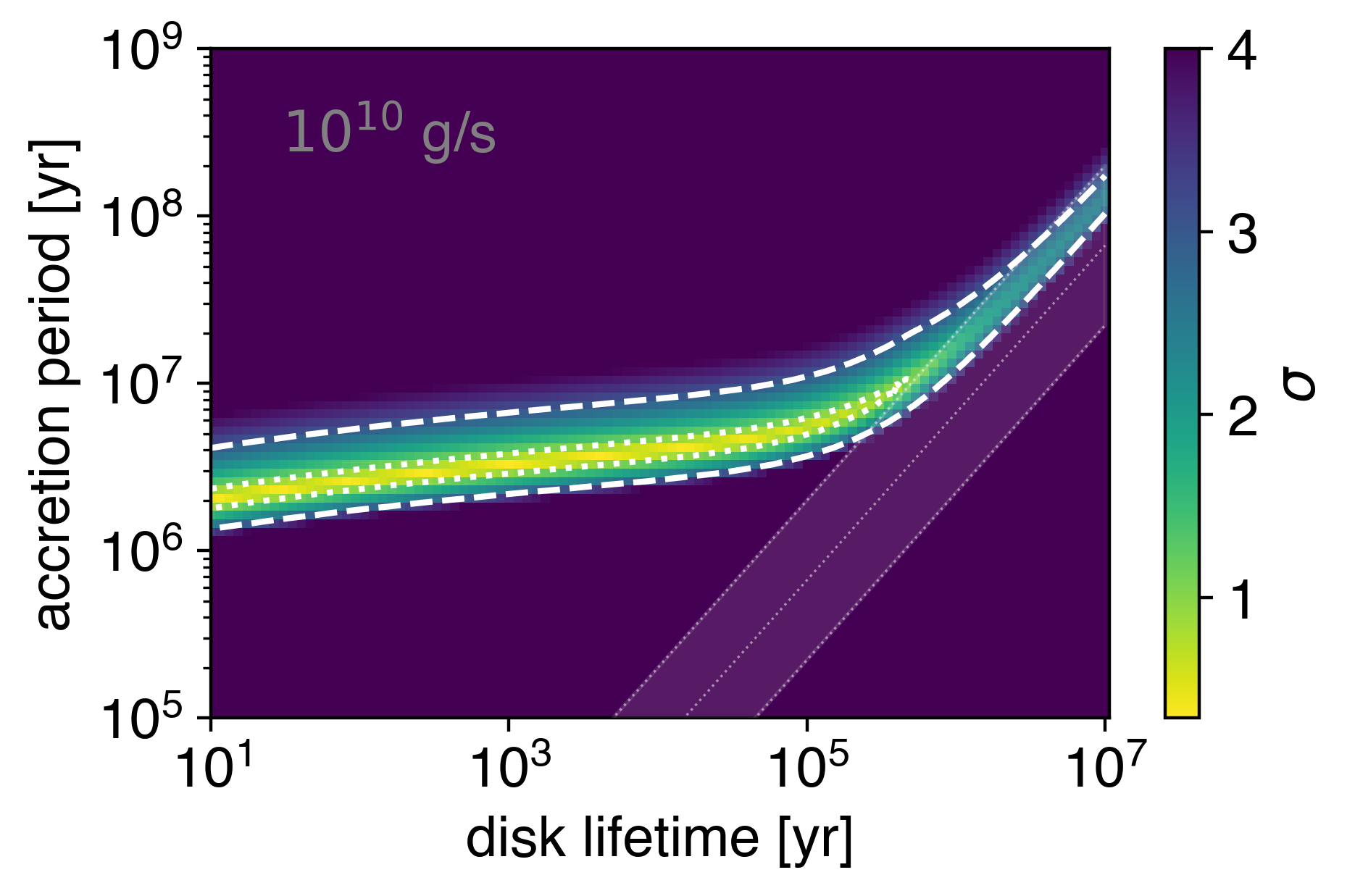}} \\
	  \subfloat{\includegraphics[width=0.35\textwidth]{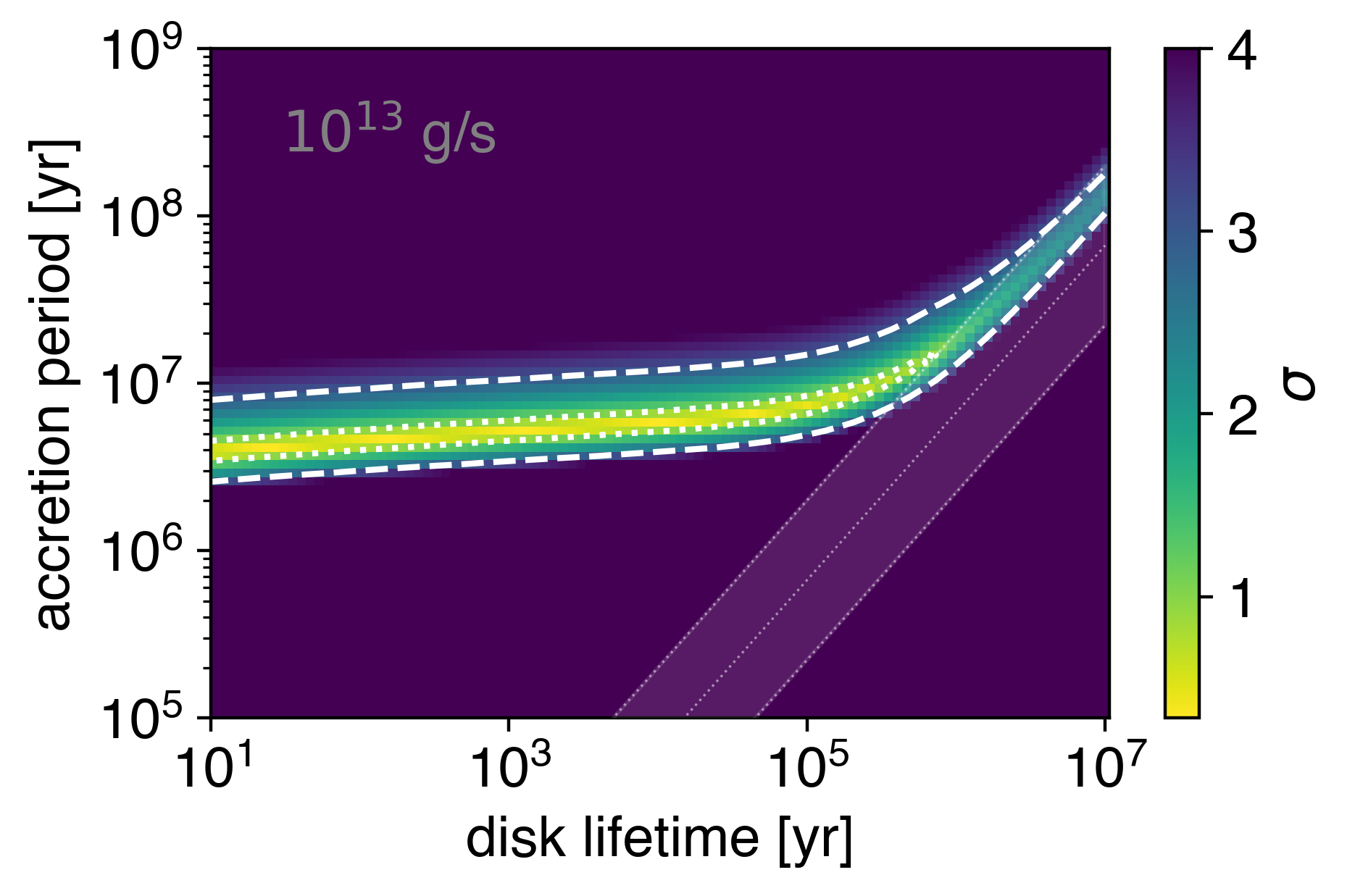}}
	\caption{Similar to the lower panel of Fig.\,\ref{fg:pop}. From top-to-bottom, the center of the log normal distribution from which the synthetic accretion rates is increased. In order, the mean accretion rate for the log normal distribution is as follows: $\langle\log \dot{M_0}\rangle=3.1$, 6.1, 10.1, and 13.1\,g/s. }
	\label{fg:pop-vary-Mdot}
\end{figure}

\subsection{Accretion rates}
For H-atmosphere white dwarfs, accretion rates inferred from optical spectroscopy span $10^3$\,$<$\,$\dot{M}$\,$<$\,$10^9$\,g/s, whilst accretion rates onto He-atmosphere white dwarfs have been inferred to reach up to $10^{10}$--$10^{11}$\,g/s. 
Within Eq.\,\ref{eq:prob} of our population synthesis model, the accretion rate for a specific white dwarf is a free parameter. However, for the purposes of our population synthesis we choose to draw from a distribution of accretion rates, motivated by observations. For each member of our synthetic white dwarf sample, the accretion rate is drawn from the log normal prescription derived by \citet{wyatt14}. We now investigate whether the results are sensitive to the choice of accretion rate distribution.

In Fig.\,\ref{fg:pop-vary-Mdot} we show the results of our population synthesis in varying the mean accretion rate of the log-normal distribution of accretion rates. From top-to-bottom, we increase the mean accretion rate, $\langle\log \dot{M_0}\rangle$, from $10^3$--$10^{13}$\,g/s. We remind the reader that \citet{wyatt14} adopted $\langle\log \dot{M_0}\rangle=10^{8.1}$\,g/s as the solution, which was used in Fig.\,\ref{fg:pop}. 
Qualitatively, we find that the higher accretion rate models show results which are not too sensitive to the absolute accretion rate. This is because the results are more sensitive to the ratio of disk lifetime and diffusion timescale. However, the results do become affected at the lower extreme of $\langle\log \dot{M_0}\rangle=3.1$\,g/s, when a larger fraction of the synthetic population have accretion rates too low to ever be detected. However, this low accretion rate scenario is largely ruled out by the observational constraints discussed above.

\begin{figure}
	\centering
	  \subfloat{\includegraphics[width=0.4\textwidth]{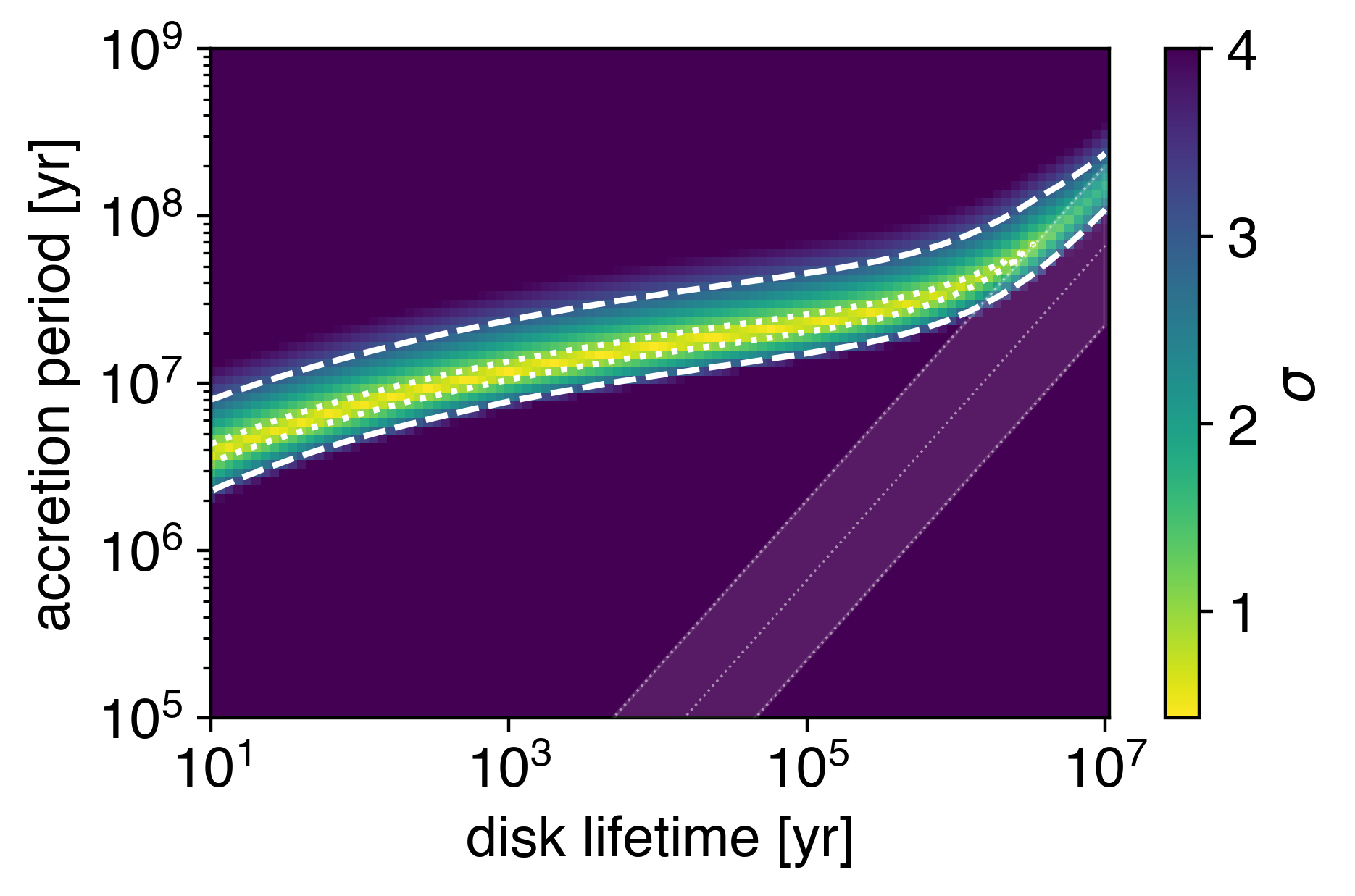}}
	\caption{Similar Fig.\,\ref{fg:pop-vary-EW}, but accounting for convective overshoot in the determination of the diffusion timescale, $\tau$, and convection zone mass, $M_{\rm cvz}$. The longer diffusion timescale in the overshoot models lead to a consistent solution at longer disk lifetimes.}
	\label{fg:pop-vary-overshoot}
\end{figure}

\subsection{Convective overshoot}
We also consider the impact of including convective overshoot in our prescription. \citet{cunningham19} derived diffusion coefficients for convective overshoot in pure-hydrogen atmosphere white dwarfs using 3D radiation-hydrodynamics. Finding between 1--4 pressure scale heights of additional mixing in the effective temperature range from 18\,000\,K (at the onset of convection) to 11\,400\,K. Cooler 3D convective overshoot simulations have not yet been possible due to computational limitations. At 11\,400\,K, approximately 1 pressure scale height of additional mixing is measured. We thus adopt the 3D diffusion timescales in the range 18\,000--11\,400\,K, and a fixed 1 pressure scale height for all cooler temperatures. Above 18\,000\,K, no correction is required, since hydrogen atmospheres at those effective temperatures are radiative, not convective.

We find that with the inclusion of overshoot we still reach a consistent solution at 1\,$\sigma$, but both the preferred disk lifetime and preferred accretion period are approximately 1 order of magnitude larger. This is explained by the larger diffusion timescales predicted by models of convective overshoot. 

Additional enhanced mixing processes have been proposed to alter inferred accretion rates. Specifically, \citet{bauer2019} showed using the stellar evolution code MESA \citep{paxton15}, that inclusion of thermohaline mixing may increase inferred accretion rates by up to 3 orders of magnitude. This result applies to H-atmosphere white dwarfs with effective temperatures above 11\,000\,K. Additionally, radiative levitation may suspend metals for longer in the atmosphere for effective temperatures above $\approx$20\,000\,K \citep{chayer95a,koester14,OuldRouis2024}. Since the majority of (metal-polluted) white dwarfs within 40 pc are cooler than this (see Fig.\,\ref{fg:obs-EW-detect}), we do not expect these effects to impact our results significantly.

\begin{figure}
	\centering
	  \subfloat{\includegraphics[width=0.4\textwidth]{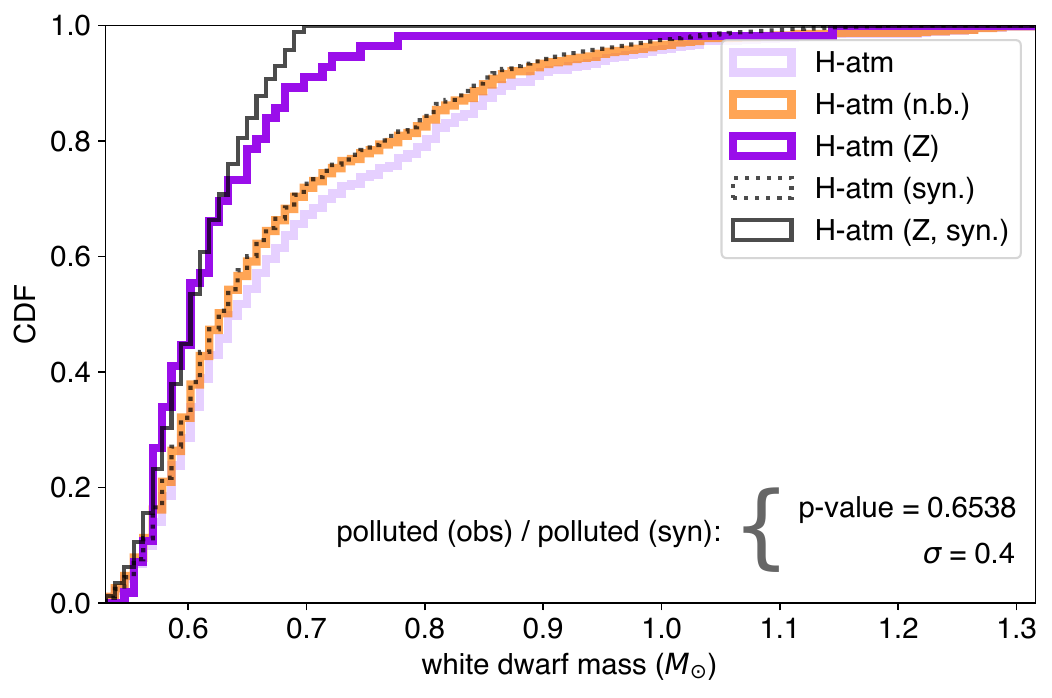}}
	\caption{Similar to the lower panel of Fig.\,\ref{fg:mass-hist}, but this time comparing to the results of \citet{OuldRouis2024}.}
	\label{fg:mass-hist-syn-planetary}
\end{figure}

\begin{figure}
	\centering
	  \subfloat{\includegraphics[width=0.46\textwidth]{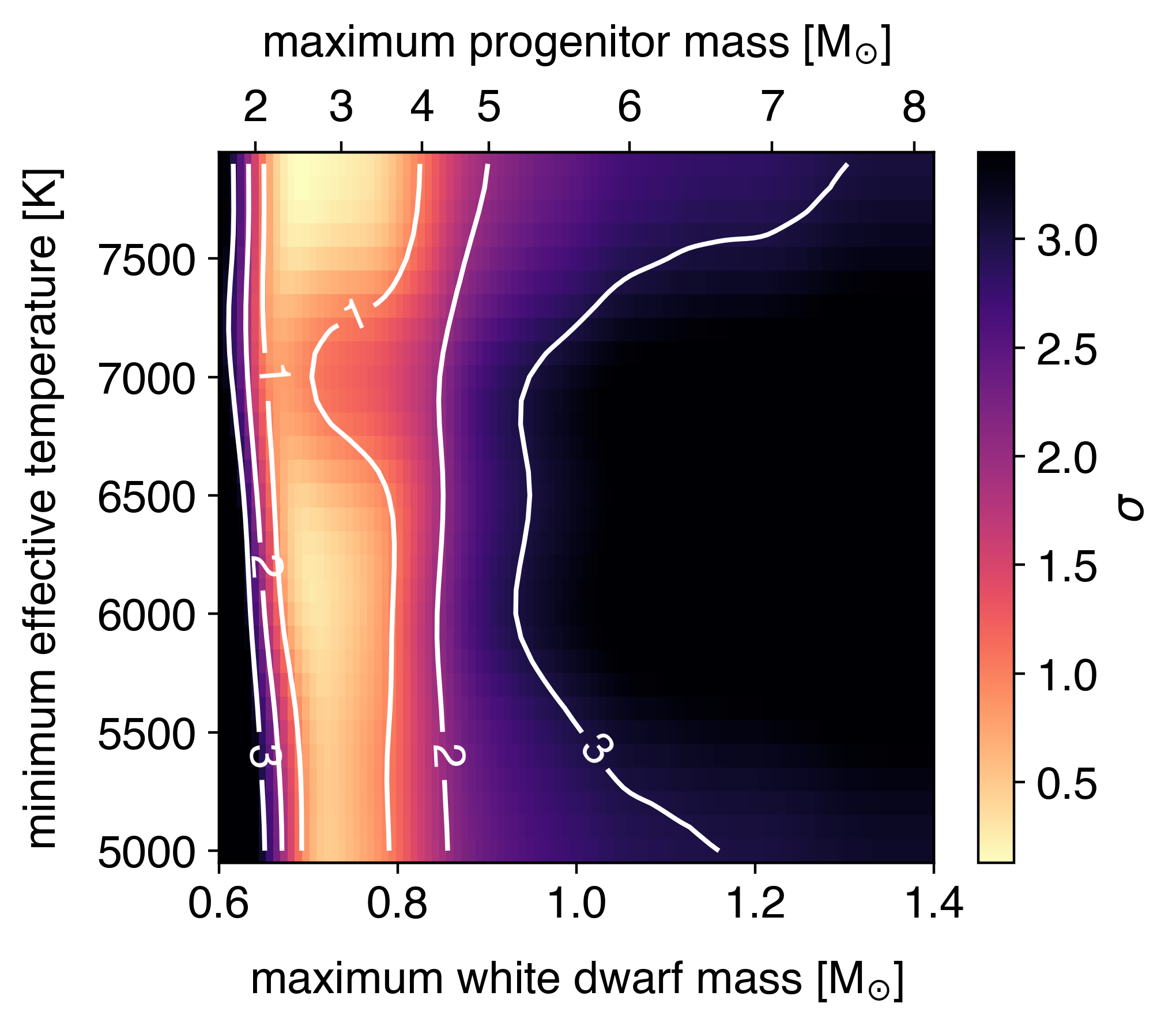}}

    \caption{Statistical result of the KS test for a range of imposed maximum polluted white dwarf masses and minimum effective temperatures, when comparing the DA and DAZ/DZA samples. We show contours of 1, 2, and 3$\sigma$ with white lines. The top axis shows the predicted progenitor (zero-age main sequence) mass by means of the IFMR of \citet{cunningham2024}.}
    
	\label{fg:mass-hist-syn-planetary-Mcut-DAZ}
\end{figure}

\subsection{Exoplanetary system formation and evolution}

Recently, a comparable dearth of high-mass ($M_{\rm wd}$\,$>$\,$0.7\,$M$_{\odot}$), metal-polluted white dwarfs was identified in a sample of 258 stars observed with HST/COS \citep{OuldRouis2024}. The authors explain this dearth as evidence of a lower rate of polluting debris in the stellar remnants of main-sequence stars more massive than $\approx$3.5\,$M_{\odot}$. Such an explanation could also be invoked to explain the dearth of high-mass metal polluted white dwarfs studied in this work. 

To make a direct comparison, in Fig.\,\ref{fg:mass-hist-syn-planetary} we impose a strict constraint on the synthetic population such that no white dwarf with a mass above 0.7\,$M_{\odot}$ has metals, while lower mass white dwarfs are simulated as before. We find that this simple, stringent limit is adequate to explain the observed DAZ/DZA mass distribution. Thus, the planetary formation/evolution model which disfavours evolved planetary systems around former intermediate-mass main sequence stars provides an alternative explanation for the data. We thus conclude that the planetary formation/evolution scenario is statistically indistinguishable from the simple periodic accretion model presented in this work, albeit with some tension on the disk lifetimes constrained from observations. 

In Fig.\,\ref{fg:mass-hist-syn-planetary-Mcut-DAZ} we explore the likelihood of a white dwarf of a certain mass hosting an evolved planetary system. For this test, we do not include any population synthesis, rather we only make use of the observational samples by applying cuts to the observed DA distribution. The colours indicate the statistical separation of the DA and DAZ/DZA 40\,pc observational samples, as determined by the KS test. On the $x$-axis is the maximum white dwarf mass that we impose on the observational DA distribution. The 1, 2, and 3$\sigma$ contours are indicated with white lines. The panel shows that the statistically-favoured mass cut is focused around 0.8\,M$_{\odot}$. On the $y$-axis, we include a minimum effective temperature cut, following the same philosophy as for the $x$-axis. We find the best-fit mass cut has no appreciable dependence on the minimum effective temperature of the sample. We thus take the results for the full sample ($T_{\rm eff}>5\,000$\,K) to be sufficient to facilitate an empirical determination of the limiting mass. We find a maximum polluted white dwarf mass of $M_{\rm wd}^{\rm max}=0.72^{+0.07}_{-0.03}$\,M$_{\odot}$, where the uncertainty corresponds to the extent of the shaded 1$\sigma$ region at $T_{\rm eff}$\,$=$\,$5000$\,K in Fig.\,\ref{fg:mass-hist-syn-planetary-Mcut-DAZ}. Adopting the IFMR of \citet{cunningham2024}, this implies that the empirical maximum zero-age main sequence (ZAMS) progenitor mass limit for the existence of evolved planetary systems is $M_{\rm ZAMS}^{\rm max}=2.9^{+0.7}_{-0.3}$\,M$_{\odot}$.  

In the following we consider two caveats in drawing a direct comparison between this study and that of \citet{OuldRouis2024}. Firstly, the underlying HST/COS sample published by \citet{sahu2023} is not complete in magnitude or volume, but exclusively targeted the brightest, warmest ($>$12,000\,K) DA white dwarfs. In contrast, almost all the metal-polluted white dwarfs in our sample are cool ($<$10,000\,K; see Fig.\,\ref{fg:obs-EW-detect}). Thus the two samples lie in almost fully independent parts of the temperature (and cooling age) parameter space. This dichotomy could lead to biases when directly comparing results from the two samples. For instance, it is possible that larger fractional mass-loss from larger mass progenitors is effective at clearing out or engulfing close-in planetary material, leading to lower rates of pollution onto high-mass white dwarfs at early times, but there could still be a large reservoir of planetary material capable of polluting more massive white dwarfs at later times. There is thus some degeneracy between lower formation/retention rates around massive white dwarfs, and lower time-average perturbation events leading to tidal disruption, accretion, and pollution.

Secondly, \citet{OuldRouis2024} conclude that there is no evidence for merger remnants among their sample based on the non-detection of white dwarfs with rapid rotation, large magnetic fields or atypical kinematics. They do, however, suggest that some planetary systems may be lost to stellar mergers that do not leave an observational signature. To overcome this hurdle, our approach is more conservative, whereby we alter the observed mass distribution, accounting for the mass-dependent merger fraction from population synthesis models \citep{temmink2020}. These models predict that between 20--45\% of white dwarfs have formed through stellar mergers, with the fraction broadly increasing as a function of white dwarf mass (see Figure 9 of \citealt{temmink2020}).

\begin{figure*}
	\centering
 \subfloat{\includegraphics[width=0.7\textwidth]{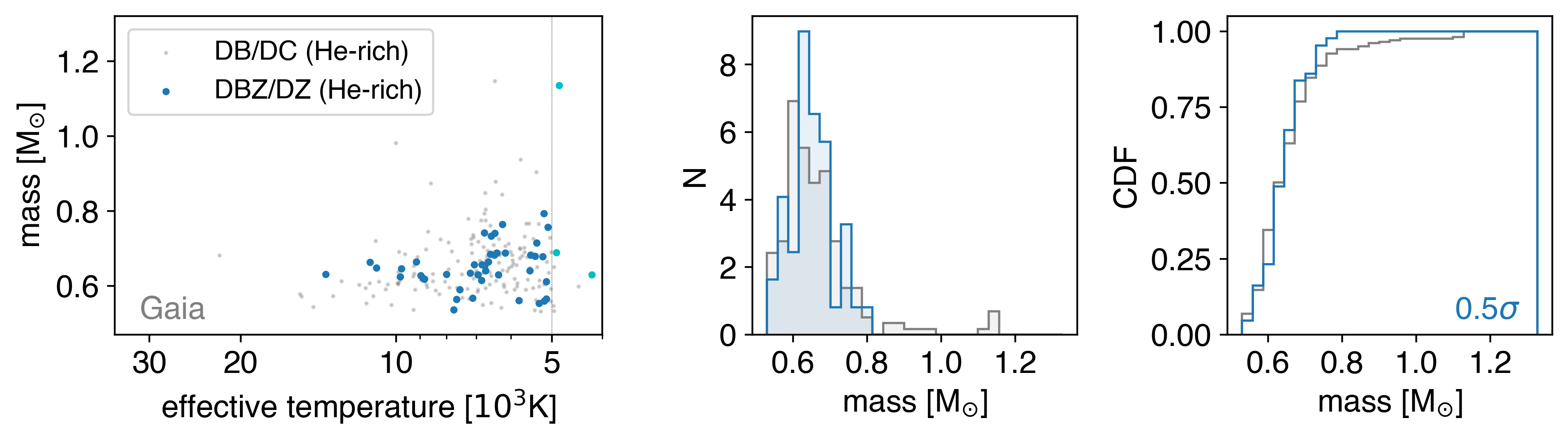}} 
 \hfill
 \subfloat{\includegraphics[width=0.27\textwidth]{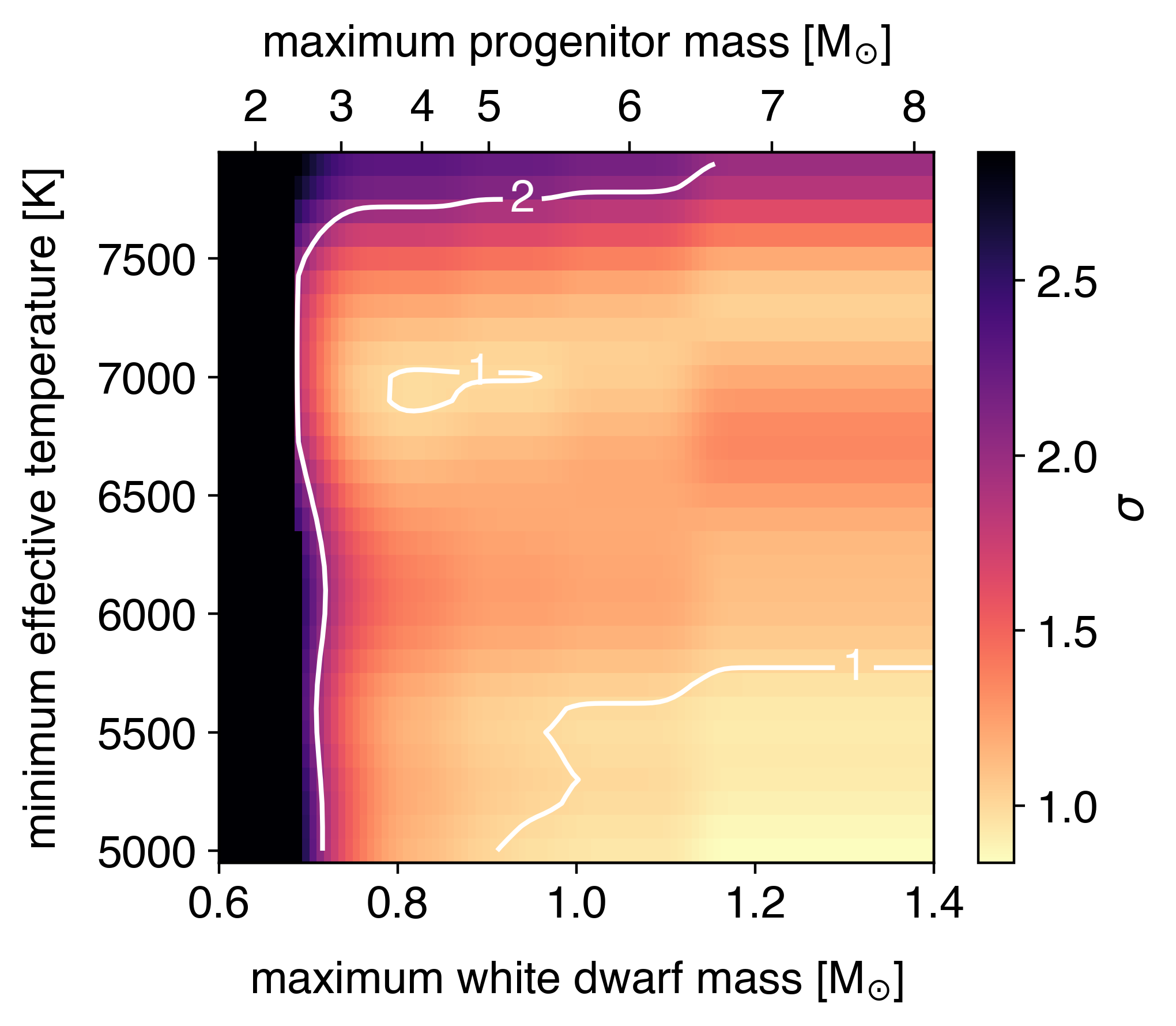}}
\\

 \subfloat{\includegraphics[width=0.7\textwidth]{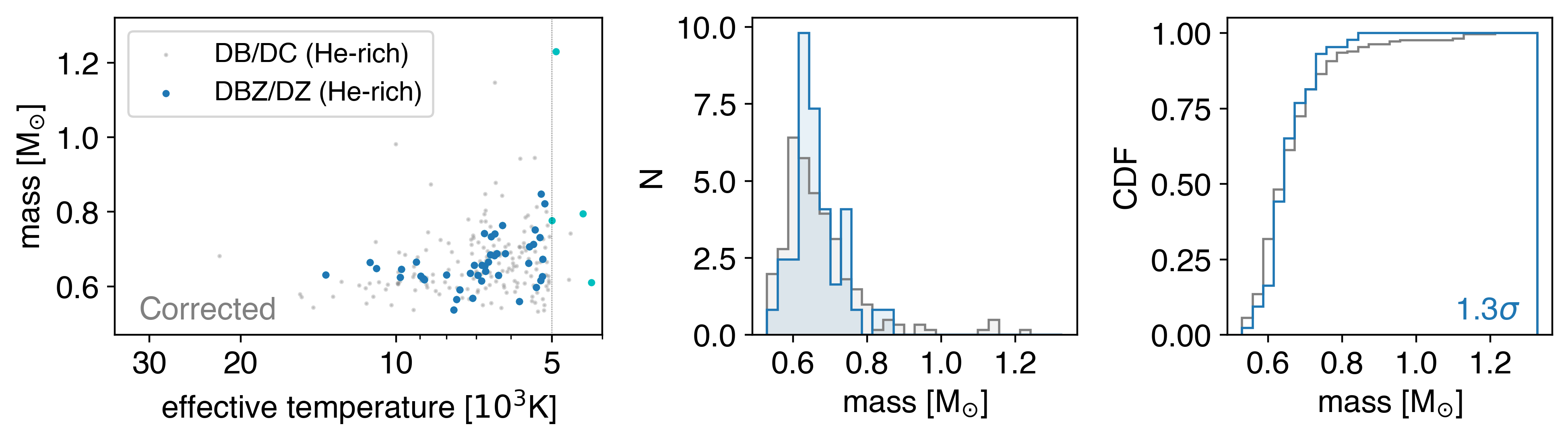}}
  \hfill
 \subfloat{\includegraphics[width=0.27\textwidth]{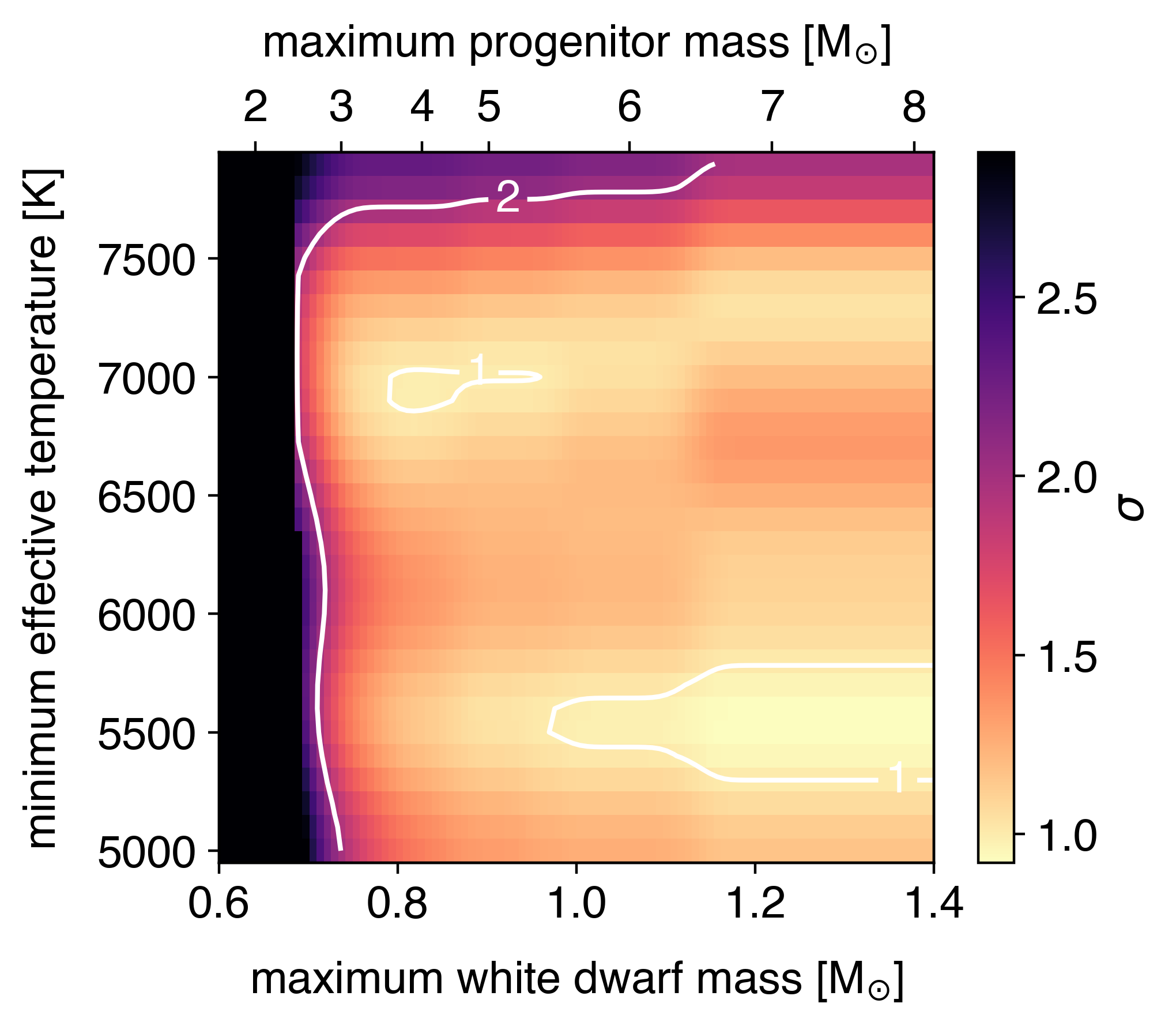}}
\\

 \subfloat{\includegraphics[width=0.7\textwidth]{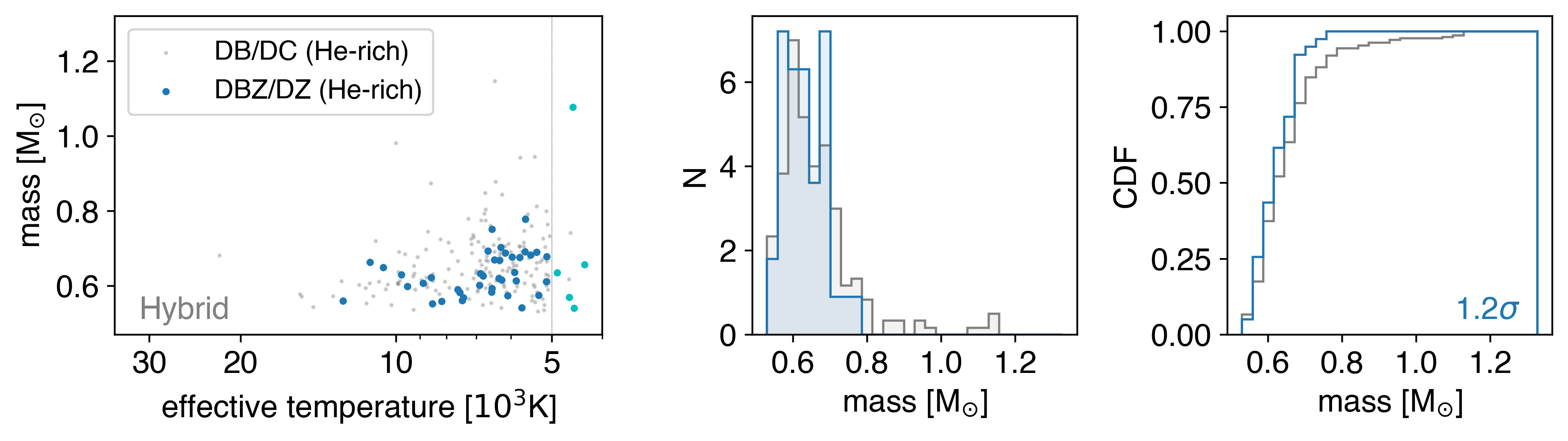}}
  \hfill
 \subfloat{\includegraphics[width=0.27\textwidth]{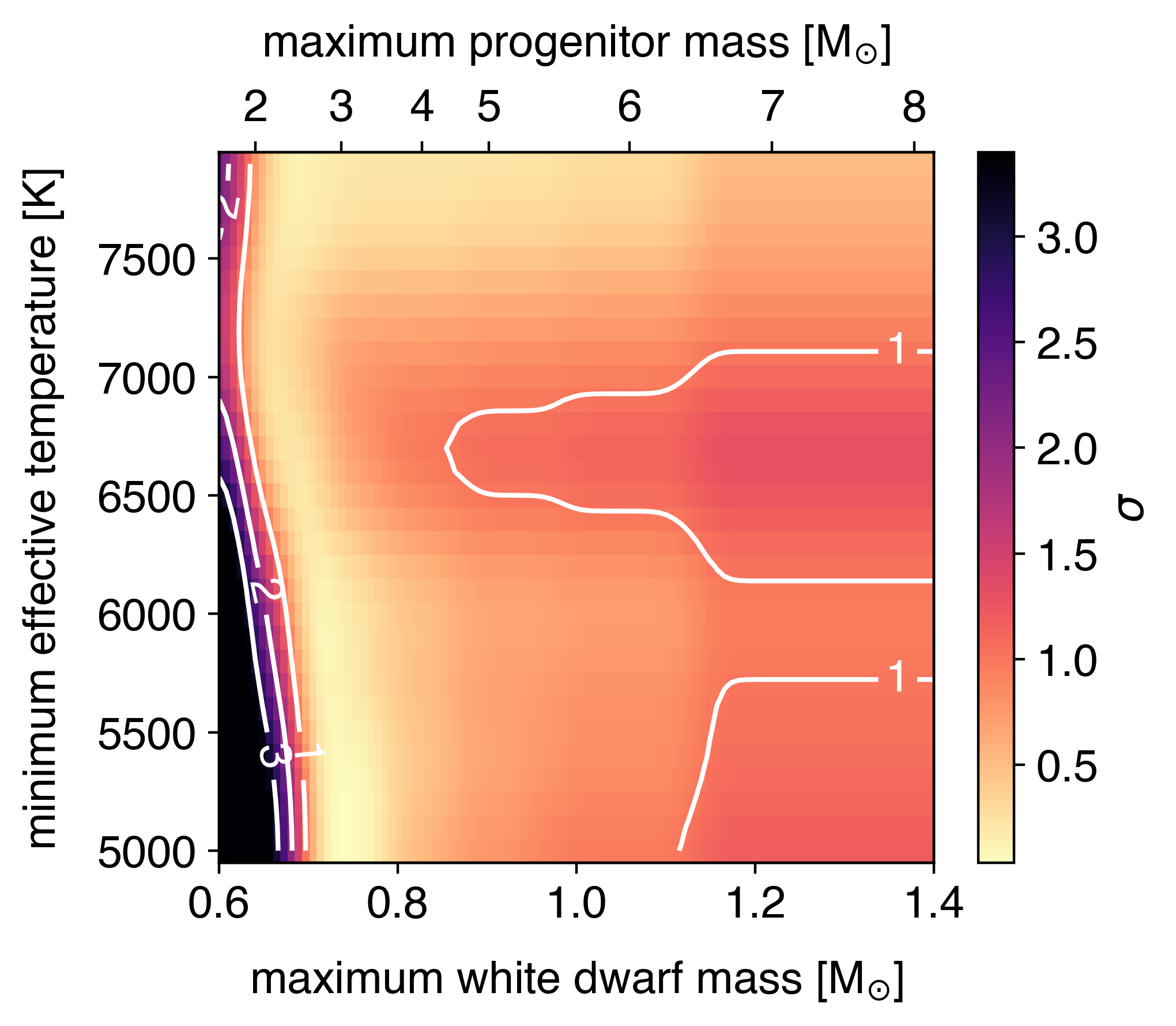}}
 
	\caption{From left-to-right (i--iv), we show; i) the white dwarf parameters, effective temperature and mass, for different He-atmosphere white dwarf samples. In grey, we show the He-atmosphere white dwarfs. In blue, we show the polluted He-atmosphere white dwarfs (DBZ/DZ/DZA) above 5000\,K. In cyan, we show the DZ white dwarfs with effective temperatures below 5000\,K. ii) With the same colours, mass distributions for the polluted (blue) and non-polluted (grey) samples shown in the left panel, with a minimum effective temperature cut of $T_{\rm eff}>5000$\,K. iii) Cumulative distribution functions for the polluted and non-polluted samples from panel ii. The statistical significance distinguishing the two distributions is written in the panel. iv) Similar to Fig.\,\ref{fg:mass-hist-syn-planetary-Mcut-DAZ}, but in the case of He-atmospheres. It shows the statistical result of the KS test in sigma units for a range of imposed maximum polluted white dwarf masses and minimum effective temperatures. The difference between the top, middle and bottom panels are the origin of the white dwarf parameters, effective temperature and mass. \textbf{Top}: White dwarf parameters are from fits to \textit{Gaia} photometry, as published by \citet{obrien2024}. \textbf{Middle}: White dwarf parameters are the ``corrected'' parameters from \citet{obrien2024}, where the correction is an ad hoc empirical correction for the ``low-mass problem'' at cool temperatures ($<$\,6000 K). \textbf{Bottom}: For the non-polluted white dwarf parameters, we take the ``corrected'' parameters (as in the middle panels). For the polluted white dwarfs, we adopt parameters from recent ``hybrid'' fits (combined spectroscopy and photometry) from the literature. These these parameters are from \citet{caron2023,OBrien2023,obrien2024,coutu19,subasavage2017}.}
	\label{fg:mass-teff-N-CDF}
\end{figure*}

\section{Helium atmospheres}
\label{sec:helium}
For comparison, we briefly consider the sample of helium atmosphere white dwarfs within 40\,pc. The reason for this is that neither the physics of planet formation and evolution, nor that of accretion and tidal disruption, are expected to depend upon the white dwarf atmospheric composition. Thus, for completeness, any explanation for the H-atmosphere distributions should be considered in the context of the He-atmosphere white dwarf sample. We provide this consideration for the two scenarios presented -- namely, i) the mass variation in planetary system formation and evolution, and ii) the episodic accretion model.

\subsection{Exoplanetary system formation and evolution}

Fig.\,\ref{fg:mass-teff-N-CDF} shows the mass-temperature distribution for the He-atmosphere sample. Significant uncertainties hamper the accurate determination of the stellar parameters for cool DZ white dwarfs, and it has been shown that the presence of metals and hydrogen can impact the derived stellar parameters \citep{dufour07}, unlike DAZ white dwarfs. This complicates the analysis compared with the H-atmosphere sample. To investigate the systematic uncertainties that could be introduced by model atmospheres, the top, middle and bottom panels make use of differently derived stellar parameters. The top panels use the \textit{Gaia} photometric parameters, as published in \citet{obrien2024}. The middle panels make use of the corrected parameters from the same source, where the correction accounts for the ``low-mass problem'' in cool DC white dwarfs. This ad hoc correction was calibrated against the DA and DC sample, and its validity for the DZ sample is broadly untested. In the bottom panel, for the polluted He-atmosphere white dwarfs (DBZ/DZ) we adopt  ``hybrid''-method parameters, based on combined fits of spectroscopic and photometric data, derived from multiple sources in the literature, but all including model atmospheres with metals. The primary sources of these parameters, in order of priority, are \citet{caron2023,OBrien2023,obrien2024,coutu19,subasavage2017}. For the non-polluted systems, we adopted the corrected parameters, as in the middle panel.

The left panels of Fig.\,\ref{fg:mass-teff-N-CDF} show the samples in effective temperature against mass.  In the second panel from the left we show the mass distributions for these samples. The next panel over shows the CDF, and we indicate the statistical significance with which the KS test rules out the polluted sample when compared to the non-polluted  sample. We find that all samples are self-consistent to within 1.5$\sigma$. For the He-atmosphere sample alone, it is thus clear that there is no statistically-significant difference between the polluted and non-polluted mass distributions. However, the purpose of this section is to consider whether the solutions invoked to reconcile the mass distributions of the polluted and non-polluted H-atmosphere white dwarf samples, can also by applied to the He-atmosphere samples to produce one globally consistent solution. In other words, in this section, we want to apply our solutions from Section\,\ref{sec:results} to the He-atmosphere samples and test whether the existing statistical agreement is maintained. 

As a test of the first solution -- the mass dependence of the planetary system formation/evolution scenario -- the right panels of Fig.\,\ref{fg:mass-teff-N-CDF} show the same analysis that was presented in Fig.\,\ref{fg:mass-hist-syn-planetary-Mcut-DAZ}. The ``hybrid'' parameters appear to favour a mass cut of 0.75\,M$_{\odot}$, fully consistent with the result from the H-atmosphere sample. We note that this result does not appear to depend on the choice of minimum effective temperature. Conversely, the \textit{Gaia} and ``corrected'' parameters appear to disfavour a mass cut, but this result has a moderate dependence on the minimum effective temperature of the samples. Nonetheless, the analysis shows that for all sets of models, the maximum polluted white dwarf mass is consistent at the 2$\sigma$ level with the results derived for the H-atmosphere sample. 

\subsection{Accretion model}
\begin{figure}
	\centering
    \subfloat{\includegraphics[width=0.2\textwidth]{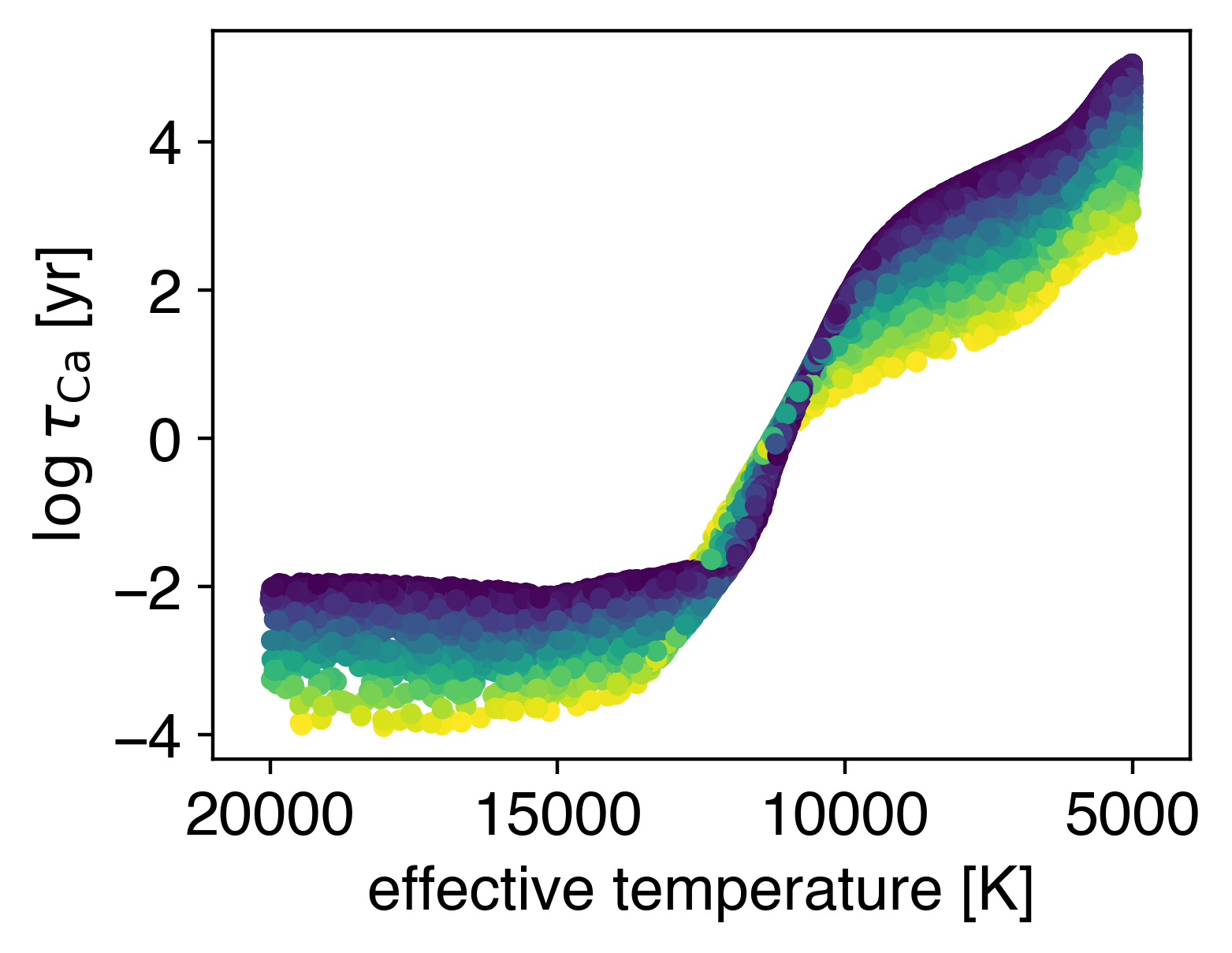}}
    \subfloat{\includegraphics[width=0.25\textwidth]{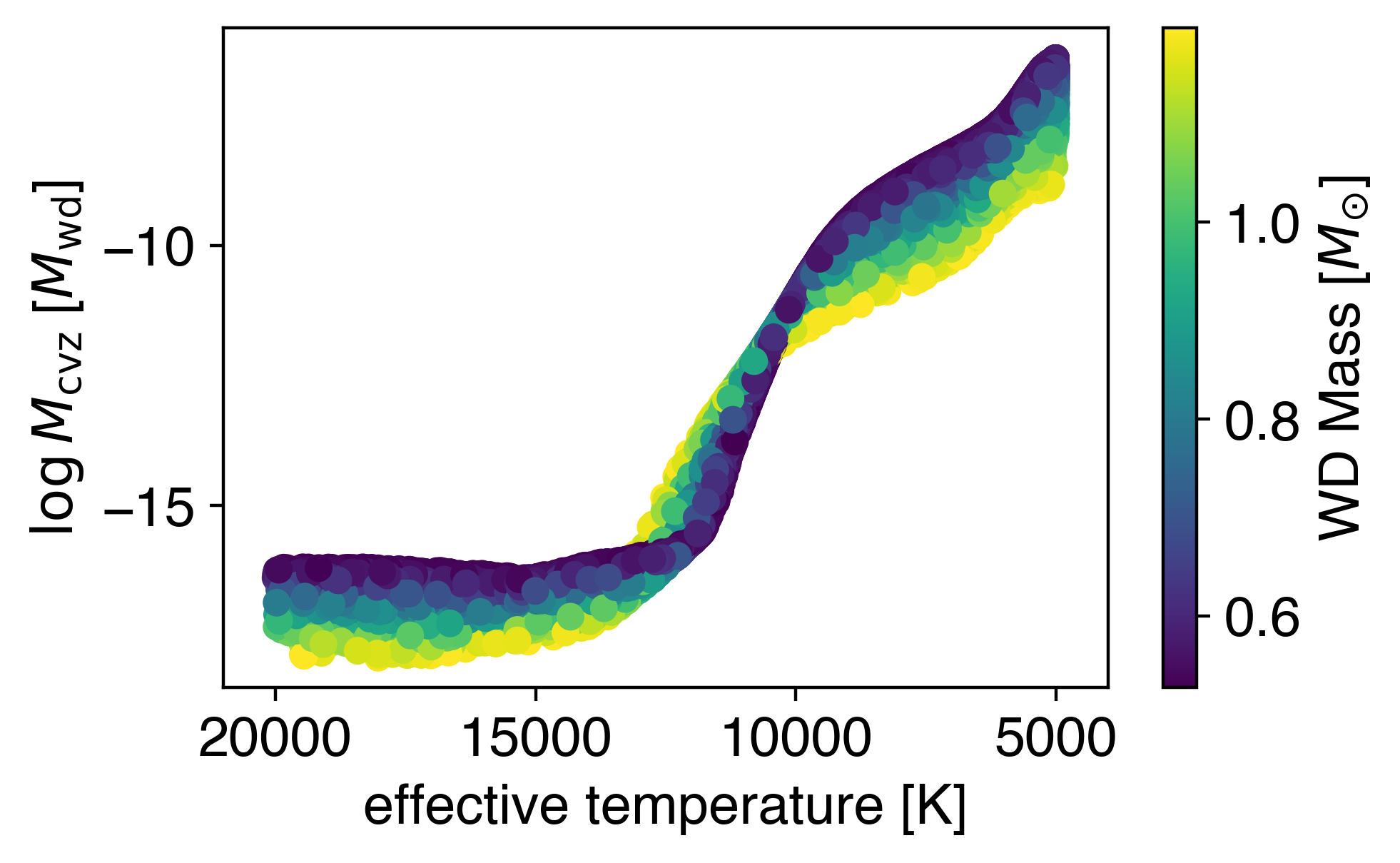}} \\
    \subfloat{\includegraphics[width=0.2\textwidth]{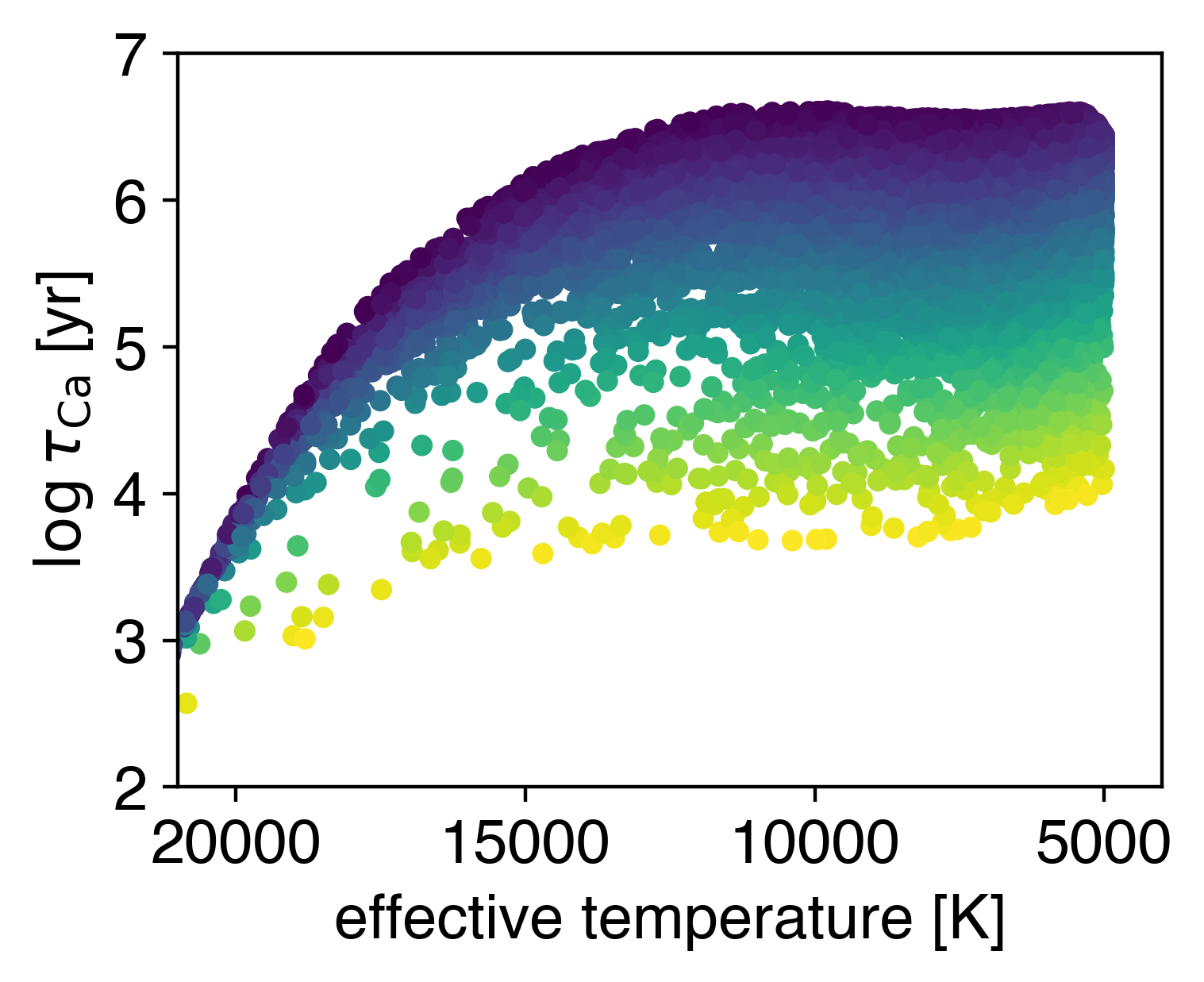}}
    \subfloat{\includegraphics[width=0.25\textwidth]{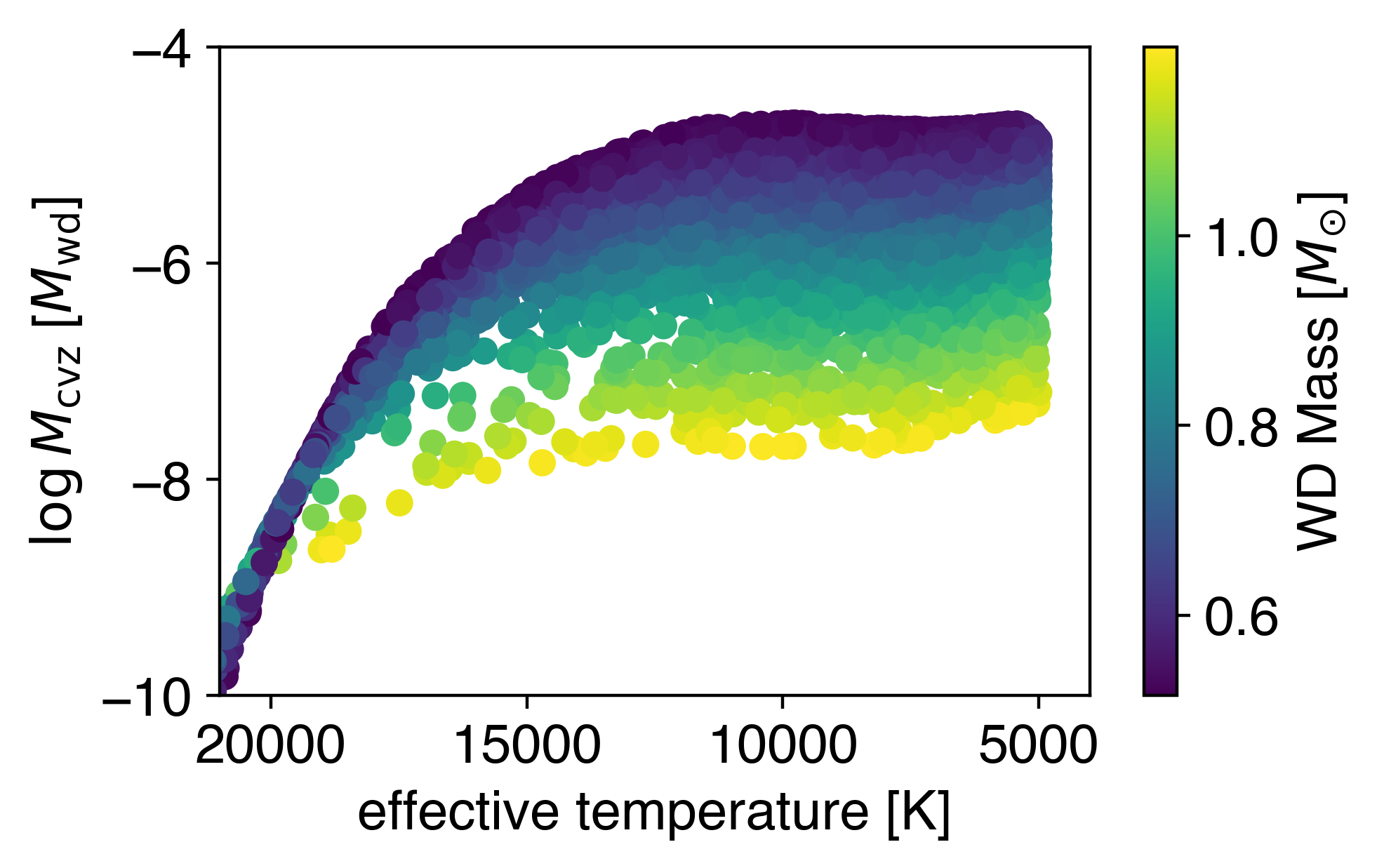}}
	\caption{\textbf{Top-left}: Diffusion timescale of trace calcium against a H-dominated atmosphere as provided in the model grids of \citet{koester2020} for the synthetic population of H-atmosphere white dwarfs. \textbf{Top-right}: Similar to the left panel, but for the convection zone mass in units of stellar mass. \textbf{Bottom}: Similar to the top two panels, but for the He-atmosphere synthetic population instead. 
The colours indicate the synthetic white dwarf mass, determined from the age and progenitor mass of each synthetic white dwarf, using the cooling models of \citet{bedard2020} and IFMR of \citet{cunningham2024}.}
	\label{fg:koester-grid-He}
\end{figure}

The diffusion timescales of He-atmosphere white dwarfs are typically at least 2 orders of magnitude larger than the H-atmosphere white dwarfs at a given effective temperature (see Fig.\,\ref{fg:koester-grid-He}). The volume-limited sample used in this study provides an unbiased population of Solar neighbourhood white dwarfs across coolings ages from 20\,Myr--10\,Gyr. Compared with UV and magnitude-limited samples, the sample includes a relatively larger fraction of cool white dwarfs.
The median temperature of the H-atmosphere and He-atmosphere samples used in this work is around 6500\,K. At this temperature, H-atmospheres are predicted to have sinking timescales on the order 10$^4$\,years, whilst He-atmosphere white dwarfs have sinking timescales of $10^6$\,years \citep{koester2020}. 
The preferred disk parameters from the DAZ accretion model are disk lifetimes around $t_{\rm d}$\,$\approx$\,$10^2$--$10^5$\,years and total accretion periods around $t_{\rm p}$\,$\approx$\,$2$\,$\times$\,$10^6$\,years, and the preferred parameters are sensitive to the diffusion timescale. 
In the following we test whether these parameters would produce a consistent description of the He-atmosphere metal-polluted sample, given the significantly longer diffusion timescales.

\begin{figure*}
	\centering
   	  \subfloat{\includegraphics[width=0.33\textwidth]{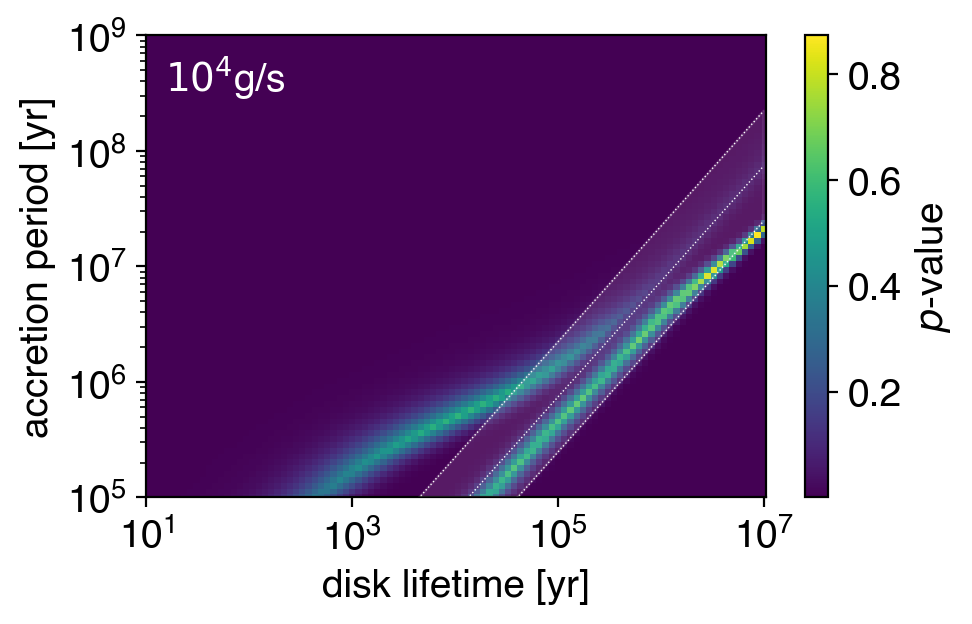}}
      \subfloat{\includegraphics[width=0.33\textwidth]{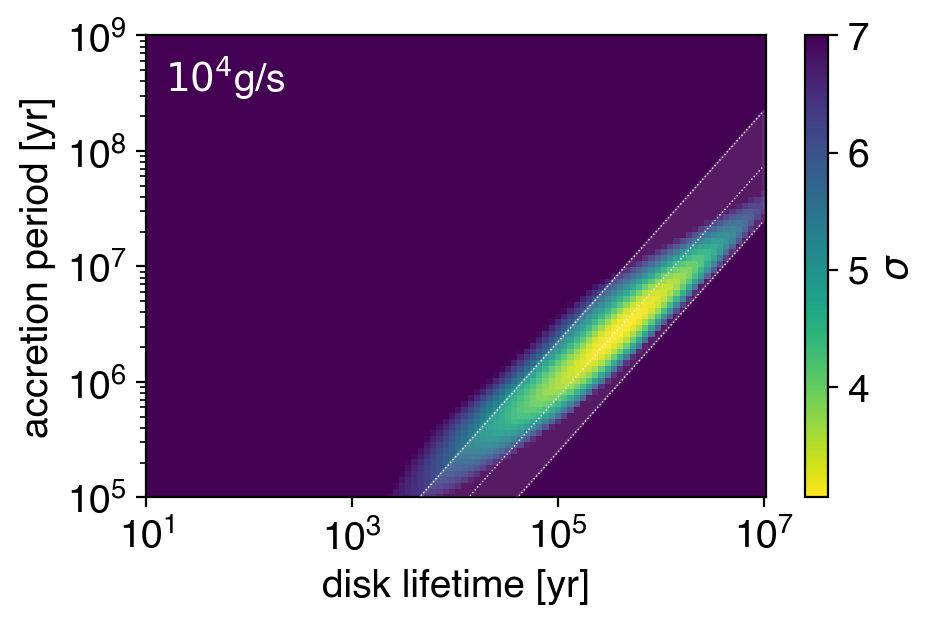}}\\
   	  \subfloat{\includegraphics[width=0.33\textwidth]{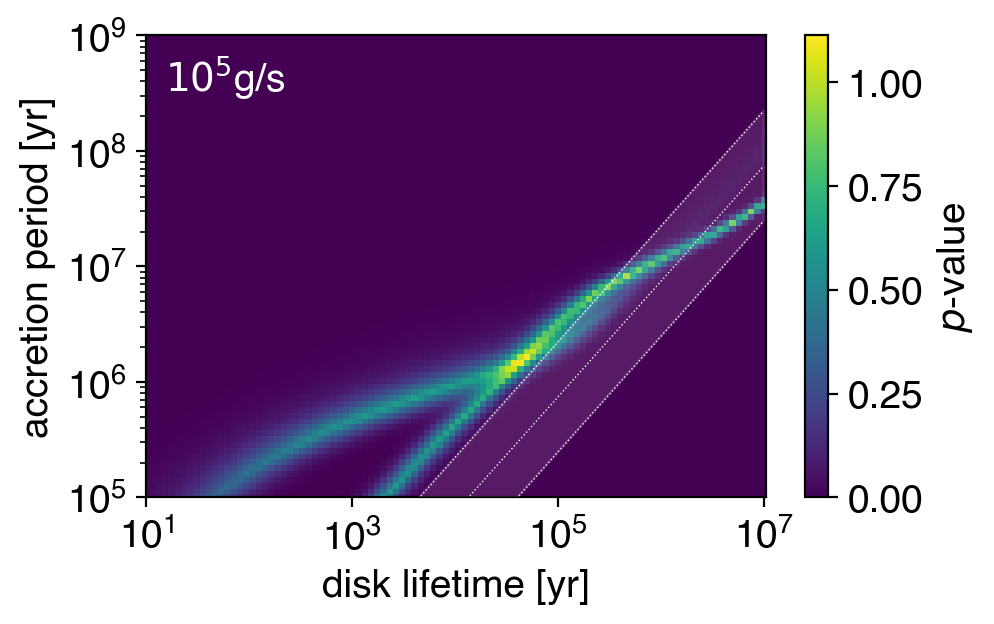}}
      \subfloat{\includegraphics[width=0.33\textwidth]{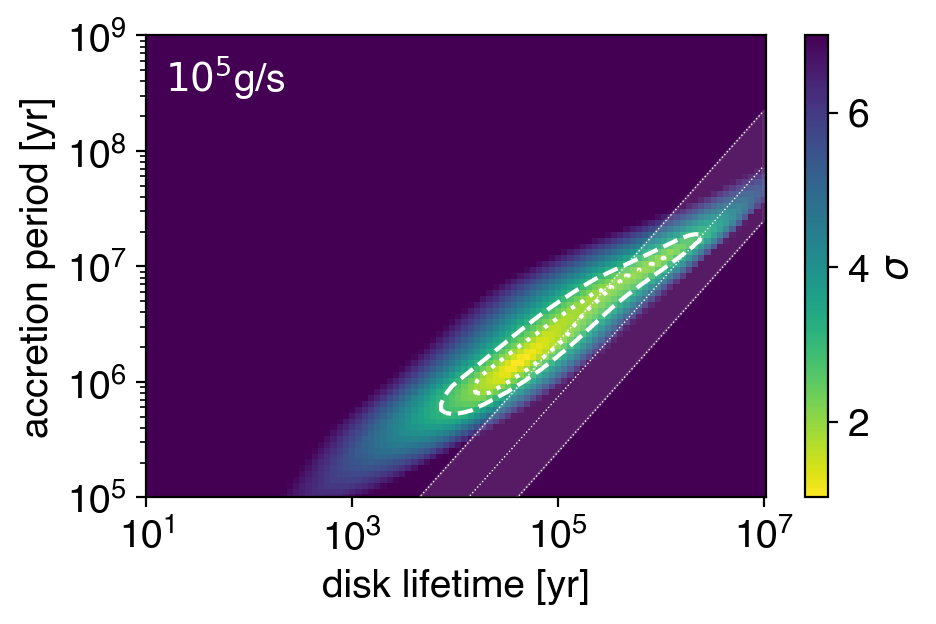}}\\
   	  \subfloat{\includegraphics[width=0.33\textwidth]{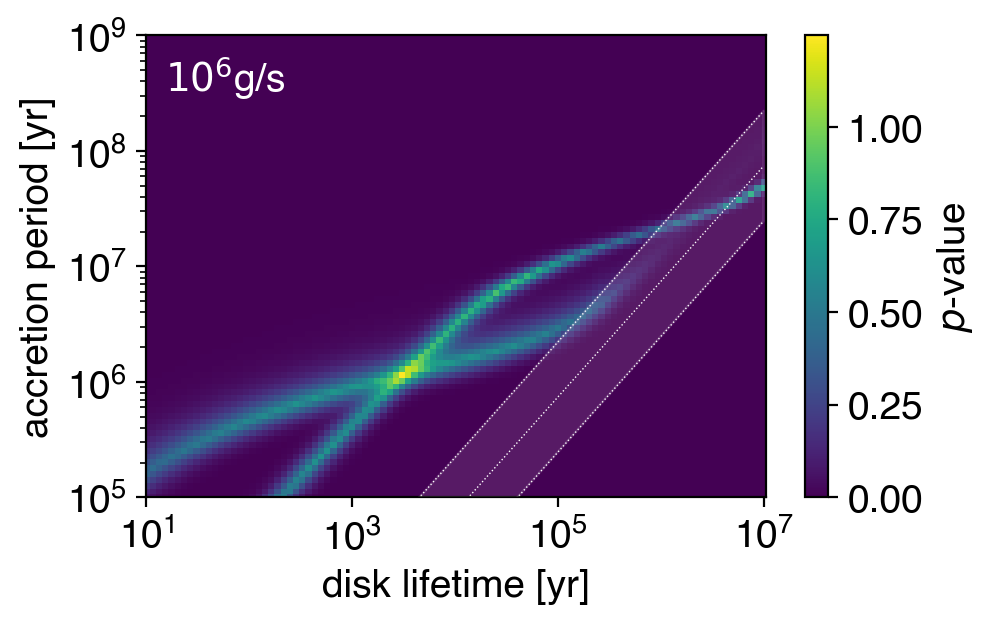}}
      \subfloat{\includegraphics[width=0.33\textwidth]{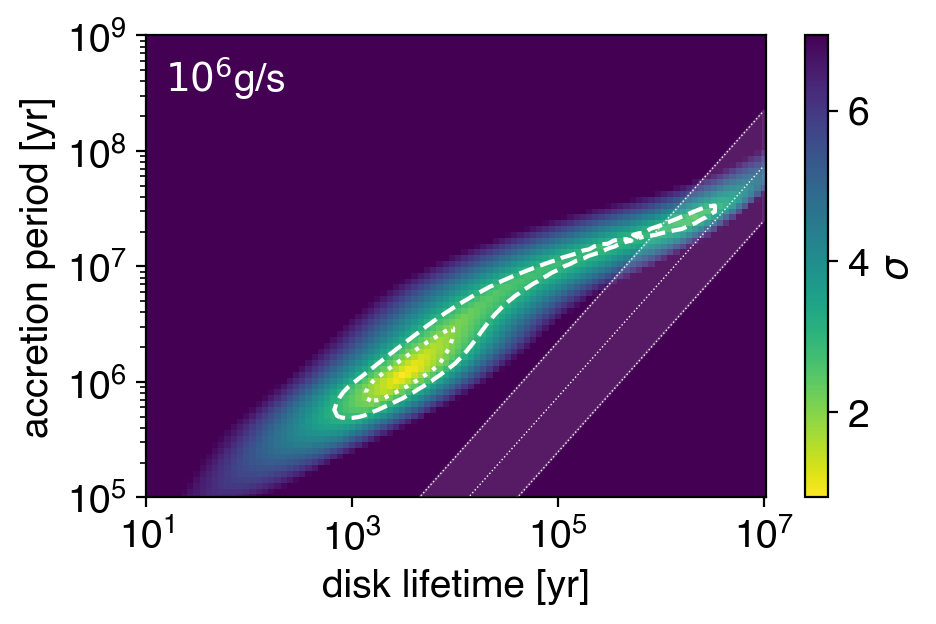}}\\
   	  \subfloat{\includegraphics[width=0.33\textwidth]{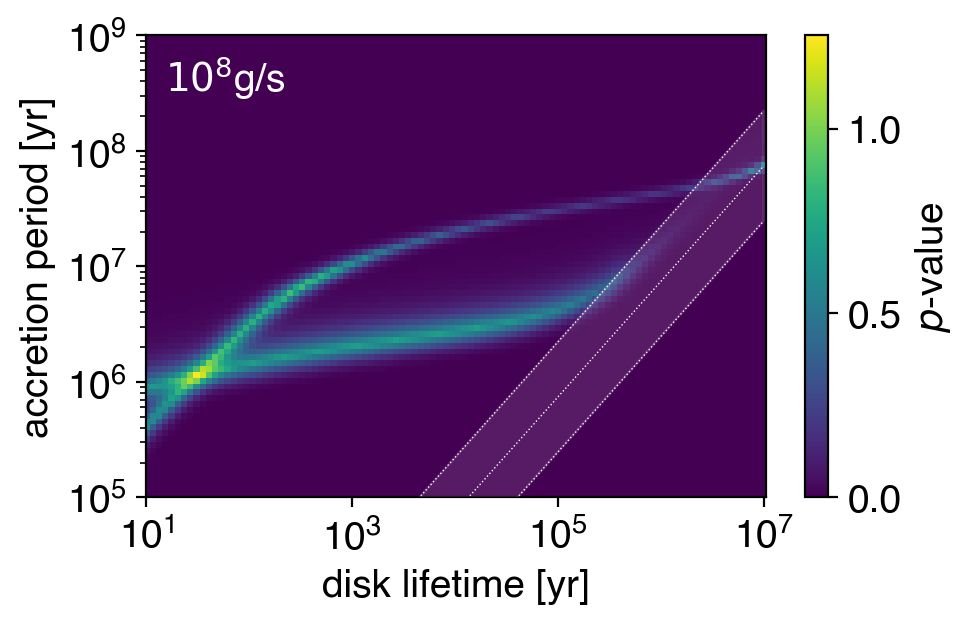}}
      \subfloat{\includegraphics[width=0.33\textwidth]{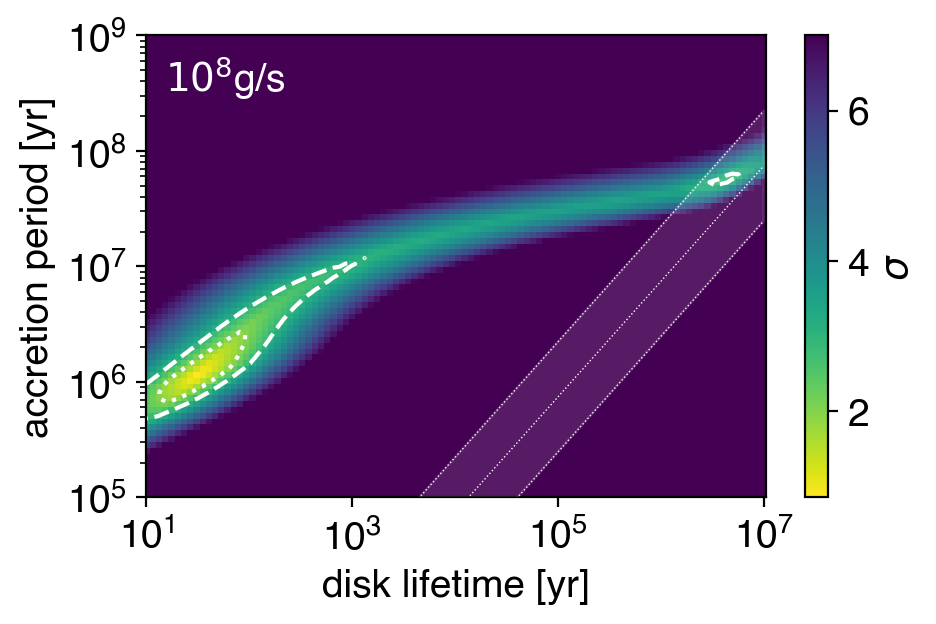}}\\
   	  \subfloat{\includegraphics[width=0.33\textwidth]{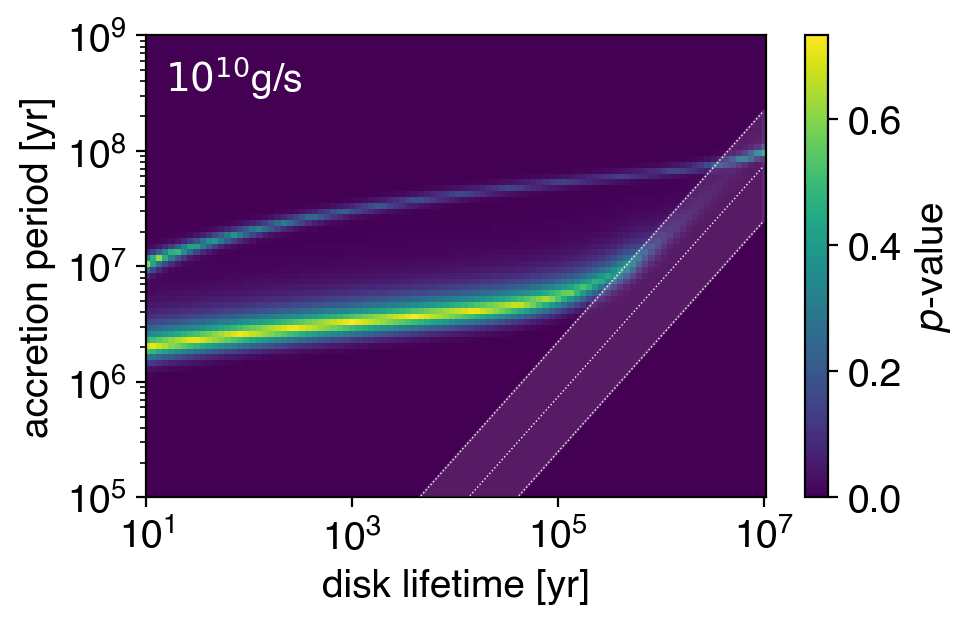}}
      \subfloat{\includegraphics[width=0.33\textwidth]{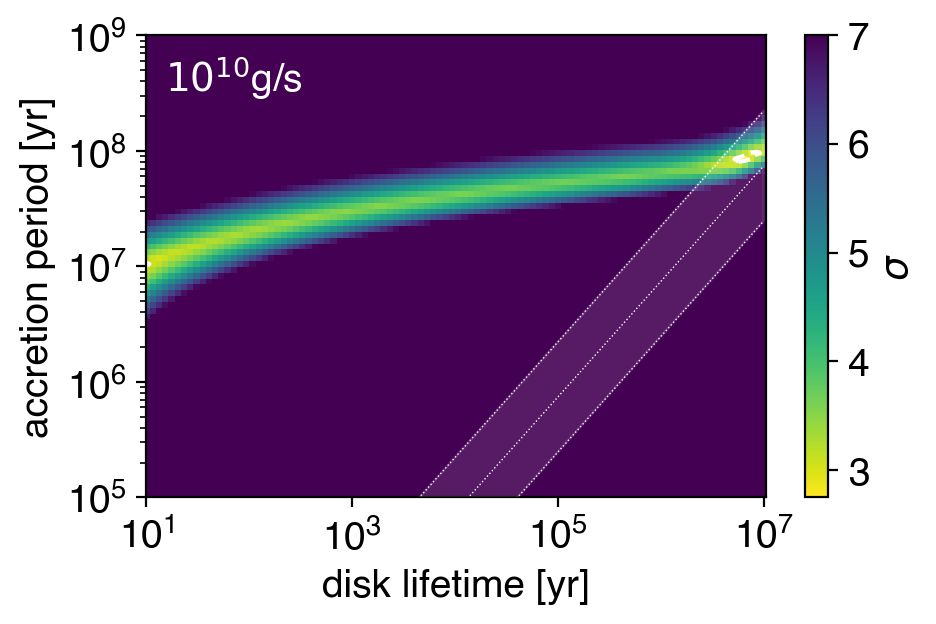}}\\
	\caption{\textbf{Left}: Additive $p$-values from the KS test for the H-atmosphere and He-atmosphere samples. \textbf{Right}: Similar to the lower panels of Fig.\,\ref{fg:pop-vary-EW}. From top-to-bottom, the center of the log normal distribution from which the synthetic accretion rates is increased. In order, the mean accretion rate for the log normal distribution is as follows: $\langle\log \dot{M_0}\rangle$ = 4.1, 5.1, 6.1, 8.1, and 10.1\,g/s. 
    For the He-atmosphere sample, we adopt the ``hybrid'' sample from Fig.\,\ref{fg:mass-teff-N-CDF}. In the right panels, we show contours of 2 and 3$\sigma$ in white dotted and dashed lines, respectively. The down turn in the He sample towards a minimum disk time originates from the significantly more massive convection zones of the He-atmosphere white dwarfs. For detailed discussion on the morphology of these distributions, see Appendix section \,\ref{sec:appendix-limits}.
 }
	\label{fg:pop-vary-Mdot-H-He}
\end{figure*}

We follow the same methodology as described in Section\,\ref{sec:PopSyn}, but this time for a non-DA sample only. Following the methodology of \citet{cunningham2024}, we refit the quantile IFMR to the mass distribution of DB/DC white dwarfs, adopting the ``corrected'' parameters from Fig.\,\ref{fg:mass-teff-N-CDF}. We then implement the same simple periodic accretion model to fit the DZ mass distribution using the ``hybrid'' parameters from Fig.\,\ref{fg:mass-teff-N-CDF}. 
The critical difference compared to H-atmospheres is in the diffusion timescales and convection zone masses, and for those we use the grids of metal-polluted He-atmosphere diffusion timescales of \citet{koester2020}. Fig.\,\ref{fg:koester-grid-He} shows the diffusion timescales and convection zone mass for trace calcium settling through a He-dominated atmosphere for the synthetic population of He-atmosphere white dwarfs. We use the same detection thresholds as in the H-atmosphere case, as we find the empirical detection thresholds from published Ca/He abundances from PEWDD \citep{williams2024} to be reasonably well described by the H-atmosphere thresholds. 

\begin{table}
\centering
\begin{tabular}{c c c c} 
 \hline
 $\langle\log\dot{M_0} [\rm{g/s}] \rangle$ & $\sigma_{\rm min}$ & $\log t_{\rm d}$\,[yr] & $\log t_{\rm p}$\,[yr] \\ [0.5ex] 
 \hline

3.1 & 6.3 & 6.2 & 6.7 \\
4.1 & 3.1 & 5.5 & 6.5 \\
\hline
5.1 & 1.0 & 4.7 & 6.2 \\
6.1 & 0.9 & 3.5 & 6.1 \\
8.1 & 0.9 & 1.5 & 6.1 \\
[1ex] 
 \hline
\end{tabular}
\caption{Table of best-fit parameters from Fig.\,\ref{fg:pop-vary-Mdot-H-He}. From top-to-bottom, the mean accretion rate increases, with the same order as in Fig.\,\ref{fg:pop-vary-Mdot-H-He}. The top two rows do not have an intercepting solution of the He and H simulations.}
\label{table:bf}
\end{table}

\subsubsection{Disk lifetimes and accretion periods}
\label{sec:discussion-disk-lifetimes}
In the left panels of Fig.\,\ref{fg:pop-vary-Mdot-H-He} we show the same analysis as in Fig.\,\ref{fg:pop-vary-Mdot}, but here in summing the $p$-values arising from both the H-atmosphere and He-atmosphere analyses. We note that the sum of $p$-values in this fashion is not fully meaningful as a probability (evidenced by values greater than 1 in some cases), but is provided for illustrative purposes to show the behaviour of the two independent (H and He) samples. In the right panels, we provide a multiplied probability, expressed in standard deviations, similar to Fig.\,\ref{fg:pop-vary-Mdot}, which we interpret as a meaningful statistical confidence. We also include the 3 and 2$\sigma$ contours. We find that there does exist a common solution at the 2$\sigma$ level for mean accretion rates between $10^5$--$10^8$\,g/s, total accretion periods of $\approx$$10^6$, and disk lifetimes in the range $10^2$--$10^5$\,years. We note that unlike in the H-atmosphere only case, the solutions in this range are now dependent on the mean accretion rate. This is due to the asymptotic behaviour of the helium simulation as it approaches an implied minimum disk lifetime. In Table\,\ref{table:bf} we summarise the best-fit parameters for the 5 simulations shown in Fig.\,\ref{fg:pop-vary-Mdot-H-He}.
There is also a second solution, consistent at the 3$\sigma$ level, with disk lifetimes around $10^6$\,years and accretion periods of $10^7$\,years. This solution is caused by the upturn in the hydrogen simulation as it enters a linear regime at disk lifetimes significantly longer than the diffusion timescales. This solution is less-favoured due to the KS test comparison against the observed polluted mass distribution. We point out that this secondary solution at longer disk lifetimes, does fall within the expected observational constraints on the disk lifetime and infrared excess fraction, but is not the statistically-favoured solution.

Table\,\ref{table:bf} summarises the best-fit solutions from the figure. Increasing from top-to-bottom, we show one row for each of the accretion rates featured in Fig.\,\ref{fg:pop-vary-Mdot-H-He}. We also show the minimum $\sigma$ reached by the model, and the corresponding disk lifetime and total accretion period. We see that accretion rates in the range $10^5$--$10^8$\,g/s give rise to solutions consistent at the 1$\sigma$ level.

The location of this preferred solution occurs where the He simulation exhibits an inflection and enters a linear regime with a steeper slope towards shorter disk lifetimes. It is only in this limiting linear regime that consistent metal-polluted fractions in the H- (8\%) and He-atmosphere (32\%) simulations are recovered. The slope in this regime can be approximated in the limit that $t_{\rm d}\ll\tau_i$ as 
\begin{align}
    t_{\rm p}&\sim t_{\rm d}\frac{\chi_{{\rm ss},i}}{\chi_{{\rm det},i}}\\
    &=t_{\rm d}\frac{\tau_i\dot{M_i}}{M_{\rm cvz}\chi_{{\rm det},i}}~,
    \label{eq:tdmin-analytic}
\end{align}
where denominator in this equation is the minimum detectable mass of metal for a white dwarf with a surface convection zone of a given mass (see Appendix Section\,\ref{sec:appendix-limits} for full derivation). 

\subsubsection{Planetesimal mass}
The inverse proportionality between this best-fit disk lifetime and the accretion rate implies that within our model there is a preferred total mass of an accreted body. This is somewhat intuitive since higher accretion rates will deplete circumstellar debris disks in shorter times.
Given that the best-fit common solution occurs at the intersection of the limiting linear relation resulting from the He simulation, defined by Eq.\,\eqref{eq:tdmin-analytic}, and the plateau in $t_{\rm p}^{\rm H}$ resulting from the H simulation, we can estimate the preferred mass of accreted planetesimal as
\begin{align}
    M_{{\rm P}} =\frac{1}{f_{i}}\frac{t_{\rm p}^{\rm H}}{\tau_i^{\rm He}}M_{\rm cvz}^{\rm He}\chi_{{\rm det},i}^{\rm He}~,
\end{align}
where $f_i$ is the fraction of the accreted body composed of element $i$, and the superscripts indicate whether the parameter comes from the H and He solution. Given that the H simulation finds a consistent accretion period of $t_{\rm p}^{\rm H}\sim 10^6$\,years, we can estimate the peak of the mass distribution of polluting planetesimals to be 
\begin{align}
    M_{{\rm P}} \sim10^{17}{\rm g}\left(\frac{0.016}{f_{i}}\right)\left(\frac{t_{\rm p}}{10^6\,\rm{yr}}\right) \left(\frac{\tau_i^{\rm He}}{10^6\,\rm{yr}}\right) \left(\frac{M_{\rm cvz}^{\rm He}}{10^{-5}\,M_{\odot}}\right)\left(\frac{\chi_{{\rm det},i}^{\rm He}}{10^{-13}}\right)~.
\end{align}

\begin{figure}
	\centering
	  \subfloat{\includegraphics[width=0.42\textwidth]{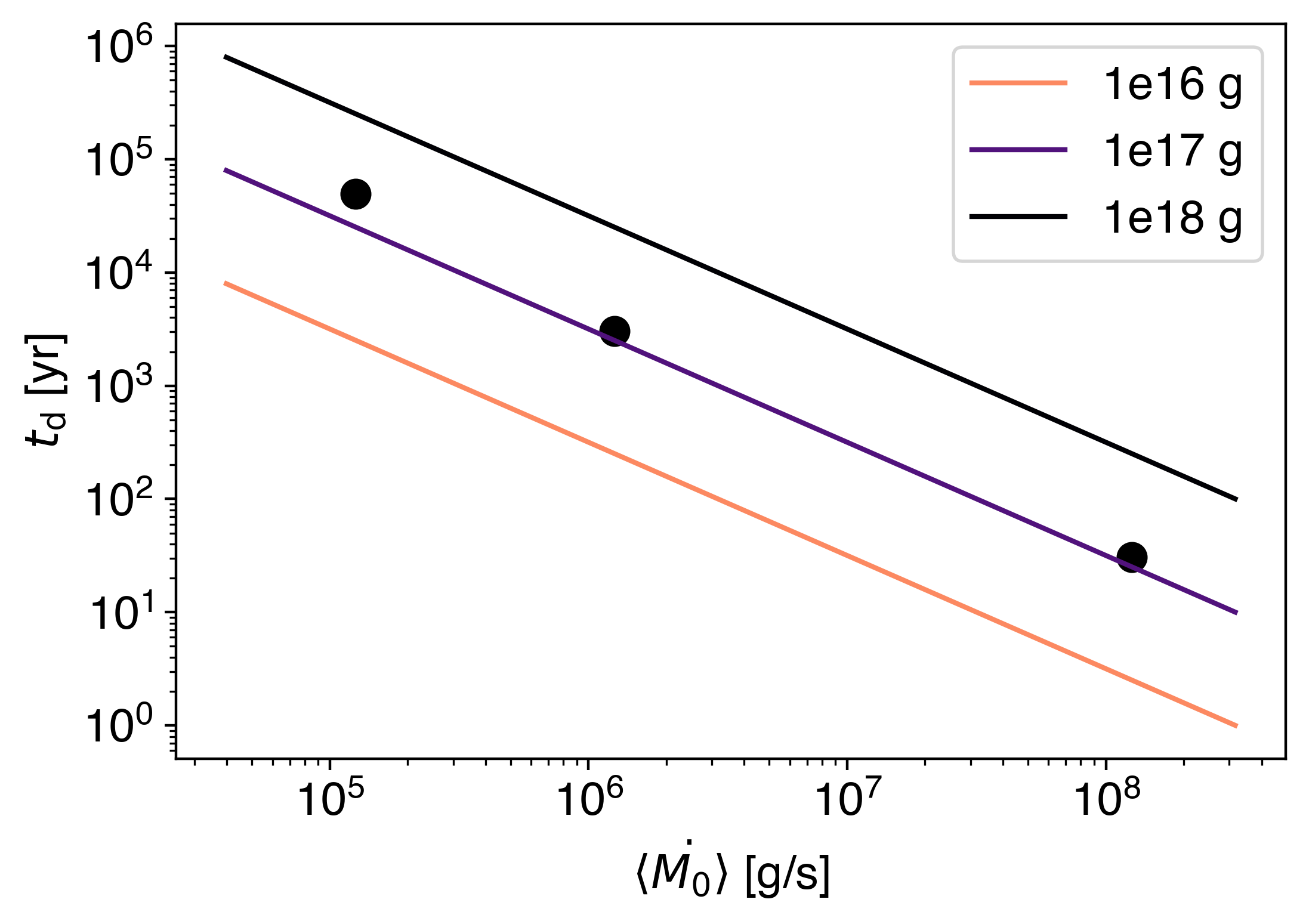}}
	\caption{Best-fit disk lifetimes for the three mean accretion rates that yielded a statistically consistent description of the observations within 2$\sigma$. We also show lines of constant total planetesimal mass.}
	\label{fg:Mdot-vs-bf-tdisk}
\end{figure}

\begin{figure}
	\centering
	  \subfloat{\includegraphics[width=0.5\textwidth]{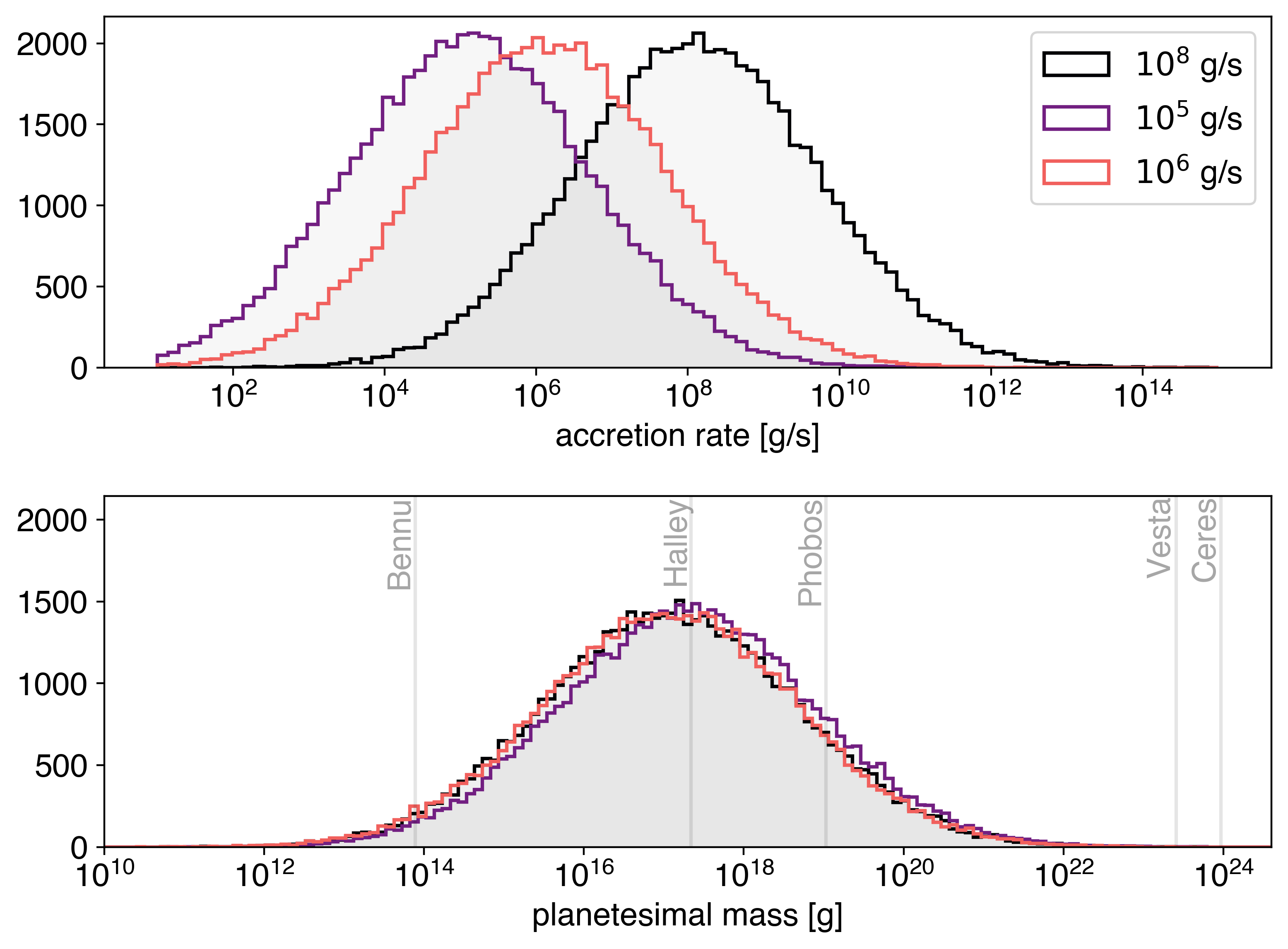}}
	\caption{\textbf{Top}: Log-normal distributions of accretion rates serving as inputs into our periodic accretion model. The distribution derived by \citet{wyatt14} has a center of $\dot{M}=8.1$ and standard deviation of $\sigma_{\dot{M}}=1.6$ and is shown in black. The accretion rates shown in this figure correspond to the top 4 panels of Fig.\,\ref{fg:pop-vary-Mdot-H-He}. \textbf{Bottom}: We take the approximate best-fit value of disk lifetime from Fig\,\ref{fg:pop-vary-Mdot-H-He}, multiplied by the accretion rate (top panel) to produce an implied mass distribution of accreted planetary bodies. }
	\label{fg:planetesimal-mass}
\end{figure}

\noindent 
In Fig.\,\ref{fg:Mdot-vs-bf-tdisk} in black dots, we show the best-fit disk lifetime for the three simulations with a solution at the 1$\sigma$ level, plotted against the mean accretion rate. We also show lines of constant average planetesimal mass. We see that the results of our three simulations at $\langle \dot{M_0}\rangle=10^{8.1}$, $10^{6.1}$, and $10^{5.1}$\,g/s follow very closely a relation with constant planetesimal mass of $M_{\rm P}=10^{17}$\,g.
In Fig.\,\ref{fg:planetesimal-mass}, we show the implied distribution of parent body masses for each log-normal distribution of accretion rates and the corresponding preferred disk lifetime resulting from our model.
The accretion rate distributions vary only in the log mean accretion rate.
The preferred mean planetesimal mass per accretion event is approximately equal to the mass of Halley's comet ($M_{\rm P}\sim10^{17}$\,g). We note that this result does not depend on the input accretion rate.
This result implies that the likely peak of the progenitor asteroid mass distribution for the parent bodies that accrete onto the observable metal-polluted white dwarf population is many orders of magnitude less than the inferred mass of the asteroid that broke up around WD\,1145+017, for which the parent body mass has been estimated to be on the order $10^{23}$\,g     \citep{rappaport2016,gurri2017,veras2017-WD1145}.
Given that the best-fit accretion periods are on the order 10$^6$\,years, our results imply that the typical white dwarf undergoes $\sim$10$^4$ accretion events over its lifetime. 

\subsection{Variable accretion rates}
In our model, we make the simplifying assumption that when accretion takes places, it does so at some constant accretion rate for the duration of the disk lifetime. This assumption is made in order to isolate the biasing effect of the white dwarf atmosphere, whilst also providing a significant computational advantage. However, we acknowledge that this is not as physical as more sophisticated models of accretion disks such as the stochastic accretion model employed by \citet{wyatt14}, or the collisional cascade models of \citet{kenyon2017a}, both of which account for the likelihood mass or size distribution of tidally-disrupted planetary bodies. Our model also omits other physics which has been accounted for in accretion models. For instance, there is a wealth of disk modeling which accounts for Poynting-Robertson drag as a dominant source of angular momentum loss \citep{brouwers2022}, the production of gas as a mechanism for angular momentum loss through dust-gas interactions \citep{okuya2023}, and models that suggest that accretion from a differentiated planetary body (e.g., a sufficiently large asteroid) may undergo asynchronous accretion \citep{brouwers2023-core-mantle}, in which the composition of the material hitting the white dwarf surface may vary over the disk lifetime. 

Observationally, there is a wealth of evidence that debris disks are not stable in time, and it is common for white dwarfs to exhibit variability in their infrared excess on the timescale of months to years \citep{swan19,swan2021,xu18}. That being said, whether infrared variability implies variability in the accretion rate is an open question. 

It is beyond the scope of this work to incorporate the detailed physics of accretion from debris disks into our model. Furthermore, our simple model presents a solution consistent with the observations of the 40\,pc sample. A larger sample with more robust number statistics in the future may necessitate the development of a more complete and physical model of the accretion physics in this accretion--diffusion scenario.

\section{Conclusions}
\label{sec:conclusions}

Our simple, episodic accretion model provides a consistent explanation for the observed mass distributions of metal-polluted, H-atmosphere and He-atmosphere white dwarfs, but only at disk lifetimes less than $\sim$\,$10^5$\,years. This result applies to the unbiased, spectroscopically-complete, volume-limited 40pc which comprises white dwarfs across a wide range of cooling ages ($\approx$20\,Myr--10\,Gyr). In developing this model, we have taken into account the dominant sources of observational and modelling biases, including the stellar merger fraction, heterogeneous resolution of the input data, and variable accretion rates. We point out that there are alternative possible explanations for the dearth of cool ($<$10\,000\,K), massive ($>$0.8\,M$_{\odot}$), metal-polluted white dwarfs, including i) hypothesised lower planetary system formation rates around intermediate-mass main sequence stars, ii) higher rates of engulfment of close-in planets around intermediate-mass stars, iii) lower time-averaged asteroid perturbation events over the lifetime of more massive white dwarfs. We demonstrate that the recent findings of \citet{OuldRouis2024}, which suggest an absence of active accretion in warm ($>$12\,000,K) DA white dwarfs with masses exceeding $0.7$\,M$_{\odot}$, offer an alternative explanation for the scarcity of high-mass, metal-polluted white dwarfs observed in this study. We refine the determination of this mass limit, and empirically determine a mass limit of $M_{\rm wd}$\,$<$\,$0.72^{+0.07}_{-0.03}$\,M$_{\odot}$.
In turn, this implies an empirical upper limit on the mass of progenitor stars to evolved planetary systems of $M_{\rm ZAMS}^{\rm max}=2.9^{+0.7}_{-0.3}$\,M$_{\odot}$.
Therefore, we conclude that the dearth of high-mass metal-polluted white dwarfs in volume limited samples could have an intrinsic (white dwarfs physics) or extrinsic explanation (real difference in their planetary system architectures).

It is well-accepted that white dwarfs do undergo episodic accretion. Thus we have demonstrated that episodic accretion will produce significant bias in the observed mass distribution of metal-polluted white dwarfs. 

\section*{Acknowledgements}
We are grateful to Karel Temmink for providing the theoretical probabilistic merger fraction data from \citet{temmink2020}. 
Support for this work was provided by NASA through the NASA Hubble Fellowship grant HST-HF2-51527.001-A awarded by the Space Telescope Science Institute, which is operated by the Association of Universities for Research in Astronomy, Inc., for NASA, under contract NAS5-26555.
This research was supported by a Leverhulme Trust Grant (ID RPG-2020-366). PET and MOB received funding from the European Research Council under the European Union’s Horizon 2020 research and innovation programme number 101002408 (MOS100PC).

\section*{Data Availability}
The observational data used in this article are published in \citet{mccleery2020,Gentile2021,OBrien2023,obrien2024}. The data from the latter is available at \url{https://cygnus.astro.warwick.ac.uk/phrtxn/}. The results presented in this study will be made available upon reasonable request to the corresponding author.




\bibliographystyle{mnras}
\bibliography{mybib}

\begin{thebibliography}{}
\makeatletter
\relax
\def\mn@urlcharsother{\let\do\@makeother \do\$\do\&\do\#\do\^\do\_\do\%\do\~}
\def\mn@doi{\begingroup\mn@urlcharsother \@ifnextchar [ {\mn@doi@} {\mn@doi@[]}}
\def\mn@doi@[#1]#2{\def\@tempa{#1}\ifx\@tempa\@empty \href {http://dx.doi.org/#2} {doi:#2}\else \href {http://dx.doi.org/#2} {#1}\fi \endgroup}
\def\mn@eprint#1#2{\mn@eprint@#1:#2::\@nil}
\def\mn@eprint@arXiv#1{\href {http://arxiv.org/abs/#1} {{\tt arXiv:#1}}}
\def\mn@eprint@dblp#1{\href {http://dblp.uni-trier.de/rec/bibtex/#1.xml} {dblp:#1}}
\def\mn@eprint@#1:#2:#3:#4\@nil{\def\@tempa {#1}\def\@tempb {#2}\def\@tempc {#3}\ifx \@tempc \@empty \let \@tempc \@tempb \let \@tempb \@tempa \fi \ifx \@tempb \@empty \def\@tempb {arXiv}\fi \@ifundefined {mn@eprint@\@tempb}{\@tempb:\@tempc}{\expandafter \expandafter \csname mn@eprint@\@tempb\endcsname \expandafter{\@tempc}}}

\bibitem[\protect\citeauthoryear{{Barber}, {Patterson}, {Kilic}, {Leggett}, {Dufour}, {Bloom}  \& {Starr}}{{Barber} et~al.}{2012}]{barber2012}
{Barber} S.~D.,  {Patterson} A.~J.,  {Kilic} M.,  {Leggett} S.~K.,  {Dufour} P.,  {Bloom} J.~S.,   {Starr} D.~L.,  2012, \mn@doi [\apj] {10.1088/0004-637X/760/1/26}, \href {https://ui.adsabs.harvard.edu/abs/2012ApJ...760...26B} {760, 26}

\bibitem[\protect\citeauthoryear{{Bauer} \& {Bildsten}}{{Bauer} \& {Bildsten}}{2019}]{bauer2019}
{Bauer} E.~B.,  {Bildsten} L.,  2019, \mn@doi [\apj] {10.3847/1538-4357/ab0028}, \href {https://ui.adsabs.harvard.edu/abs/2019ApJ...872...96B} {872, 96}

\bibitem[\protect\citeauthoryear{{B{\'e}dard}, {Bergeron}, {Brassard}  \& {Fontaine}}{{B{\'e}dard} et~al.}{2020}]{bedard2020}
{B{\'e}dard} A.,  {Bergeron} P.,  {Brassard} P.,   {Fontaine} G.,  2020, \mn@doi [\apj] {10.3847/1538-4357/abafbe}, \href {https://ui.adsabs.harvard.edu/abs/2020ApJ...901...93B} {901, 93}

\bibitem[\protect\citeauthoryear{{Bonsor} \& {Wyatt}}{{Bonsor} \& {Wyatt}}{2010}]{bonsor2010}
{Bonsor} A.,  {Wyatt} M.,  2010, \mn@doi [\mnras] {10.1111/j.1365-2966.2010.17412.x}, \href {https://ui.adsabs.harvard.edu/abs/2010MNRAS.409.1631B} {409, 1631}

\bibitem[\protect\citeauthoryear{{Brouwers}, {Bonsor}  \& {Malamud}}{{Brouwers} et~al.}{2022}]{brouwers2022}
{Brouwers} M.~G.,  {Bonsor} A.,   {Malamud} U.,  2022, \mn@doi [\mnras] {10.1093/mnras/stab3009}, \href {https://ui.adsabs.harvard.edu/abs/2022MNRAS.509.2404B} {509, 2404}

\bibitem[\protect\citeauthoryear{{Brouwers}, {Bonsor}  \& {Malamud}}{{Brouwers} et~al.}{2023}]{brouwers2023-core-mantle}
{Brouwers} M.~G.,  {Bonsor} A.,   {Malamud} U.,  2023, \mn@doi [\mnras] {10.1093/mnras/stac3316}, \href {https://ui.adsabs.harvard.edu/abs/2023MNRAS.519.2646B} {519, 2646}

\bibitem[\protect\citeauthoryear{{Caron}, {Bergeron}, {Blouin}  \& {Leggett}}{{Caron} et~al.}{2023}]{caron2023}
{Caron} A.,  {Bergeron} P.,  {Blouin} S.,   {Leggett} S.~K.,  2023, \mn@doi [\mnras] {10.1093/mnras/stac3733}, \href {https://ui.adsabs.harvard.edu/abs/2023MNRAS.519.4529C} {519, 4529}

\bibitem[\protect\citeauthoryear{{Chayer}, {Fontaine}  \& {Wesemael}}{{Chayer} et~al.}{1995}]{chayer95a}
{Chayer} P.,  {Fontaine} G.,   {Wesemael} F.,  1995, \mn@doi [\apjs] {10.1086/192184}, \href {http://adsabs.harvard.edu/abs/1995ApJS...99..189C} {99, 189}

\bibitem[\protect\citeauthoryear{{Coutu}, {Dufour}, {Bergeron}, {Blouin}, {Loranger}, {Allard}  \& {Dunlap}}{{Coutu} et~al.}{2019}]{coutu19}
{Coutu} S.,  {Dufour} P.,  {Bergeron} P.,  {Blouin} S.,  {Loranger} E.,  {Allard} N.~F.,   {Dunlap} B.~H.,  2019, \mn@doi [\apj] {10.3847/1538-4357/ab46b9}, \href {https://ui.adsabs.harvard.edu/abs/2019ApJ...885...74C} {885, 74}

\bibitem[\protect\citeauthoryear{{Cukanovaite}, {Tremblay}, {Freytag}, {Ludwig}, {Fontaine}, {Brassard}, {Toloza}  \& {Koester}}{{Cukanovaite} et~al.}{2019}]{cukanovaite19}
{Cukanovaite} E.,  {Tremblay} P.~E.,  {Freytag} B.,  {Ludwig} H.~G.,  {Fontaine} G.,  {Brassard} P.,  {Toloza} O.,   {Koester} D.,  2019, \mn@doi [\mnras] {10.1093/mnras/stz2656}, \href {https://ui.adsabs.harvard.edu/abs/2019MNRAS.490.1010C} {490, 1010}

\bibitem[\protect\citeauthoryear{{Cukanovaite}, {Tremblay}, {Toonen}, {Temmink}, {Manser}, {O'Brien}  \& {McCleery}}{{Cukanovaite} et~al.}{2023}]{cukanovaite2023}
{Cukanovaite} E.,  {Tremblay} P.~E.,  {Toonen} S.,  {Temmink} K.~D.,  {Manser} C.~J.,  {O'Brien} M.~W.,   {McCleery} J.,  2023, \mn@doi [\mnras] {10.1093/mnras/stad1020}, \href {https://ui.adsabs.harvard.edu/abs/2023MNRAS.522.1643C} {522, 1643}

\bibitem[\protect\citeauthoryear{{Cunningham}, {Tremblay}, {Freytag}, {Ludwig}  \& {Koester}}{{Cunningham} et~al.}{2019}]{cunningham19}
{Cunningham} T.,  {Tremblay} P.-E.,  {Freytag} B.,  {Ludwig} H.-G.,   {Koester} D.,  2019, \mn@doi [\mnras] {10.1093/mnras/stz1759}, \href {https://ui.adsabs.harvard.edu/abs/2019MNRAS.488.2503C} {488, 2503}

\bibitem[\protect\citeauthoryear{{Cunningham}, {Tremblay}, {Gentile Fusillo}, {Hollands}  \& {Cukanovaite}}{{Cunningham} et~al.}{2020}]{cunningham20}
{Cunningham} T.,  {Tremblay} P.-E.,  {Gentile Fusillo} N.~P.,  {Hollands} M.,   {Cukanovaite} E.,  2020, \mn@doi [\mnras] {10.1093/mnras/stz3638}, \href {https://ui.adsabs.harvard.edu/abs/2020MNRAS.492.3540C} {492, 3540}

\bibitem[\protect\citeauthoryear{{Cunningham} et~al.,}{{Cunningham} et~al.}{2021}]{cunningham2021}
{Cunningham} T.,  et~al., 2021, \mn@doi [\mnras] {10.1093/mnras/stab553}, \href {https://ui.adsabs.harvard.edu/abs/2021MNRAS.503.1646C} {503, 1646}

\bibitem[\protect\citeauthoryear{{Cunningham}, {Wheatley}, {Tremblay}, {G{\"a}nsicke}, {King}, {Toloza}  \& {Veras}}{{Cunningham} et~al.}{2022}]{cunningham2022}
{Cunningham} T.,  {Wheatley} P.~J.,  {Tremblay} P.-E.,  {G{\"a}nsicke} B.~T.,  {King} G.~W.,  {Toloza} O.,   {Veras} D.,  2022, \mn@doi [\nat] {10.1038/s41586-021-04300-w}, \href {https://ui.adsabs.harvard.edu/abs/2022Natur.602..219C} {602, 219}

\bibitem[\protect\citeauthoryear{{Cunningham}, {Tremblay}  \& {W. O'Brien}}{{Cunningham} et~al.}{2024}]{cunningham2024}
{Cunningham} T.,  {Tremblay} P.-E.,   {W. O'Brien} M.,  2024, \mn@doi [\mnras] {10.1093/mnras/stad3275}, \href {https://ui.adsabs.harvard.edu/abs/2024MNRAS.527.3602C} {527, 3602}

\bibitem[\protect\citeauthoryear{{Dufour} et~al.,}{{Dufour} et~al.}{2007}]{dufour07}
{Dufour} P.,  et~al., 2007, \mn@doi [\apj] {10.1086/518468}, \href {https://ui.adsabs.harvard.edu/abs/2007ApJ...663.1291D} {663, 1291}

\bibitem[\protect\citeauthoryear{{Dupuis}, {Fontaine}, {Pelletier}  \& {Wesemael}}{{Dupuis} et~al.}{1992}]{dupuis1992}
{Dupuis} J.,  {Fontaine} G.,  {Pelletier} C.,   {Wesemael} F.,  1992, \mn@doi [\apjs] {10.1086/191728}, \href {https://ui.adsabs.harvard.edu/abs/1992ApJS...82..505D} {82, 505}

\bibitem[\protect\citeauthoryear{{Farihi}}{{Farihi}}{2016}]{farihi2016review}
{Farihi} J.,  2016, \mn@doi [\nar] {10.1016/j.newar.2016.03.001}, \href {https://ui.adsabs.harvard.edu/abs/2016NewAR..71....9F} {71, 9}

\bibitem[\protect\citeauthoryear{{Gentile Fusillo} et~al.,}{{Gentile Fusillo} et~al.}{2019}]{ngf19}
{Gentile Fusillo} N.~P.,  et~al., 2019, \mn@doi [\mnras] {10.1093/mnras/sty3016}, \href {http://adsabs.harvard.edu/abs/2019MNRAS.482.4570G} {482, 4570}

\bibitem[\protect\citeauthoryear{{Gentile Fusillo} et~al.,}{{Gentile Fusillo} et~al.}{2021}]{Gentile2021}
{Gentile Fusillo} N.~P.,  et~al., 2021, \mn@doi [\mnras] {10.1093/mnras/stab2672}, \href {https://ui.adsabs.harvard.edu/abs/2021MNRAS.508.3877G} {508, 3877}

\bibitem[\protect\citeauthoryear{{Girven}, {Brinkworth}, {Farihi}, {G{\"a}nsicke}, {Hoard}, {Marsh}  \& {Koester}}{{Girven} et~al.}{2012}]{girven12}
{Girven} J.,  {Brinkworth} C.~S.,  {Farihi} J.,  {G{\"a}nsicke} B.~T.,  {Hoard} D.~W.,  {Marsh} T.~R.,   {Koester} D.,  2012, \mn@doi [\apj] {10.1088/0004-637X/749/2/154}, \href {https://ui.adsabs.harvard.edu/abs/2012ApJ...749..154G} {749, 154}

\bibitem[\protect\citeauthoryear{{Green} et~al.,}{{Green} et~al.}{2012}]{green2012COS}
{Green} J.~C.,  et~al., 2012, \mn@doi [\apj] {10.1088/0004-637X/744/1/60}, \href {https://ui.adsabs.harvard.edu/abs/2012ApJ...744...60G} {744, 60}

\bibitem[\protect\citeauthoryear{{Gurri}, {Veras}  \& {G{\"a}nsicke}}{{Gurri} et~al.}{2017}]{gurri2017}
{Gurri} P.,  {Veras} D.,   {G{\"a}nsicke} B.~T.,  2017, \mn@doi [\mnras] {10.1093/mnras/stw2293}, \href {https://ui.adsabs.harvard.edu/abs/2017MNRAS.464..321G} {464, 321}

\bibitem[\protect\citeauthoryear{{Harrison}, {Bonsor}  \& {Madhusudhan}}{{Harrison} et~al.}{2018}]{harrison2018}
{Harrison} J. H.~D.,  {Bonsor} A.,   {Madhusudhan} N.,  2018, \mn@doi [\mnras] {10.1093/mnras/sty1700}, \href {https://ui.adsabs.harvard.edu/abs/2018MNRAS.479.3814H} {479, 3814}

\bibitem[\protect\citeauthoryear{{Hollands}, {Tremblay}, {G{\"a}nsicke}, {Gentile-Fusillo}  \& {Toonen}}{{Hollands} et~al.}{2018}]{hollands18b}
{Hollands} M.~A.,  {Tremblay} P.~E.,  {G{\"a}nsicke} B.~T.,  {Gentile-Fusillo} N.~P.,   {Toonen} S.,  2018, \mn@doi [\mnras] {10.1093/mnras/sty2057}, \href {https://ui.adsabs.harvard.edu/abs/2018MNRAS.480.3942H} {480, 3942}

\bibitem[\protect\citeauthoryear{{Hurley}, {Pols}  \& {Tout}}{{Hurley} et~al.}{2000}]{hurley2000}
{Hurley} J.~R.,  {Pols} O.~R.,   {Tout} C.~A.,  2000, \mn@doi [\mnras] {10.1046/j.1365-8711.2000.03426.x}, \href {https://ui.adsabs.harvard.edu/abs/2000MNRAS.315..543H} {315, 543}

\bibitem[\protect\citeauthoryear{{Johnston}, {Pani{\'c}}  \& {Liu}}{{Johnston} et~al.}{2024}]{johnston2024}
{Johnston} H.~F.,  {Pani{\'c}} O.,   {Liu} B.,  2024, \mn@doi [\mnras] {10.1093/mnras/stad3254}, \href {https://ui.adsabs.harvard.edu/abs/2024MNRAS.527.2303J} {527, 2303}

\bibitem[\protect\citeauthoryear{{Jura}}{{Jura}}{2008}]{jura2008}
{Jura} M.,  2008, \mn@doi [\aj] {10.1088/0004-6256/135/5/1785}, \href {https://ui.adsabs.harvard.edu/abs/2008AJ....135.1785J} {135, 1785}

\bibitem[\protect\citeauthoryear{{Kenyon} \& {Bromley}}{{Kenyon} \& {Bromley}}{2017}]{kenyon2017a}
{Kenyon} S.~J.,  {Bromley} B.~C.,  2017, \mn@doi [\apj] {10.3847/1538-4357/aa7b85}, \href {https://ui.adsabs.harvard.edu/abs/2017ApJ...844..116K} {844, 116}

\bibitem[\protect\citeauthoryear{{Koester}}{{Koester}}{2009}]{koester09}
{Koester} D.,  2009, \mn@doi [\aap] {10.1051/0004-6361/200811468}, \href {http://adsabs.harvard.edu/abs/2009A%26A...498..517K} {498, 517}

\bibitem[\protect\citeauthoryear{{Koester}}{{Koester}}{2010}]{koester2010}
{Koester} D.,  2010, \memsai, \href {https://ui.adsabs.harvard.edu/abs/2010MmSAI..81..921K} {81, 921}

\bibitem[\protect\citeauthoryear{{Koester}, {G{\"a}nsicke}  \& {Farihi}}{{Koester} et~al.}{2014}]{koester14}
{Koester} D.,  {G{\"a}nsicke} B.~T.,   {Farihi} J.,  2014, \mn@doi [\aap] {10.1051/0004-6361/201423691}, \href {http://adsabs.harvard.edu/abs/2014A%26A...566A..34K} {566, A34}

\bibitem[\protect\citeauthoryear{{Koester}, {Kepler}  \& {Irwin}}{{Koester} et~al.}{2020}]{koester2020}
{Koester} D.,  {Kepler} S.~O.,   {Irwin} A.~W.,  2020, \mn@doi [\aap] {10.1051/0004-6361/202037530}, \href {https://ui.adsabs.harvard.edu/abs/2020A&A...635A.103K} {635, A103}

\bibitem[\protect\citeauthoryear{{Kunitomo}, {Ida}, {Takeuchi}, {Pani{\'c}}, {Miley}  \& {Suzuki}}{{Kunitomo} et~al.}{2021}]{kunitomo2021}
{Kunitomo} M.,  {Ida} S.,  {Takeuchi} T.,  {Pani{\'c}} O.,  {Miley} J.~M.,   {Suzuki} T.~K.,  2021, \mn@doi [\apj] {10.3847/1538-4357/abdb2a}, \href {https://ui.adsabs.harvard.edu/abs/2021ApJ...909..109K} {909, 109}

\bibitem[\protect\citeauthoryear{{Limoges}, {Bergeron}  \& {L{\'e}pine}}{{Limoges} et~al.}{2015}]{limoges15}
{Limoges} M.~M.,  {Bergeron} P.,   {L{\'e}pine} S.,  2015, \mn@doi [\apjs] {10.1088/0067-0049/219/2/19}, \href {https://ui.adsabs.harvard.edu/abs/2015ApJS..219...19L} {219, 19}

\bibitem[\protect\citeauthoryear{{McCleery} et~al.,}{{McCleery} et~al.}{2020}]{mccleery2020}
{McCleery} J.,  et~al., 2020, \mn@doi [\mnras] {10.1093/mnras/staa2030}, \href {https://ui.adsabs.harvard.edu/abs/2020MNRAS.499.1890M} {499, 1890}

\bibitem[\protect\citeauthoryear{{McDonough} \& {Sun}}{{McDonough} \& {Sun}}{1995}]{mcdonough1995}
{McDonough} W.~F.,  {Sun} S.~s.,  1995, \mn@doi [Chemical Geology] {10.1016/0009-2541(94)00140-4}, \href {https://ui.adsabs.harvard.edu/abs/1995ChGeo.120..223M} {120, 223}

\bibitem[\protect\citeauthoryear{{O'Brien} et~al.,}{{O'Brien} et~al.}{2023}]{OBrien2023}
{O'Brien} M.~W.,  et~al., 2023, \mn@doi [\mnras] {10.1093/mnras/stac3303}, \href {https://ui.adsabs.harvard.edu/abs/2023MNRAS.518.3055O} {518, 3055}

\bibitem[\protect\citeauthoryear{{O'Brien} et~al.,}{{O'Brien} et~al.}{2024}]{obrien2024}
{O'Brien} M.~W.,  et~al., 2024, \mn@doi [\mnras] {10.1093/mnras/stad3773}, \href {https://ui.adsabs.harvard.edu/abs/2024MNRAS.527.8687O} {527, 8687}

\bibitem[\protect\citeauthoryear{{Okuya}, {Ida}, {Hyodo}  \& {Okuzumi}}{{Okuya} et~al.}{2023}]{okuya2023}
{Okuya} A.,  {Ida} S.,  {Hyodo} R.,   {Okuzumi} S.,  2023, \mn@doi [\mnras] {10.1093/mnras/stac3522}, \href {https://ui.adsabs.harvard.edu/abs/2023MNRAS.519.1657O} {519, 1657}

\bibitem[\protect\citeauthoryear{{Ould Rouis} et~al.,}{{Ould Rouis} et~al.}{2024}]{OuldRouis2024}
{Ould Rouis} L.~B.,  et~al., 2024, \mn@doi [arXiv e-prints] {10.48550/arXiv.2410.06335}, \href {https://ui.adsabs.harvard.edu/abs/2024arXiv241006335O} {p. arXiv:2410.06335}

\bibitem[\protect\citeauthoryear{{Paquette}, {Pelletier}, {Fontaine}  \& {Michaud}}{{Paquette} et~al.}{1986}]{paquette86a}
{Paquette} C.,  {Pelletier} C.,  {Fontaine} G.,   {Michaud} G.,  1986, \mn@doi [\apjs] {10.1086/191111}, \href {http://adsabs.harvard.edu/abs/1986ApJS...61..177P} {61, 177}

\bibitem[\protect\citeauthoryear{{Paxton} et~al.,}{{Paxton} et~al.}{2015}]{paxton15}
{Paxton} B.,  et~al., 2015, \mn@doi [\apjs] {10.1088/0067-0049/220/1/15}, \href {https://ui.adsabs.harvard.edu/abs/2015ApJS..220...15P} {220, 15}

\bibitem[\protect\citeauthoryear{{Rappaport}, {Gary}, {Kaye}, {Vanderburg}, {Croll}, {Benni}  \& {Foote}}{{Rappaport} et~al.}{2016}]{rappaport2016}
{Rappaport} S.,  {Gary} B.~L.,  {Kaye} T.,  {Vanderburg} A.,  {Croll} B.,  {Benni} P.,   {Foote} J.,  2016, \mn@doi [\mnras] {10.1093/mnras/stw612}, \href {https://ui.adsabs.harvard.edu/abs/2016MNRAS.458.3904R} {458, 3904}

\bibitem[\protect\citeauthoryear{{Sahu} et~al.,}{{Sahu} et~al.}{2023}]{sahu2023}
{Sahu} S.,  et~al., 2023, \mn@doi [\mnras] {10.1093/mnras/stad2663}, \href {https://ui.adsabs.harvard.edu/abs/2023MNRAS.526.5800S} {526, 5800}

\bibitem[\protect\citeauthoryear{{Salpeter}}{{Salpeter}}{1955}]{salpeter1955}
{Salpeter} E.~E.,  1955, \mn@doi [\apj] {10.1086/145971}, \href {https://ui.adsabs.harvard.edu/abs/1955ApJ...121..161S} {121, 161}

\bibitem[\protect\citeauthoryear{{Subasavage} et~al.,}{{Subasavage} et~al.}{2017}]{subasavage2017}
{Subasavage} J.~P.,  et~al., 2017, \mn@doi [\aj] {10.3847/1538-3881/aa76e0}, \href {https://ui.adsabs.harvard.edu/abs/2017AJ....154...32S} {154, 32}

\bibitem[\protect\citeauthoryear{{Swan}, {Farihi}  \& {Wilson}}{{Swan} et~al.}{2019}]{swan19}
{Swan} A.,  {Farihi} J.,   {Wilson} T.~G.,  2019, \mn@doi [\mnras] {10.1093/mnrasl/slz014}, \href {https://ui.adsabs.harvard.edu/abs/2019MNRAS.484L.109S} {484, L109}

\bibitem[\protect\citeauthoryear{{Swan}, {Kenyon}, {Farihi}, {Dennihy}, {G{\"a}nsicke}, {Hermes}, {Melis}  \& {von Hippel}}{{Swan} et~al.}{2021}]{swan2021}
{Swan} A.,  {Kenyon} S.~J.,  {Farihi} J.,  {Dennihy} E.,  {G{\"a}nsicke} B.~T.,  {Hermes} J.~J.,  {Melis} C.,   {von Hippel} T.,  2021, \mn@doi [\mnras] {10.1093/mnras/stab1738}, \href {https://ui.adsabs.harvard.edu/abs/2021MNRAS.506..432S} {506, 432}

\bibitem[\protect\citeauthoryear{{Tassoul}, {Fontaine}  \& {Winget}}{{Tassoul} et~al.}{1990}]{tassoul90}
{Tassoul} M.,  {Fontaine} G.,   {Winget} D.~E.,  1990, \mn@doi [\apjs] {10.1086/191420}, \href {http://adsabs.harvard.edu/abs/1990ApJS...72..335T} {72, 335}

\bibitem[\protect\citeauthoryear{{Temmink}, {Toonen}, {Zapartas}, {Justham}  \& {G{\"a}nsicke}}{{Temmink} et~al.}{2020}]{temmink2020}
{Temmink} K.~D.,  {Toonen} S.,  {Zapartas} E.,  {Justham} S.,   {G{\"a}nsicke} B.~T.,  2020, \mn@doi [\aap] {10.1051/0004-6361/201936889}, \href {https://ui.adsabs.harvard.edu/abs/2020A&A...636A..31T} {636, A31}

\bibitem[\protect\citeauthoryear{{Tremblay} et~al.,}{{Tremblay} et~al.}{2020}]{tremblay2020}
{Tremblay} P.~E.,  et~al., 2020, \mn@doi [\mnras] {10.1093/mnras/staa1892}, \href {https://ui.adsabs.harvard.edu/abs/2020MNRAS.497..130T} {497, 130}

\bibitem[\protect\citeauthoryear{{Veras}}{{Veras}}{2016}]{veras16}
{Veras} D.,  2016, \mn@doi [Royal Society Open Science] {10.1098/rsos.150571}, \href {http://adsabs.harvard.edu/abs/2016RSOS....350571V} {3, 150571}

\bibitem[\protect\citeauthoryear{{Veras} \& {Heng}}{{Veras} \& {Heng}}{2020}]{veras2020-disk-lifetime}
{Veras} D.,  {Heng} K.,  2020, \mn@doi [\mnras] {10.1093/mnras/staa1632}, \href {https://ui.adsabs.harvard.edu/abs/2020MNRAS.496.2292V} {496, 2292}

\bibitem[\protect\citeauthoryear{{Veras} \& {Scheeres}}{{Veras} \& {Scheeres}}{2020}]{veras2020-yorp}
{Veras} D.,  {Scheeres} D.~J.,  2020, \mn@doi [\mnras] {10.1093/mnras/stz3565}, \href {https://ui.adsabs.harvard.edu/abs/2020MNRAS.492.2437V} {492, 2437}

\bibitem[\protect\citeauthoryear{{Veras}, {Jacobson}  \& {G{\"a}nsicke}}{{Veras} et~al.}{2014}]{veras2014}
{Veras} D.,  {Jacobson} S.~A.,   {G{\"a}nsicke} B.~T.,  2014, \mn@doi [\mnras] {10.1093/mnras/stu1926}, \href {https://ui.adsabs.harvard.edu/abs/2014MNRAS.445.2794V} {445, 2794}

\bibitem[\protect\citeauthoryear{{Veras}, {Carter}, {Leinhardt}  \& {G{\"a}nsicke}}{{Veras} et~al.}{2017}]{veras2017-WD1145}
{Veras} D.,  {Carter} P.~J.,  {Leinhardt} Z.~M.,   {G{\"a}nsicke} B.~T.,  2017, \mn@doi [\mnras] {10.1093/mnras/stw2748}, \href {https://ui.adsabs.harvard.edu/abs/2017MNRAS.465.1008V} {465, 1008}

\bibitem[\protect\citeauthoryear{{Veras}, {Tremblay}, {Hermes}, {McDonald}, {Kennedy}, {Meru}  \& {G{\"a}nsicke}}{{Veras} et~al.}{2020}]{veras2020-mass}
{Veras} D.,  {Tremblay} P.-E.,  {Hermes} J.~J.,  {McDonald} C.~H.,  {Kennedy} G.~M.,  {Meru} F.,   {G{\"a}nsicke} B.~T.,  2020, \mn@doi [\mnras] {10.1093/mnras/staa241}, \href {https://ui.adsabs.harvard.edu/abs/2020MNRAS.493..765V} {493, 765}

\bibitem[\protect\citeauthoryear{{Williams}, {Gaensicke}, {Swan}, {O'Brien}, {Izquierdo}, {Cutolo}  \& {Cunningham}}{{Williams} et~al.}{2024}]{williams2024}
{Williams} J.,  {Gaensicke} B.,  {Swan} A.,  {O'Brien} M.,  {Izquierdo} P.,  {Cutolo} A.-M.,   {Cunningham} T.,  2024, \mn@doi [arXiv e-prints] {10.48550/arXiv.2409.16046}, \href {https://ui.adsabs.harvard.edu/abs/2024arXiv240916046W} {p. arXiv:2409.16046}

\bibitem[\protect\citeauthoryear{{Wilson}, {Farihi}, {G{\"a}nsicke}  \& {Swan}}{{Wilson} et~al.}{2019}]{wilson2019}
{Wilson} T.~G.,  {Farihi} J.,  {G{\"a}nsicke} B.~T.,   {Swan} A.,  2019, \mn@doi [\mnras] {10.1093/mnras/stz1050}, \href {https://ui.adsabs.harvard.edu/abs/2019MNRAS.487..133W} {487, 133}

\bibitem[\protect\citeauthoryear{{Wyatt}, {Farihi}, {Pringle}  \& {Bonsor}}{{Wyatt} et~al.}{2014}]{wyatt14}
{Wyatt} M.~C.,  {Farihi} J.,  {Pringle} J.~E.,   {Bonsor} A.,  2014, \mn@doi [\mnras] {10.1093/mnras/stu183}, \href {http://adsabs.harvard.edu/abs/2014MNRAS.439.3371W} {439, 3371}

\bibitem[\protect\citeauthoryear{{Xu} et~al.,}{{Xu} et~al.}{2018}]{xu18}
{Xu} S.,  et~al., 2018, \mn@doi [\mnras] {10.1093/mnras/stx3023}, \href {http://adsabs.harvard.edu/abs/2018MNRAS.474.4795X} {474, 4795}

\bibitem[\protect\citeauthoryear{{Zotos} \& {Veras}}{{Zotos} \& {Veras}}{2020}]{zotos2020}
{Zotos} E.~E.,  {Veras} D.,  2020, \mn@doi [\aap] {10.1051/0004-6361/202037514}, \href {https://ui.adsabs.harvard.edu/abs/2020A&A...637A..14Z} {637, A14}

\bibitem[\protect\citeauthoryear{{Zuckerman}, {Koester}, {Reid}  \& {H{\"u}nsch}}{{Zuckerman} et~al.}{2003}]{zuckerman03}
{Zuckerman} B.,  {Koester} D.,  {Reid} I.~N.,   {H{\"u}nsch} M.,  2003, \mn@doi [\apj] {10.1086/377492}, \href {http://adsabs.harvard.edu/abs/2003ApJ...596..477Z} {596, 477}

\bibitem[\protect\citeauthoryear{{Zuckerman}, {Melis}, {Klein}, {Koester}  \& {Jura}}{{Zuckerman} et~al.}{2010}]{zuckerman10}
{Zuckerman} B.,  {Melis} C.,  {Klein} B.,  {Koester} D.,   {Jura} M.,  2010, \mn@doi [\apj] {10.1088/0004-637X/722/1/725}, \href {http://adsabs.harvard.edu/abs/2010ApJ...722..725Z} {722, 725}

\makeatother
\end{thebibliography}


\appendix

\section{Derivation of mathematical model}
\label{appendixA}

Following the framework developed by \citet{paquette86a, dupuis1992, koester09}, the solution to the differential equation governing the evolution of the surface abundance of a specific metal, $\chi_i$, is given as follows:
\begin{equation}
\chi_i(t) = \chi_i(t_0) e^{-t/\tau_i} + \frac{\tau_i \dot{M_i}}{M_{\rm cvz}} \left(1-e^{-t/\tau_i} \right)~,
\end{equation}
where $M_i$ and $\tau_i$ are the accretion rate and diffusion timescale of a specific metal, and $M_{\rm cvz}$ is the mass inside the convection zone. The time evolution of the surface abundance in our model can be characterised into times for which accretion is occurring ($t_0<t<t_{\rm d}$) and times when accretion is not occurring ($t_{\rm d}<t<t_p$). By imposing continuity across the regime when accretion is on and when it is off, the temporal evolution of the surface abundance of a given element is as follows:

\smallskip
\noindent
For $t_0\leq (t~\rm{mod}~t_{\rm p}) < t_{\rm d}$:
\begin{equation}  
\chi_i(t)= \chi_{0,i}e^{-t/\tau_i} + \chi_{{\rm ss},i} \left(1-e^{-t/\tau_i} \right)~,
\end{equation}

\smallskip
\noindent 
For $t_{\rm d} \leq (t~\rm{mod}~t_{\rm p}) < t_p$:
\begin{align}  
\chi_i(t) &= \chi_{0,i}e^{-t/\tau_i} + \chi_{{\rm ss},i}\left(1-e^{-t_{\rm d}/\tau_i} \right)e^{-(t-t_{\rm d})/\tau_i}\\
&=\left[ \chi_{0,i} + \chi_{{\rm ss},i}\left( e^{t_{\rm d}/\tau_i} -1 \right)\right]e^{-t/\tau_i}~.
\label{eq:chi-declining}
\end{align}
The modulo term in the above conditions represents the periodic boundary conditions used in our model. 

In Eq.\,\eqref{eq:chi-declining}, the steady state surface abundance, $\chi_{{\rm ss},i}$ is given by
\begin{equation}
    \chi_{{\rm ss},i} = \frac{\tau_i \dot{M_i}}{M_{\rm cvz}}~,
    \label{eq:chiSS}
\end{equation}
where $\dot{M_i}$ is the elemental specific accretion rate, which relates to the total accretion rate, $\dot{M}$, as 
\begin{equation}
    \dot{M_i} = f_{i}\dot{M}~.
    \label{eq:app-Mdoti}
\end{equation}
Throughout this study we focus solely on calcium as the proxy for accretion. This is largely because the vast majority of the metal-polluted white dwarfs detected in ground-based observational samples (including the 40\,pc sample studied in this work) are discovered via Ca\,{\sc h\,\&\,k} absorption. We make the assumption that the accreted material has bulk Earth composition, which corresponds to $f_{\rm Ca}$\,$=$\,$0.016$ in the equation above.

The times for which metals are detectable are defined by the times for which the surface abundance satisfies $\chi_i > \chi_{{\rm det},i}$. In the no-accretion phase, this time can be analytically determined via Eq.\,\ref{eq:chi-declining}.

\smallskip
\noindent 
For $t=t_{\rm z}$:
\begin{align}
    \chi_i(t_{\rm z}) &\equiv \chi_{{\rm det},i} \\
    &=\left[ \chi_{0,i} + \chi_{{\rm ss},i}\left( e^{t_{\rm d}/\tau_i} -1 \right)\right]e^{-t_{\rm z}/\tau_i}~.
\end{align}

\smallskip
\noindent 
Rearranging for $t_{\rm z}$ we can write
\begin{equation}
    t_{\rm z} = t_{\rm d} +  \tau_{i}\ln\left( \frac{\chi_{0,i}e^{-t_{\rm d}/\tau_i} + \chi_{{\rm ss}, i} \left( 1 - e^{-t_{\rm d}/\tau_i}\right)}{\chi_{{\rm det},i}} \right) \, .
    \label{eq:app-tz}
\end{equation}
This equation produces a valid solution, only if the maximum abundance exceeds the detection threshold. The maximum abundance at long disk lifetimes will be given by the steady state abundance. If the disk lifetime is shorter than the diffusion timescale, the maximum abundance will be given by the abundance at time $t=t_{\rm d}$, i.e.,
\begin{align}  
\chi_{{\rm max},i} &= \chi_{0,i}e^{-t_{\rm d}/\tau_i} + \chi_{{\rm ss},i} \left(1-e^{-t_{\rm d}/\tau_i} \right)~.
\label{eq:app-chi-max}
\end{align}
Thus, Eq.\,\eqref{eq:app-tz} is valid if $\chi_{{\rm max},i}>\chi_{{\rm det},i}$.

Now we must compute the time, $t_{\rm z}^*$, at which metals become detectable in the accreting phase ($t_0\leq (t~\rm{mod}~t_{\rm p}) < t_{\rm d}$). We will enforce periodic boundary conditions such that $\chi_i(t=t_0)\equiv\chi_i(t=t_{\rm p})$ for each accretion cycle. Thus, we must first find the abundance at $t_{\rm p}$.

\smallskip
\noindent 
For $t=t_{\rm p}$:
\begin{align}
    \chi_i(t_{\rm p}) &\equiv \chi_{{\rm p},i} \equiv \chi_{0,i}\notag\\
    &=\chi_{{\rm ss},i} \frac{\left(1-e^{-t_{\rm d}/\tau_i} \right)}{\left(1-e^{-t_{\rm p}/\tau_i} \right)}e^{-(t_{\rm p}-t_{\rm d})/\tau_i}\notag\\
    &=\chi_{{\rm ss},i} \left(\frac{e^{t_{\rm d}/\tau_i} -1}{e^{t_{\rm p}/\tau_i} -1}\right)
    \label{eq:chi0}~.
\end{align}

\smallskip
\noindent 
Now that we have the "initial" abundance, we want to find the earliest time at which $\chi_i(t)=\chi_{{\rm det},i}$.

\smallskip
\noindent 
For $t=t_{\rm z}^*$:
\begin{align}
    \chi_i(t_{\rm z}^*) &\equiv \chi_{{\rm det},i} \\
    &=\chi_{0,i} e^{-t_{\rm z}^*/\tau_i} + \chi_{{\rm ss},i}\left(1- e^{-t_{\rm z}^*/\tau_i}\right) 
\end{align}

Rearranging for $t_{\rm z}^*$, we can write
\begin{equation}
    t_{\rm z}^*=\tau_i\ln\left(\frac{\chi_{0,i} - \chi_{{\rm ss},i}}{\chi_{{\rm det},i}-\chi_{{\rm ss},i}}\right) \, .
\label{eq:app-tzstar}
\end{equation}
This equation is also only valid in the regime where $\chi_{{\rm max},i}>\chi_{{\rm det},i}$.

\smallskip
\noindent
The time-averaged probability of a given white dwarf having detectable metals is defined as $P_{\rm det}\equiv (t_{\rm z}-t_{\rm z}^*)/t_{\rm p}$, and can be expressed using Eqs.\,\eqref{eq:app-tz}\,\&\,\eqref{eq:app-tzstar} as

\begin{align}
    P_{\rm det} &\equiv \frac{t_{\rm z}-t_{\rm z}^*}{t_{\rm p}} \\
    &= \frac{\tau_i}{t_{\rm p}}\ln\left( \frac{\left[\chi_{0,i} + \chi_{{\rm ss}, i} \left( e^{t_{\rm d}/\tau_i} - 1\right)\right]\left(\chi_{{\rm det},i}-\chi_{{\rm ss},i}\right)}{\chi_{{\rm det},i}\left(\chi_{0,i} - \chi_{{\rm ss},i}\right)} \right)~.
    \label{eq:app-Pdet-full}
\end{align}

{\noindent}Since the times $t_{\rm z}$ and $t_{\rm z}^{*}$ are only defined for $\chi_{{\rm max},i}>\chi_{{\rm det},i}$, with $\chi_{{\rm max},i}$ defined by Eq.\,\eqref{eq:app-chi-max} and $\chi_{{\rm det},i}$ determined from model atmosphere predictions of the metal absorption line equivalent width, the Eq.\,\eqref{eq:app-Pdet-full} is also only valid for $\chi_{{\rm max},i}>\chi_{{\rm det},i}$.

\smallskip
\noindent
This equation comprises three independent variables; the disk lifetime ($t_{\rm d}$), total accretion period ($t_{\rm p}$), and total accretion rate ($\dot{M}$), where $\dot{M}$ is found within Eq.\,\ref{eq:chiSS}. There are an additional three parameters with values set by the specific white dwarf parameters (effective temperature and surface gravity). Those three parameters are the diffusion timescale for a given element ($\tau_i$), convection zone mass ($M_{\rm cvz}$), and observational detection limit ($\chi_{{\rm det},i}$).

\subsection{Limits}
\label{sec:appendix-limits}
In this section we consider some limiting behaviours of the detection probability as defined by Eq.\,\eqref{eq:app-Pdet-full}. This approach allows to extract relations of the accretion period as a function of disk lifetime. There are three key features that define the morphology of the solution, namely; i) the plateau at near-constant $t_{\rm p}$, ii) the up turn at long disk lifetimes, and iii) the down turn at short disk lifetimes. We consider these three features in the following.

\subsubsection{Plateau}
\noindent
In the limit that $t_{\rm p} \gg \tau_i,t_{\rm d}$  and $\chi_{0,i} \to 0$,
Eq.\,\eqref{eq:app-Pdet-full} can be rearranged to express the accretion period $t_{\rm p}$ as an analytic function of the disk lifetime, $t_{\rm d}$, diffusion timescale, $\tau_i$, detection threshold, $\chi_{\rm det}$, and (through $\chi_{{\rm ss},i}$) accretion rate, $\dot{M}_i$, and convection zone mass, $M_{\rm cvz}$, as
\begin{equation}
    t_{\rm p} = \frac{ t_{\rm d} + \tau_i \ln\left(\frac{(\chi_{{\rm ss},i} - \chi_{{\rm det},i})(1-e^{-t_{\rm d}/\tau_i})}{\chi_{{\rm det},i}}\right)}{P_{\rm det}}\, .
    \label{eq:tp-analytic}
\end{equation}
If all white dwarfs in the synthetic population share the same parameters, then the metal-polluted fraction, $f_{\rm z}$ is equal to the detection probability. In Fig.\,\ref{fg:pop-analytic} we show the solution to the above equation for the H- and He-atmosphere populations. We adopt the approximate parameters ($\tau_i$, $M_{\rm cvz}$, and $\chi_{\rm det}$) for the canonical white dwarf in each sample with $M_{\rm wd}$$\,$$\approx$\,0.65\,M$_{\odot}$ and $T_{\rm eff}$$\,$$\approx$$\,$$5000$\,K. We also adopt an accretion rate of $10^8$\,g/s. 

\subsubsection{Up-turn at long disk lifetimes}
In the limit that $t_{\rm d}\gg \tau_i$, the accretion period expressed in Eq.\,\eqref{eq:tp-analytic}, becomes linear with disk lifetime, such that 
\begin{equation}
    t_{\rm p} = \frac{t_{\rm d}}{P_{\rm det}}~.
\end{equation}
Since the mean detection probability across the full population is exactly the metal-polluted fraction, this equation can be equivalently written in terms of the metal-polluted fraction, $f_{\rm z}$, as
\begin{equation}
    t_{\rm p} = \frac{t_{\rm d}}{f_{\rm z}}\, .
    \label{eq:app-tp-analytic-limit-long-fz}
\end{equation}
We show this relation in Fig.\,\ref{fg:pop-analytic} for the H-atmosphere simulation, in the regime where $t_{\rm d}\gg \tau$ and $\chi_{{\rm ss},i}\gg \chi_{{\rm det},i}$. Although it is not shown in the figure, the He-atmosphere simulation also tends to the above relation for $t_{\rm d}>10^7$\,years. The longer disk times required for this relation to hold in the He sample is due to the longer diffusion timescales of He-atmosphere white dwarfs.

\subsubsection{Down-turn at short disk lifetimes}
In the He-atmosphere analysis, a turn down is apparent for disk lifetimes shorter than 100\,years, and we identify that the specific shape of this slope is the primary functional behaviour driving the point of intersection with the H simulation. We note this steeper linear regime occurs for the He simulation in the limit that $t_{\rm d}\ll\tau$.

\begin{figure}
	\centering

 	\subfloat{\includegraphics[width=1.\columnwidth]{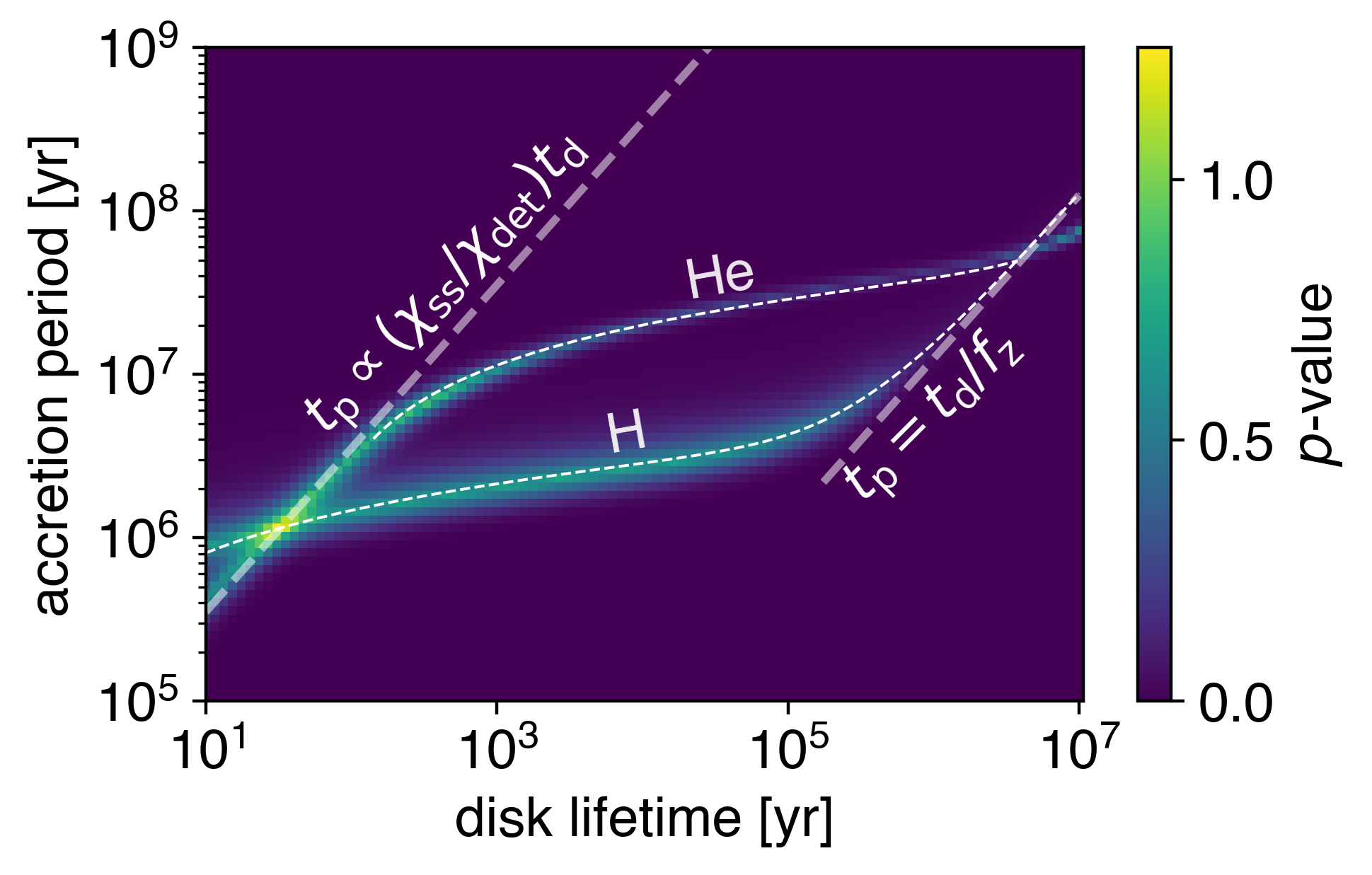}}
	\caption{The colours show the summed $p$-values from Fig.\,\ref{fg:pop-vary-Mdot-H-He}, for the population synthesis accretion model of the H- and He-atmosphere samples. The dotted white lines show the analytic accretion period as a function of disk lifetime from Eq.\,\eqref{eq:app-Pdet-full}. We show two curves, one of the H-atmosphere sample with a metal-polluted fraction of $f_{\rm z}=0.08$, and the other for the He-atmosphere sample with metal-polluted fraction 0.32. For the analytic curves, we adopt the atmosphere parameters for the canonical white dwarf (most frequent mass and temperature), and a single accretion rate of $10^8$\,g/s. The dashed white lines indicate two linear limits of our model. The linear relation at short disk lifetimes corresponds to Eq.\eqref{eq:tp-analytic-steep}. The linear relation at long disk lifetimes corresponds to Eq.\eqref{eq:app-tp-analytic-limit-long-fz} for the H sample polluted fraction of $f_{\rm z}=0.08$.}
	\label{fg:pop-analytic}
\end{figure}

In the limit of short disk lifetime, we can expand Eq.~\eqref{eq:chi0} in the small parameter $t_{\rm d}/\tau_i$ to obtain
\begin{equation}
\chi_{0,i} \to \frac{t_{\rm d}}{t_{\rm p}} \chi_{{\rm ss},i}~.
\end{equation}
We can use this while also expanding Eq.~\eqref{eq:app-tz} in $t_{\rm d}/\tau_i \ll 1$ to obtain
\begin{equation}
t_{\rm z} \to \tau_{i}\ln\left[ \frac{\chi_{{\rm ss}, i}}{\chi_{{\rm det},i}}\left(\frac{t_{\rm d}}{t_{\rm p}} + \frac{t_{\rm d}}{\tau_i} \right) \right]~.
\end{equation}
Because $t_{\rm z}^* < t_{\rm d}$, we take $t_{\rm z}^* \to 0$ to write
\begin{equation}
    P_{\rm det} \to \frac{t_{\rm z}}{t_{\rm p}}
    \to \frac{\tau_{i}}{t_{\rm p}}\ln\left[ \frac{\chi_{{\rm ss}, i}}{\chi_{{\rm det},i}}\left(\frac{t_{\rm d}}{t_{\rm p}} + \frac{t_{\rm d}}{\tau_i} \right) \right]~.
\end{equation}
Solving for $t_{\rm d}$ then yields
\begin{equation}
    t_{\rm d} \approx
     \left(\frac{t_{\rm p}\tau_i}{t_{\rm p} + \tau_i}\right) \frac{\chi_{{\rm det},i}}{\chi_{{\rm ss}, i}}\exp\left(P_{\rm det}\frac{t_{\rm p}}{\tau_{i}}\right)
\end{equation}
For long enough diffusion timescales that the accretion period satisfies $t_{\rm p} \lesssim \tau_i$ (applicable to WDs with He atmospheres), the exponential term will be order unity, and so this reduces to a useful order-of-magnitude estimate for the accretion period
\begin{align}
    t_{\rm p}&\sim t_{\rm d}\frac{\chi_{{\rm ss},i}}{\chi_{{\rm det},i}}\\
    &=t_{\rm d}\frac{\tau_i\dot{M_i}}{M_{\rm cvz}\chi_{{\rm det},i}}~,
    \label{eq:tp-analytic-steep}
\end{align}
We show this relation in Fig.\,\ref{fg:pop-analytic} with the white dashed line on the left of the plot.
On the other hand, if the accretion period is instead much longer than the diffusion timescales, we obtain
\begin{equation}
    t_{\rm p} \to \frac{\tau_i}{P_{\rm det}} \ln \left( \frac{t_{\rm d}}{\tau_i} \frac{\chi_{{\rm ss}, i}}{\chi_{{\rm det},i}} \right)~.
\end{equation}

We note that there are in fact two intercepting solutions for the H and He samples. One solution occurs where the He sample is in the down-turn regime at short disk lifetimes, and the other occurs where the H sample enters the linear regime at long disk lifetimes. If the only constraint available was the observed polluted fraction, these solutions would be degenerate. However, the KS test comparing the observed and synthetic polluted mass distributions leads to the long disk lifetime solution being ruled out. It is for this reason that in Fig.\,\ref{fg:pop-vary-Mdot-H-He} the solutions are not seen to overlap at the long disk lifetime solution.

\bsp	
\label{lastpage}
\end{document}